\documentclass[fleqn,10pt]{article}

\pdfoutput=1

\usepackage{graphicx}
\usepackage{latexsym}
\usepackage{amsbsy}
\usepackage{amsfonts}
\usepackage{amssymb}
\usepackage{amscd}
\usepackage{amsmath}
\usepackage{color}
\usepackage{cite}
\usepackage{rotating}

\usepackage{caption}
\usepackage{cases}
\usepackage{pst-all}

\usepackage[utf8]{inputenc} 
\usepackage[T1]{fontenc}

\newcommand{\sq}{\hbox{\rlap{$\sqcap$}$\sqcup$}}
\newcommand{\qed}{\ifmmode\sq\else{\unskip\nobreak\hfil
  \penalty50\hskip1em\null\nobreak\hfil\sq
  \parfillskip=0pt\finalhyphendemerits=0\endgraf}\fi{}}

\def\cha{\kern .6em{\sqcup \kern -1.12em \sqcup}\kern .6em}

\def\D{{\rm d}}
\def\E{{\rm e}}
\def\I{{\rm i}}
\def\vec{\boldsymbol}
\def\q{\rm} %POUR ECRIRE LES QUATERNIONS DE BASES EN ROMAIN
\def\qv{\bf} % IDEM MAIS EN GRAS
\def\rap{\!\!\!\! } %POUR RAPPROCHER LES MEMBRES D'UNE EQUATION DANS eqnarray
\def\its{\sl}
\def\ket#1{|#1\rangle}

\def\CD{\hbox{
$\bigcirc {\kern-.75em { 
\hbox{$\buildrel > \over {\buildrel \over{}}$}}}$}}
\def\CG{\hbox{
$\bigcirc {\kern-.75em { 
\hbox{$\buildrel < \over {\buildrel \over{}}$}}}$}}
\def\boxlim#1#2{\buildrel\hbox{\scriptsize $#1$}\over {\hbox{\scriptsize $#2$}}}%SERT A LA DEFINITION SUIVANTE
\def\doublelimite#1#2{\lim_{\boxlim{#1}{#2}}}%PERMET DE METTRE DEUX LIGNES SOUS LE SIGNE LIMITE

\mathchardef\Gamma="0100
\mathchardef\Delta="0101
\mathchardef\Theta="0102
\mathchardef\Lambda="0103
\mathchardef\Xi="0104
\mathchardef\Pi="0105
\mathchardef\Sigma="0106
\mathchardef\Upsilon="0107
\mathchardef\Phi="0108
\mathchardef\Psi="0109
\mathchardef\Omega="010A
\def\appendix{\par
  \setcounter{section}{0}%
  \def\thesection{Appendix \Alph{section}}}
\newtheorem{theorem}{Theorem}[section]{\bf}{\sl}
\newtheorem{proposition}{Proposition}[section]{\bf}{\its}
{\bf}{\it} 
{\bf}{\it} 
{\sl}{\rm}
\newtheorem{remark}{Remark}[section]{\sl}{\rm}
\renewcommand{\thefootnote}{\fnsymbol{footnote}}
\begin{document}

\title{Quaternions and  rotations: applications to \\ Minkowski's four-vectors, electromagnetic waves\\ and  polarization optics}

\date{}
\maketitle

\begin{center}

\vskip -1.2cm {
  \renewcommand{\thefootnote}{}
  
 {\bf   Pierre Pellat-Finet%\footnote{\hskip -.53cm Laboratoire de Math\'ematiques de Bretagne Atlantique (LMBA) UMR CNRS 6205,

%\noindent Universit\'e de Bretagne Sud, CS 60573, 56017 Vannes, France.

%\noindent pierre.pellat-finet@univ-ubs.fr     \hfill \today
%}
}
}
\setcounter{footnote}{0}

\medskip
    {\sl \small Université Bretagne Sud,  UMR CNRS 6205, LMBA, F-56000 Vannes, France

       pierre.pellat-finet@univ-ubs.fr  
}
 
\end{center}

\vskip .2cm
\begin{center}
  \begin{minipage}{12.3cm}
    \hrulefill

\smallskip
{\small
  {\bf Abstract.} Rotations on the 3-dimensional Euclidean vector-space can be represented by real quaternions, as was shown by Hamilton. 
Introducing complex quaternions  allows us to extend the result to elliptic and hyperbolic rotations on the Minkowski space, that is, to proper Lorentz rotations.
Another generalization deals with complex rotations on the complex-quaternion algebra.
Appropriate quaternionic expressions of differential operators lead to a quaternionic form of Maxwell equations;  the quaternionic expressions of an electromagnetic-field  in two Galilean frames in relative motion are linked by a complex rotation;
some relativistic invariants of electromagnetic waves are deduced, including their polarization states
  and their degrees of polarization. Quaternionic forms of proper Lorentz rotations are applied to polarization optics along with  illustrative examples. Equivalences between relativistic effects and crystal optics effects are mentioned. A derivation of the triangle inequality on the Minkowski space is given; the inequality is illustrated by some properties of partially polarized lightwaves. 
}

{\small

\bigskip
\noindent {\bf Content}  

\smallskip

\noindent 1. Introduction\dotfill \pageref{sect1} \\
\noindent 2. Quaternions, rotations and Minkowski space\dotfill \pageref{sect2}\\
\noindent 3. Electromagnetism\dotfill \pageref{sect3} \\
\noindent 4. Electromagnetic waves\dotfill\pageref{sect4}\\
\noindent 5. Quaternionic representation of polarization optics\dotfill \pageref{sect5}\\
\noindent 6. Some equivalences between special theory of relativity and polarization optics\dotfill \pageref{sect6}\\
\noindent 7. Triangle inequality and applications\dotfill \pageref{sect7}\\
\noindent 8. Conclusion\dotfill \pageref{conc}\\
\noindent Appendix A. Quaternionic trigonometry\dotfill \pageref{appenA}\\
\noindent Appendix B. Proof of $V={\q e}_u\,\sin (\psi /2)$ (Sect.\ \ref{sect258})\dotfill \pageref{appenB}\\
\noindent Appendix C. Mappings $q\,\longmapsto\, u\,q\,u^*$ are rotations on ${\mathbb H}_{\rm c}$\dotfill \pageref{appenC}\\
\noindent Appendix D.$\!$ Two lemmas on unit pure quaternions\dotfill \pageref{appenD}\\
\noindent Appendix E. Regular rotations: eigenquaternions and eigenvalues\dotfill \pageref{appenE}\\
\noindent Appendix F. Singular Lorentz rotations: eigenminquats\dotfill \pageref{appenF}\\
\noindent Appendix G.$\!$ Equivalent birefringence and optical activity\dotfill \pageref{appenG}\\
\noindent References\dotfill \pageref{ref2}

}
\hrulefill
\end{minipage}
\end{center}

\newpage

{\small
%*******************
  \section*{Notations}\label{not}
%*******************

\begin{tabbing}
Largeurpl1234567 \= \kill
${\mathbb R}$ \> Field of real numbers  \\
${\mathbb C}$ \> Field of complex numbers  \\
${\mathbb E}_3$ \> 3--dimensional Euclidean vector-space \\
${\mathbb H}$ \> Algebra of real quaternions (Hamilton's quaternions) \\
${\mathbb H}_{\rm c}$ \> Algebra of complex quaternions (Hamilton's biquaternions) \\
\hskip -.23cm $\left.\begin{array}{l}({\mathbb R}^4,Q) \\ ({\mathbb R}\oplus {\mathbb E}_3,Q)\end{array}\right\}$ \> Minkowski  space ($Q$ denotes the Lorentz quadratic form, or Lorentz metric) \\
${\mathbb M}$ \> Vector space of minquats \\
%$\isomor$ \> Isomorphic \\   %\cong
$\I$ \> Complex number such that $\I^2=-1$\\
$\{ {\q e}_0,{\q e}_1,{\q e}_2,{\q e}_3\}$\> Canonic basis of ${\mathbb H}_{\rm c}$ (also canonic basis of ${\mathbb H}$) \\
${\q e}_n$ \> Real unit pure quaternion (${{\q e}_n}^{\! 2}=-{\q e}_0=-1$) \\
$\overline z$ \> Complex conjugate of z \\
$q^*$ \> Quaternionic (or Hamilton) conjugate of $q$ \\
$N(q)$\> Norm of quaternion $q$  ($N(q)=q\,q^*$)\\
$q\vec\cdot r$ \> Dot  (or scalar) product of quaternions $q$ and $r$ \\
$\vec V$ \> Vector of the 3--dimensional Euclidean vector-space ${\mathbb E}_3$ \\
$V$ \> Pure quaternion representing vector $\vec V$ \\
$\|\vec V\|$ \> Euclidean norm of vector $\vec V$\\
$\langle \vec V,  \vec W\rangle $ \> Euclidean scalar product of vectors $\vec V$ and $\vec V$\\
$\vec V\hskip .21ex\vec \times \hskip .15ex \vec W$\> Vector (or cross) product of vectors $\vec V$ and $\vec W$\\
$V\,\vec \times \,W $ \> Vector (or cross)  product of pure quaternions $V$ and $W$\\
$\vec{OP}$ \> Vector with origin $O$ and endpoint $P$\\
$\hskip -.2ex \vec \nabla$ \> Nabla operator \\
$\nabla$ \> Quaternionic nabla operator \\
${\rm c}$ \> Light velocity (in vaccum)\\
\end{tabbing}
\vskip -.3cm
\noindent For coordinates, for vector or quaternion components, a Latin index, like $j$ in $A_j$ or in $x_j$, takes the values $1,2,3$; a Greek index, like $\mu$ in $x_\mu$, takes the values $0,1,2,3$.

\noindent We use $0$ as a universal symbol; for example, we generally write $\vec V=0$ in place of $\vec V=\vec 0$, to indicate that $\vec V$ is a zero-vector.

}

%**********************************
\section{Introduction}\label{sect1}
%**********************************

In his famous {\em A treatise on electricity and magnetism}, published in 1873,  Maxwell used quaternions, invented  by Hamilton in 1843,  to represent electromagnetic fields. Nevertheless, the parallel development of vector calculus by Grossmann and Cayley led Heaviside to eventually express Maxwell equations with the help of divergence and curl operators, and utilizing quaternions for such applications was almost forgotten. Some years later, around 1912, quaternions were
employed by  Silberstein \cite{Con,Sil1,Sil2} for expressing the special theory of relativity; but for subsequent developments and  descriptions of the theory, they were prefered by vector calculus, once more. Quantum mechanics put matrix calculus in the foreground and reinforced the use of vectors in physics, that is, linear algebra in its standard form (no quaternions!). As a result, quaternions are not very known to physicists, who generally prefer to employ vector or matrix calculus. To be complete, we should mention an original way of expressing Lorentz transformations with quaternions, proposed by Dirac \cite{Dir}; and some attempts to apply quaternions to a space-time description of quantum phenomena by Kwal \cite{Kwa1,Kwa2} and  in polarization optics in a graphical way by Cernosek \cite{Cer,The}. Nevertheless, in 1972, J. L. Synge used quaternions in Relativity again by introducing minquats \cite{Syn}, that is, complex quaternions suited to  the Minkowski  space. Independently, in 1982--1983, complex quaternions (that are equivalent to Synge's minquats) were shown to be fruitful in polarization optics too \cite{PPF0,PPF2,PPF5}, with effective applications to polarization mode dispersion as encountered in optical communications \cite{Bruy,PPF11}.  

The above is focused on applications of quaternions to physics. But of course quaternions have also been of interest in mathematics since 1843, in particular in arithmetic \cite{Vig}.
For a broad mathematical overview of quaternion algebras, let us mention the recent book of Voight \cite{Voi}.

Maybe there is  a new interest today in using quaternions for various applications (other than mathematics), such as cristallography  and chemistry \cite{Alt}, molecular modelling \cite{Kam}, quantum physics \cite{Adl}, mechanics \cite{Mah} and relativistic physics \cite{Gir}. Quaternions are also useful  in robotics and for video game softwares: indeed, their hability to express rotations in the 3--dimensional Euclidean space  has been shown to avoid the gimbal lock \cite{Ozc}.

In this article, we will mainly focus on a subset of complex quaternions, the subset of minquats \cite{Syn}, which is well suited to representing 4--dimensional vectors of the Minkowski  space, that is, ${\mathbb R}^4$ equipped with the Lorentz quadratic form. Moreover, proper Lorentz rotations form the group SO$_+(1,3)$, which operates on the Minkowski  space, and can also be represented by complex quaternions. Those structures play and important and well-known role in the special theory of relativity \cite{Syn}, and offer some interest in polarization optics too \cite{PPF3,PPF4,PPF6,PPF7,PPF10}. 

A part of the present article  constitutes an attempt to show how quaternions can  be used for expressing Maxwell equations in a very synthetical way and for deriving how electromagnetic fields are transformed under  Lorentz boosts \cite{Gir}. The results are then illustrated by proving
the relativistic invariance of polarizations of lightwaves, including their degrees of polarization.

Another part of the article deals with a quaternionic representation of polarization optics and complements some previous studies on the subject \cite{PPF2,PPF5,PPF10}. Polarization states are represented by minquats and polarization operators by complex quaternions, which generally correspond to proper Lorentz rotations on the Minkowski  space of minquats.

We eventually point out some effects that take place in Relativity and in polarization optics and that are similar to each other, as already shown in a previous article \cite{PPF10}. The deep reason or their similarity, indeed, is that they are physical translations in the relativistic  or in the polarization optics areas, of geometrical properties of the Minkowski  space.  In the present article we make an effort to study the geometrical behaviours of the Minkowski  space first, and then to examine how they appear in Relativity and in polarization optics.

%******************************************************************
\section{Quaternions, rotations and Minkowski  space}\label{sect2}
%******************************************************************

\subsection{Quaternions}%********************************************************

\subsubsection{Complex quaternion algebra}

The space ${\mathbb C}^4$ is a four-dimensional complex vector-space whose canonical basis is denoted $\{{\q e}_0,{\q e}_1,{\q e}_2,{\q e}_3\}$, where
\begin{equation}
{\q e}_0=(1,0,0,0)\,,\hskip .5cm
{\q e}_1=(0,1,0,0)\,,\hskip .5cm
{\q e}_2=(0,0,1,0)\,,\hskip .5cm
{\q e}_3=(0,0,0,1)\,.\end{equation} 
We introduce a {\bf quaternionic product}, as defined in Table 1, where the first element of the product is written in the left column and the second element in the upper row (for example: ${\q e}_1{\q e}_2={\q e}_3=-{\q e}_2{\q e}_1$).
\begin{table}[h]
\begin{center}
\begin{tabular}{c|rrrr}
 &${\q e}_0$& ${\q e}_1$&${\q e}_2$&${\q e}_3$ \\
  \hline
${\q e}_0$&${\q e}_0$ &${\q e}_1$& ${\q e}_2$ &${\q e}_3$\\
${\q e}_1$&${\q e}_1$ &$-{\q e}_0$& ${\q e}_3$ &$-{\q e}_2$\\
${\q e}_2$&${\q e}_2$ &$-{\q e}_3$& $-{\q e}_0$ &${\q e}_1$\\
${\q e}_3$&${\q e}_3$ &${\q e}_2$& $-{\q e}_1$ &$-{\q e}_0$\\
\end{tabular}
\caption{\small Quaternionic product.}\label{table1}
\end{center}
\end{table}

The quaternionic product is distributive with respect to addition; it is non-commu\-ta\-tive. When equipped with the quaternionic product, the vector space  ${\mathbb C}^4$ becomes the {\bf complex quaternion algebra}, denoted by ${\mathbb H}_{\rm c}$. (Complex quaternions were called bi-quaternions by Hamilton.) Thus, a (complex) quaternion is written as
\begin{equation}
q=q_0{\q e}_0+q_1{\q e}_1+q_2{\q e}_2+q_3{\q e}_3\,,\end{equation}
where the $q_\mu$'s are complex numbers. An explicit expression of the product $q\,r$ of quaternions $q$ and $r$ is given later.

If $\overline z$ denotes the complex conjugate of the complex number $z$, the {\bf complex conjugate} of the quaternion $q$ above  is $\overline{q}$, defined by
\begin{equation}
  \overline{q}=\overline{q}_0{\q e}_0+\overline{q}_1{\q e}_1+\overline{q}_2{\q e}_2+\overline{q}_3{\q e}_3\,.\end{equation}
For every quaternion $q$ and every quaternion $r$, we have $\overline{q\,r}=\overline q\,\overline r$.

The {\bf Hamilton conjugate} of $q$ is $q^*$, such that
\begin{equation}
  q^*=q_0{\q e}_0-q_1{\q e}_1-q_2{\q e}_2-q_3{\q e}_3\,.\end{equation}
It can be proved that $(qr)^*=r^*q^*$, for all quaternions $q$ and $r$.
Complex and Hamilton conjugations commute: $(\overline{q})^*=\overline{q^*}$, for every quaternion $q$.

The {\bf scalar part} of $q$ is $q_0{\q e}_0$ and its {\bf vector part} is $q_1{\q e}_1+q_2{\q e}_2+q_3{\q e}_3$. The quaternion $q$ is a {\bf pure quaternion},  if $q_0=0$. A quaternion $q$ is a pure quaternion if, and only if, $q^*=-q$. Every quaternion $q$ can be written
\begin{equation}
  q=q_0{\q e}_0+V\,,\end{equation}
where $V$ is a pure  quaternion.

The (quaternionic) {\bf norm} of $q$ is $N(q)$, defined by \cite{Deh}
\begin{equation}
N(q)=qq^*=(q_0)^2+(q_1)^2+(q_2)^2+(q_3)^2\,.\end{equation}
Generally $N(q)$ is a complex number ($N$ is not a norm in the sense of mathematical analysis). We note that $N(q)=N(q^*)$ and $N(\overline{q})=\overline{q}\;\overline{q}^*=\overline{q}\,\overline{q^*}=\overline{q\,q^*}=\overline{N(q)}$.
The norm is multiplicative:
\begin{equation}
  N(q\,r)=N(q)N(r)\,,\end{equation}
and such that $ N(q\,r)=N(r\,q)$, for all quaternions $q$ and $r$.

The {\bf dot product} (or {\bf scalar product}) of quaternions is the {\bf polar form} of the quaternionic norm, defined for $q=q_0{\q e}_0+q_1{\q e}_1+q_2{\q e}_2+q_3{\q e}_3$ and $r=r_0{\q e}_0+r_1{\q e}_1+r_2{\q e}_2+r_3{\q e}_3$  by
\begin{equation}
q\vec\cdot r=q_0r_0+q_1r_1+q_2r_2+q_3r_3\,,\end{equation}
so that $N(q)=q\vec\cdot q$.
The dot product of quaternions is commutative: $q\vec\cdot r=r\vec\cdot q$\,; it is such that
\begin{equation}
  q\vec \cdot r={1\over 2}\bigl[N(q+r)-N(q)-N(r)\bigr]\,.\label{eq8}\end{equation}
For $\mu$ and $\nu$ in $\{0,1,2,3\}$, we have ${\q e}_\mu\vec\cdot{\q e}_\mu =1$, and ${\q e}_\mu\vec\cdot{\q e}_\nu =0$, for $\nu\ne \mu$. Quaternions $q$ and $r$ are said to be {\bf orthogonal} if, and only if, $q\vec\cdot r=0$.

 A quaternion $u$ is {\bf unitary} if $N(u)=1$, that is, if $uu^*=1$ ($u$ is also called a {\bf unit quaternion}).
If $N(q)\ne 0$, the quaternion $q$ has an {\bf inverse} $q^{-1}$ for the quaternionic product:
\begin{equation}
q^{-1}={q^*\over N(q)}\,.\end{equation}
Some quaternions, for example ${\q e}_0+\I\, {\q e}_1$, are {\bf zero divisors} and have no inverses, since their norms are zero. They will be helpful in representing null 4-vectors in Relativity as well as completely polarized states and polarizers in polarization optics.

\begin{remark} {\rm According to some authors \cite{Syn}, a quaternion $u$ is unitary if $N(u)=\pm 1$. Then a quaternion $u$ is a positive unit quaternion, if $N(u)=1$, and a negative unit quaternion, if $N(u)=-1$. A negative unit quaternion $u$ ($N(u)=-1$) takes necessarily the form $u=\I\, v$, where $v$ is a positive unit quaternion ($N(v)=1$).
    For example ${\q e}_\mu$ ($\mu\in\{0,1,2,3\}$) is a positive unit quaternion, whereas $\I\,{\q e}_\mu$ is a negative unit quaternion.

    Unit quaternions as previously defined in this article are then positive unit quaternions.
    }
\end{remark}

\subsubsection{Practical analytic derivations}\label{sect212}%****************************

We write the product of quaternions $q$ and $r$ as
\begin{equation}
q\,r=(q_0{\q e}_0+q_1{\q e}_1+q_2{\q e}_2+q_3{\q e}_3)(r_0{\q e}_0+r_1{\q e}_1+r_2{\q e}_2+r_3{\q e}_3)\,,\end{equation}
and the result is obtained by applying Table \ref{table1} and using the  distributivity of the quaternionic product with respect to addition.  Explicitly we obtain
\begin{eqnarray}
  q\,r\rap&=&\rap(q_0r_0 -q_1r_1  -q_2r_2 -q_3 r_3)\,{\q e}_0
+(q_0r_1+q_1r_0+q_2r_3-q_3r_2)\,{\q e}_1\nonumber \\
&& \hskip 1cm + (q_0r_2+q_2r_0-q_1r_3+q_3r_1)\,{\q e}_2 +
(q_0r_3+q_3r_0+q_1r_2-q_2r_1)\,{\q e}_3\,.\label{eq10n}
\end{eqnarray}

\begin{figure}[h]%$$$$$$$$$$$$$$$$$$$$$$$$$$$$$$$$$$$$$$$$$$$$$$$$$$$$$$$
  \begin{center}
  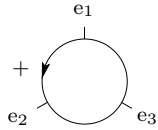
  \caption{\small For $\{ i,j,k\}= \{1,2,3\}$,
    we have ${\q e}_i{\q e}_j={\q e}_k$, if we pass from ${\q e}_i$ to ${\q e}_j$ in a direct rotation (whose angle is $2\pi /3$, if oriented according to the arrow); and ${\q e}_i{\q e}_j=-{\q e}_k$ otherwise. For example ${\q e}_2{\q e}_3={\q e}_1$, and ${\q e}_1{\q e}_3=-{\q e}_2$.\label{fig0}}
  \end{center}
    \end{figure}

Figure \ref{fig0} provides a mnemonic way to attribute the appropriate sign to a product of the form $q_ir_j{\q e}_i{\q e}_j$, when $i\ne j$ (both $i$ and $j$ run over 1, 2, 3).

%$$$$$$$$$$$$$$$$$$$$$$$$$$$$$$$$$$$$$$$$$$$$$$$$$$$$$

Nevertheless, for actual derivations, we also use the following method. We adopt the notation
\begin{equation}
q=\left[\begin{matrix}q_0\\ q_1\\ q_2\\ q_3\\\end{matrix}\right]\,,\label{eq9}\end{equation}
and we call the right part of Eq.\ (\ref{eq9})  the {\bf column form} or column representation of $q$. We avoid to call it ``matrix representation'', since we will introduce a product of columns that is not the matrix product. (A true matrix-representation of quaternions will be introduced in the next section.)

We write the product of $q$ and $r$ in the form 
\begin{equation}
q\,r=\begin{bmatrix}q_0\\ q_1\\ q_2\\ q_3\\\end{bmatrix}
\begin{bmatrix}r_0\\ r_1\\ r_2\\ r_3\\\end{bmatrix}=
\begin{bmatrix}q_0r_0 -q_1r_1  -q_2r_2 -q_3 r_3\\
q_0r_1+q_1r_0+q_2r_3-q_3r_2\\
q_0r_2+q_2r_0-q_1r_3+q_3r_1\\
q_0r_3+q_3r_0+q_1r_2-q_2r_1\\\end{bmatrix}
\,,\label{eq10}\end{equation}
which should be understood as an operational writing for obtaining the product of two quaternions; it is not a matrix product!

The process for obtaining the product of two quaternions is as follows. The first line of the column form of $q\,r$, that is, the ${\q e}_0$--component of $q\,r$,  is obtained by mutiplying  column forms of $q$ and $r$ line by line, and setting appropriate signs, as shown in Eq.\ (\ref{eq10a}), in which components to be multiplied are linked by a straight segment; explicitly 
\begin{equation}%$$$$$$$$$$$$$$$--1--$$$$$$$$$$$$$$$$$$$$$$$$$$$$$
  \left[\begin{matrix}q_0\\ q_1\\ q_2\\ q_3\\\end{matrix}\right]
  \begin{picture}(20,0)
    \put(0,18){\line(5,0){20}}
    \put(0,6){\line(5,0){20}}
    \put(0,-6){\line(5,0){20}}
     \put(0,-18){\line(5,0){20}}
  \end{picture}
  \left[\begin{matrix}r_0\\ r_1\\ r_2\\ r_3\\\end{matrix}\right]
  \begin{matrix}+\\ -\\ -\\ -\\\end{matrix} \hskip .3cm \mbox{provides}
\left[\begin{matrix}q_0r_0 -q_1r_1  -q_2r_2 -q_3 r_3\\
    \mbox{---}  \\
    \mbox{---} \\
    \mbox{---} \\
  \end{matrix}\right]
\,.\label{eq10a}\end{equation}

The second line of $q\,r$ (that is, the ${\q e}_1$--component of $q\,r$), is obtained from the cross product of the first and second lines of column forms of $q$ and $r$, and from the cross product of the third and fourth lines, and setting appropriate signs as follows, so that  

\begin{equation}%$$$$$$$$$$$$$$$--2--$$$$$$$$$$$$$$$$$$$$$$$$$$$$$
  \left[\begin{matrix}q_0\\ q_1\\ q_2\\ q_3\\\end{matrix}\right]
  \begin{picture}(20,20)
    \put(0,18){\line(5,-3){20}}
    \put(0,6){\line(5,3){20}}
    \put(0,-6){\line(5,-3){20}}
     \put(0,-18){\line(5,3){20}}
  \end{picture}
  \left[\begin{matrix}r_0\\ r_1\\ r_2\\ r_3\\\end{matrix}\right]
  \begin{matrix}+\\ +\\ -\\ +\\\end{matrix}
     \hskip .3cm \mbox{provides}
\left[\begin{matrix} \mbox{---}\\
q_0r_1+q_1r_0+q_2r_3-q_3r_2\\
\mbox{---}\\
\mbox{---}\\
  \end{matrix}\right]
\,.\label{eq10b}\end{equation}
(Signs depend on indices, according to Table \ref{table1}.)

A similar process is applied for the third and fourth lines, and
\begin{equation}%$$$$$$$$$$$$$$$--3--$$$$$$$$$$$$$$$$$$$$$$$$$$$$$
  \left[\begin{matrix}q_0\\ q_1\\ q_2\\ q_3\\\end{matrix}\right]
  \begin{picture}(20,0)
    \put(0,18){\line(5,-6){20}}
    \put(0,6){\line(5,-6){20}}
    \put(0,-6){\line(5,6){20}}
     \put(0,-18){\line(5,6){20}}
  \end{picture}
  \left[\begin{matrix}r_0\\ r_1\\ r_2\\ r_3\\\end{matrix}\right]
  \begin{matrix}+\\ +\\ +\\ -\\\end{matrix}
     \hskip .3cm \mbox{provides}
\left[\begin{matrix} \mbox{---}\\
    \mbox{---}
\\
q_0r_2+q_2r_0-q_1r_3+q_3r_1\\
\mbox{---}
\\\end{matrix}\right]
\,,\label{eq10c}\end{equation}

\begin{equation}%$$$$$$$$$$$$$$$--4--$$$$$$$$$$$$$$$$$$$$$$$$$$$$$
  \left[\begin{matrix}q_0\\ q_1\\ q_2\\ q_3\\\end{matrix}\right]
  \begin{picture}(18,20)
    \put(0,18){\line(1,-2){18}}
    \put(-1,6){\line(5,-3){20}}
    \put(-1,-6){\line(5,3){20}}
     \put(0,-18){\line(1,2){18}}
  \end{picture}
  \left[\begin{matrix}r_0\\ r_1\\ r_2\\ r_3\\\end{matrix}\right]
  \begin{matrix}+\\ -\\ +\\ +\\\end{matrix}
     \hskip .3cm \mbox{provides}
\left[\begin{matrix} \mbox{---}\\
\mbox{---}\\
\mbox{---}\\
q_0r_3+q_3r_0+q_1r_2-q_2r_1\\\end{matrix}\right]
\,.\label{eq10d}\end{equation}%$$$$$$$$$$$$$$$$$$$$$$$$$$$$$$$$$$$$$$$

The rules to obtain the appropriate sign of a product $q_\mu r_\nu$  or $q_i r_j$ are as follows:
\begin{itemize}
\item Products $q_0r_\nu$  and $q_\mu r_0$ are given the sign $+$.
\item Products $q_ir_j$ ($ij\ne 0$):
  \begin{itemize}
  \item If $i = j$, then $q_i r_j=q_i r_i$ is given the sign $-$.
    \item If $i\ne j$, let $k\in\{1,2,3\}$ with $k\ne i$ and $k\ne j$. Then $q_ir_j$ is given the sign $+$ if $[i,j,k]$ is a circular permutation of $[1,2,3]$ (see also Fig.\ \ref{fig0}).
    \end{itemize}
\end{itemize}

%**********************************************
%PAULI
%**********************************************
\subsubsection{Matrix representation}
The Pauli matrices $\sigma_\mu$ ($\mu=0,1,2,3$) are
  \begin{equation}
    \sigma_0=\begin{pmatrix}1 & 0 \cr 0 & 1\end{pmatrix}\,,\hskip .5cm
    \sigma_1=\begin{pmatrix}1 & 0 \cr 0 & -1\end{pmatrix}\,,\hskip .5cm
    \sigma_2=\begin{pmatrix}0 & 1 \cr 1 & 0\end{pmatrix}\,,\hskip .5cm
    \sigma_3=\begin{pmatrix}0 & -\I \cr \I  & 0\end{pmatrix}\,.
    \end{equation}
    The following correspondence between quaternions and  Pauli matrices
  \begin{equation}
    {\q e}_0\longmapsto \sigma_0\,,\hskip 1cm {\q e}_k\longmapsto -\I\sigma_k\,,\hskip .3cm k=1,2,3,
    \end{equation}
  defines a quadratic isomorphism (denoted $\sigma$) between the complex quaternion algebra ${\mathbb H}_c$ and the $2\times2$--complex-matrix algebra. We have
  \begin{equation}
    q=q_0{\q e}_0+q_1{\q e}_1+q_2{\q e}_2+q_3{\q e}_3\longmapsto \sigma (q) =\begin{pmatrix} q_0-\I q_1 & -q_3-\I q_2 \cr
    q_3 -\I q_2 & \;\;\,q_0 +\I q_1\end{pmatrix}\,,\end{equation}
  and
  \begin{equation}
    N(q)={\rm det}\,\sigma (q)\,,\end{equation}
where  ${\rm det}\,\sigma (q)$ denotes the determinant of $\sigma (q)$.

  The inverse isomorphism is given by
  \begin{equation}
    \begin{pmatrix}\sigma_{11} & \sigma_{12} \cr
      \sigma_{21} & \sigma_{22}\end{pmatrix}\longmapsto q={\sigma_{11}+\sigma_{22}\over 2}{\q e}_0
    +\I {\sigma_{11}-\sigma_{22}\over 2} \,{\q e}_1
    +\I {\sigma_{12}+\sigma_{21}\over 2}\,{\q e}_2
    +{\sigma_{21}-\sigma_{12}\over 2}\,{\q e}_3\,.\label{eq23a}
    \end{equation}

  Actual calculations on quaternions can be developed by using the previous isomorphism. Explicitly, if $q=q_0{\q e}_0+q_1{\q e}_1+q_2{\q e}_2+q_3{\q e}_3\longmapsto \sigma (q)$ and $r=r_0{\q e}_0+r_1{\q e}_1+r_2{\q e}_2+r_3{\q e}_3\longmapsto \sigma (r)$, then
  \begin{equation}
    q\,r\longmapsto \sigma (qr)=\sigma (q) \sigma (r) =\begin{pmatrix} q_0-\I q_1 & -q_3-\I q_2 \cr
    q_3 -\I q_2 & \;\;\,q_0 +\I q_1\end{pmatrix}
    \begin{pmatrix} r_0-\I r_1 & -r_3-\I r_2 \cr
      r_3 -\I r_2 & \;\;\,r_0 +\I r_1\end{pmatrix}\,,
      \end{equation}
      and we obtain the following equations
   \begin{eqnarray}
   \sigma (qr)_{11}\!\!\!&=& q_0r_0-q_1r_1-q_2r_2-q_3r_3-\I (q_0r_1+q_1r_0+q_2r_3-q_3r_2)\,, \\
   \sigma (qr)_{12}\!\!\!&=& \!\!\!\!\!-q_0r_3-q_1r_2+q_2r_1 -q_3r_0 - \I (q_0r_2-q_1r_3+q_2r_0+q_3r_1)\,, \\
   \sigma (qr)_{21}\!\!\!&=&  q_0r_3+q_1r_2-q_2r_1 + q_3r_0 -\I (q_0r_2-q_1r_3+q_2r_0+q_3r_1 )\,,\\
   \sigma (qr)_{22}\!\!\!&=& q_0r_0-q_1r_1-q_2r_2-q_3r_3+ \I (q_0r_1+q_1r_0+q_2r_3-q_3r_2)\,,
   \end{eqnarray}
which lead to Eq.\ (\ref{eq10n}) once more, if we apply Eq.\ (\ref{eq23a}).

Finally we point out that quaternions can also be represented by $4\times 4$--complex matrices (Dirac matrices) \cite{Syn,Deh}.
%**********************************************
%FIN DE PAULI
%**********************************************

\subsubsection{Real quaternion algebra}%******************************
A quaternion is a {\bf real quaternion} if its components $q_\mu$ are real numbers. Real quaternions are also called {\bf Hamilton's quaternions}. They form a real algebra, denoted by ${\mathbb H}$. If ${\mathbb E}_3$  denotes the 3-dimensional Euclidean vector space (that is, ${\mathbb R}^3$ equipped with the Euclidean norm), we identify ${\mathbb H}$, considered as a four-dimensional real vector-space, with ${\mathbb R}\oplus{\mathbb E}_3$, so that every three-dimensional  vector $\vec V$ can be seen as a real pure quaternion $V$. If $\|\vec V\|$ denotes the Euclidean norm of vector $\vec V$ in ${\mathbb E}_3$, we have %3-dimensional Euclidean space, we have
\begin{equation} 
N(V)=\|\vec V\|^2\,.\end{equation}
Sometimes we write $\vec V$ in place of $V$ (and vice versa), that is, we consider $\vec V$ as a real pure quaternion: if $\{{\qv e}_1,\vec {\qv e}_2,\vec {\qv e}_3\}$ is an orthonormal basis of the three-dimensional vector space ${\mathbb E}_3$,
the quaternion that corresponds to vector $\vec V=V_1\vec {\qv e}_1+V_2\vec {\qv e}_2+V_3\vec {\qv e}_3$ is $V=V_1{\q e}_1+V_2{\q e}_2+V_3{\q e}_3$. We also write $\|V\|$ for $\|\vec V\|$, so that $\|V\|^2=N(V)$. 

If $\vec V$ and $\vec W$ are two vectors of the  three-dimensional Euclidean space ${\mathbb E}_3$, we denote by $\langle\vec V,\vec W\rangle$ their Euclidean scalar product. If $V$ and $W$ are the corresponding quaternions of $\vec V$ and $\vec W$, we have
\begin{equation}
V\vec \cdot W=\langle\vec V,\vec W\rangle\,.\end{equation}

If $\vec V\vec \times \vec W$ denotes the {\bf vector product} (or cross product) of $\vec V$ and $\vec W$ (in ${\mathbb E}_3$), we define $V{\vec \times} W$ as being the quaternion corresponding to $\vec V \vec \times \vec W$. By abuse we write $V \vec\times  W= \vec V \vec\times \vec W$, so that 
\begin{equation}
V\,W=- V\vec\cdot  W \,{\q e}_0+ V \vec \times W =-\langle\vec V, \vec W\rangle \,{\q e}_0+\vec V \vec\times \vec W\,,\label{eq13}
\end{equation}
as we  deduce from Eq.\ (\ref{eq10}), if we set $q_0=r_0=0$, $q_i=V_i$ and $r_j=W_j$.

The identification of ${\mathbb H}$ with ${\mathbb R}\oplus{\mathbb E}_3$ leads us to write ${\q e}_0=1$. For example Eqx.\ (\ref{eq13}) is also written
\begin{equation}
V\,W=- V\vec\cdot  W + V\vec \times W\,.\label{eq30a}
\end{equation}

A real unit pure quaternion is denoted ${\q e}_n$ and corresponds to a unitary vector $\vec {\qv e}_n$ of the 3-dimensional Euclidean vector space. We have ${{\q e}_n}^{\! 2}=-{\q e}_0=-1$, and ${\q e}_n$ appears to be a quaternionic generalization of the complex number $\I$. The subscript $n$ in ${\q e}_n$ is not a variable index: explicitly we have
    \begin{equation}
      {\q e}_n=n_1{\q e}_1+n_2{\q e}_2+n_3{\q e}_3\,,\hskip .5cm (n_1)^2+(n_2)^2 +(n_3)^2=1\,.\label{eq32q}\end{equation}
    On the other hand, we write ${\q e}_i$, $i=1,2,3$ (or ${\q e}_j$, $j=1,2,3$, or ${\q e}_k$) to designate ${\q e}_1$, ${\q e}_2$ or ${\q e}_3$. For example   Eqs.\ (\ref{eq32q}) can be written
    \begin{equation}
      {\q e}_n=\sum_{i=1}^{i=3}n_i{\q e}_i\,,\hskip .5cm  \sum_{i=1}^{i=3}(n_i)^2=1\,.\end{equation}
    Of course it may happen that ${\q e}_n={\q e}_1$, etc.

    Eventually,  if ${\q e}_m$ and ${\q e}_n$ are real unit pure quaternions, then ${\q e}_m\vec\cdot{\q e}_n=\langle{\qv e}_m,{\qv e}_n\rangle =\cos\alpha$, where $\alpha$, the angle between ${\qv e}_m$ and ${\qv e}_n$, may be thought of like the {\bf angle between  quaternions} ${\q e}_m$ and ${\q e}_n$.

    \begin{remark} {\rm The algebra ${\mathbb H}$ is a division algebra, that is, the set of real quaternions, equipped with  addition and the quaternionic product, is a division ring: it does not contain zero divisors.  The algebra ${\mathbb H}$ is a ${\mathbb R}$--algebra and is not a subalgebra of ${\mathbb H}_{\rm c}$, which is a ${\mathbb C}$--algebra.}
      \end{remark}

\subsubsection{Cross product of complex pure quaternions and intrinsic derivations}\label{sect215}%******************************
If we set $q_0=0=r_0$ in Eq.\ (\ref{eq10}), we obtain
\begin{equation}
q\,r=\left[\begin{matrix}0\\ q_1\\ q_2\\ q_3\\\end{matrix}\right]
\left[\begin{matrix}0\\ r_1\\ r_2\\ r_3\\\end{matrix}\right]=
\left[\begin{matrix}-q_1r_1  -q_2r_2 -q_3 r_3\\
q_2r_3-q_3r_2\\
q_3r_1-q_1r_3\\
q_1r_2-q_2r_1\\\end{matrix}\right]
\,.\label{eq10e}\end{equation}
We remark that  the scalar part of $q\, r$ is opposite to the dot product of $q$ and $r$, already defined and denoted $q\vec\cdot r$. If the $q_i$'s and the $r_i$'s are real numbers, the vector part of $q\,r$ is identical to the quaternion associated with the vector product of $(q_1,q_2,q_3)$ and $(r_1,r_2,r_3)$ in ${\mathbb E}_3$.  We then define the {\bf vector product} (or cross product) of complex pure quaternions $V$ and $W$ by
\begin{equation}
  V\vec \times W=V\,W+V\vec\cdot W\,{\q e}_0\,.\label{eq35a}\end{equation}
The vector product $V\vec \times W$ is a pure quaternion. (Since ${\q e}_0\equiv 1$, we also write the right member of Eq.\ (\ref{eq35a}) as $V\,W+V\vec\cdot W$.) Equation (\ref{eq35a}) is a generalization of Eq.\ (\ref{eq30a}), since its holds for complex pure quaternions, not only for real ones.

In particular, if $V'$ and $W'$ are real pure quaternions, and if $V=aV'$ and $W=bW'$, where $a$ and $b$ are complex numbers, we have
\begin{equation}
  V\vec \times W= (aV')\vec \times (bW')=ab \;V'\!\vec \times W'\,.\label{eq37}\end{equation}

\bigskip
\noindent{\bf Intrinsic derivations.} A consequence of Eq.\ (\ref{eq35a}) is that some derivations with quaternions can be developed without referring to the canonical basis, as done in Sect.\ \ref{sect212}. Indeed, let $q$ and $r$ be two complex quaternions, written $q=q_0+ V$ and $r=r_0 +W$, where $q_0$ and $r_0$ are complex numbers and  $V$ and $W$ complex pure quaternions. Then
\begin{equation}
  q\,r= (q_0+ V)(r_0+ W)=q_0r_0+q_0W+r_0V+V\, W =q_0r_0-V\vec\cdot W+q_0W+r_0V+V\vec \times W\,.\label{eq37a}\end{equation}
The result expressed by Eq.\ (\ref{eq37a}) is intrinsic, since it does not explicitly refer to the basis $\{ {\q e}_1,{\q e}_2, {\q e}_3\}$ of pure quaternions. (Components on ${\q e}_0$ still play a special role.)

\subsubsection*{Examples of dot and cross products of complex quaternions.}
\begin{enumerate}
  \item Let $q={\q e}_1+\I\, {\q e}_2$, and $r={\q e}_2+\I\, {\q e}_3$ (both $q$ and $r$ are pure quaternions).
Then
\begin{eqnarray}
  &&\hskip -.7cm N(q)=-({\q e}_1+\I \,{\q e}_2)({\q e}_1+\I \,{\q e}_2)=0=N(r)\,,\\
  &&\hskip -.7cm q\vec\cdot r=({\q e}_1+\I\, {\q e}_2)\vec\cdot({\q e}_2+\I\, {\q e}_3)=\I\;\; (\mbox{or}\;\; \I\,{\q e}_0)\,\\
 && \hskip -.7cm q\vec\times r=({\q e}_1+\I\, {\q e}_2)\vec \times ({\q e}_2+\I\, {\q e}_3)=-{\q e}_1-\I\, {\q e}_2+{\q e}_3\,,\\
  &&\hskip -.7cm q\,r=({\q e}_1+\I\, {\q e}_2)({\q e}_2+\I\, {\q e}_3)=-\I -{\q e}_1-\I\, {\q e}_2+{\q e}_3=-q\vec\cdot r+ q\vec \times r\,.\end{eqnarray}

\item  Let $q=1+\I\, {\q e}_q$ and $r=1+\I\, {\q e}_r $, where ${\q e}_q$ and ${\q e}_r$ are real unit pure quaternions (such that ${{\q e}_q}^{\! 2}={{\q e}_r}^{\! 2}=-1$). Then
\begin{eqnarray}
  && \hskip -.7cm N(q)=(1+\I \,{\q e}_q)(1-\I\, {\q e}_q)=0=N(r)\,,\\
  && \hskip -.7cmq\vec\cdot r=(1+\I\, {\q e}_q)\vec\cdot(1+\I\, {\q e}_r)=1-{\q e}_q\vec\cdot{\q e}_r\,,\\
  && \hskip -.7cm q\,r=(1+\I \,{\q e}_q)(1+\I\, {\q e}_r)=1+{\q e}_q\vec\cdot {\q e}_r+\I\,{\q e_q}+\I\,{\q e_r}-{\q e}_q\vec \times {\q e}_r\,.\end{eqnarray}
Since $q$ and $r$ are not pure quaternions, the vector product does not make sense for them.
If ${\q e}_r={\q e}_q$, then  $q\, r=q^2= 2(1+\I \,{\q e}_q)$, and $q\vec\cdot r=q\vec\cdot q=0$. Since $q\vec\cdot q=0$ for $q\ne 0$, the dot product for complex quaternions is degenerate.
\end{enumerate}

\subsubsection{Quaternionic exponential}%*******************************************

The {\bf exponential} of a quaternion $q$ is defined by
\begin{equation}
\E^q=\exp q= 1+q+{q^2\over 2!}+ {\dots} +{q^j\over j!}+\dots\label{eq32a}\end{equation}

If $q={\q e}_n\varphi$, with $\varphi$ a real number and  ${\q e}_n$ a real unit pure quaternion, since ${{\q e}_n}^{\! 2}=-1$, we have ${{\q e}_n}^{\! 2 j}=(-1)^j$ and ${{\q e}_n}^{\! 2 j+1}=(-1)^j{\q e}_n$, so that
\begin{equation}
  \E^{{\q e}_n\varphi}=\exp {\q e}_n\varphi =
\sum_{j=0}^{+\infty} (-1)^j{\varphi ^{2j}\over (2j)!}+{\q e}_n\sum_{j=0}^{+\infty} (-1)^j{\varphi ^{2j+1}\over (2j+1)!}
  =\cos\varphi +{\q e}_n\sin\varphi\,.\label{eq33a}\end{equation}
If $q=\I \,{\q e}_n\delta$ ($\delta$ a real number),
then
\begin{equation}
  \E^{\I {\q e}_n\delta}=\exp \I\,{\q e}_n\delta =\cos\I\delta +{\q e}_n\sin\I\delta =\cosh \delta +\I\, {\q e}_n\sinh\delta\,.\label{eq34a}\end{equation}
Equation (\ref{eq34a}) can be deduced from Eq.\ (\ref{eq33a})  by replacing $\varphi$ with $\I\delta$. Both $\exp {\q e}_n\varphi$ and $\exp \I\,{\q e}_n\delta$ are unit quaternions.

In \ref{appenA} we provide some quaternionic trigonometric-formulas.
In particular, we point out that Eqs.\ (\ref{eq33a}) and (\ref{eq34a}) may take  surprising forms: for exemple, if ${\q e}_m$ and ${\q e}_n$ are two orthogonal real unit pure quaternions (${\q e}_m\vec\cdot {\q e}_n=0$), then  $N({\q e}_m+\I{\q e}_n)=-({\q e}_m+\I{\q e}_n)^2=0$, and according to Eq.\ (\ref{eq32a}), for every real number $\varphi$
\begin{equation}
  \exp [({\q e}_m+\I\,{\q e}_n)\varphi ]=1+({\q e}_m+\I\,{\q e}_n)\varphi\,,\hskip .5cm\mbox{and}\;\;\;
  N\bigl[1+({\q e}_m+\I\,{\q e}_n)\varphi\bigr]=1\,.\label{eq35}\end{equation}
Such a unitary quaternion is useful in the special theory of relativity to represent singular proper Lorentz rotations \cite{Syn} and in crystal optics to represent singular-axis crystals  \cite{PPF10} (see Sect. \ref{sect6}).

\bigskip
\noindent{\bf Polar form of a real quaternion.} Let $q=q_0+V$ be a real quaternion, with $V\ne 0$ : $q_0$ is a real number and $V$ is a real pure quaternion. Then $
N(q)={q_0}^{\! 2}+  \|V\|^2>0$. Let ${\q e}_q=V/\|V\|$ and $\theta\in ]-\pi,\pi ]$ be such that
\begin{equation}
  \cos\theta ={q_0\over\sqrt{N(q)}}\,,\hskip 1cm {\q e}_q\sin\theta ={V\over \sqrt{N(q)}}\,.\end{equation} 
Then $q$ can be written
\begin{equation}
  q=\sqrt{N(q)}\,(\cos \theta +{\q e}_q\sin\theta )=\sqrt{N(q)}\,\exp {\q e}_q\theta\,.\label{eq51a}
\end{equation}
The right part of Eq.\ (\ref{eq51a}) is the polar form of $q$ (to be compared with the polar form of a complex number $z$, written in the form $z=|z|\exp \I\theta$). The quaternion ${\q e}_q$ is a real unit pure quaternion. We remark that only the sign of ${\q e}_q\sin\theta$ is determined. If ${\q e}'_q=-{\q e}_q$ and $\theta '=-\theta$, we also have $q=\sqrt{N(q)}\,\exp {\q e}_q'\,\theta'$.

%***************************************
%ROTE3  (ROTATIONS DE L'ESPACE EUCLIDIEN
%***************************************

\subsection{Rotations on the 3--dimensional Euclidean vector space ${\mathbb E}_3$}\label{sect22}
Hamilton invented quaternions to represent rotations on  the 3--dimensional Euclidean vector space ${\mathbb E}_3$ in the same way as complex numbers of the form $\exp\I\theta$ represent rotations on the 2--dimensional Euclidean plane (isomorphic to the complex plane).

\subsubsection{Quaternionic expression of a rotation on ${\mathbb E}_3$}

The result obtained by Hamilton is expressed as follows. (In this section considered quaternions are real quaternions.)

\begin{theorem}
  {\its
    Let ${\qv e}_n$ be a unit vector in ${\mathbb E}_3$. Let vector $\vec V'$ be the image of vector $\vec V$ under the rotation of angle $\varphi$ around the axis ${\qv e}_n$ in ${\mathbb E}_3$, and let $V'$, $V$ and ${\q e}_n$ be the associated quaternions. Then
\begin{equation}
  V'=\E^{{\q e}_n\varphi /2}\,V\,\E^{-{\q e}_n\varphi/2}\,.\label{eq52q}\end{equation}
(The angle $\varphi$ is positive if the sense of rotation is the trigonometric sense for an observer who sees the vector ${\qv e}_n$ pointed towards him.)
}
\end{theorem}

\medskip
\noindent{\its Proof.} We denote 3--dimensional vectors by their associated quaternions. Let  ${\cal P}$ be the plane orthogonal to ${\q e}_n$ and passing through the origin $O$. Let $V_\perp$ be the (orthogonal) projection of $V$ on ${\cal P}$. Then
\begin{equation}
  V=\|V\|\,{\q e}_n\cos\alpha +V_\perp\,,\end{equation}
where $\alpha$ is the angle between ${\q e}_n$ and $V$, and $V_\perp$ is perpendicular to ${\q e}_n$. Since ${\q e}_n$ and $\exp ({\q e}_n\varphi/2)$ commute,
we have
\begin{equation}
\E^{{\q e}_n\varphi /2}\,\bigl(\|V\|\,{\q e}_n\cos\alpha \bigr)\,\E^{-{\q e}_n\varphi/2}=\|V\|\,{\q e}_n\cos\alpha\,.\end{equation}
We also derive
\begin{eqnarray}
  \E^{{\q e}_n\varphi /2}\,V_\perp \,\E^{-{\q e}_n\varphi/2}\rap&=&\rap\left(\cos{\varphi\over 2}+{\q e}_n\sin{\varphi\over 2}\right)
  \, V_\perp\,\left(\cos{\varphi\over 2}-{\q e}_n\sin{\varphi\over 2}\right)\nonumber \\
  &=& \rap \left( V_\perp\cos{\varphi\over 2}+{\q e}_n\,V_\perp\sin{\varphi\over 2}\right)\left(\cos{\varphi\over 2}-{\q e}_n\sin{\varphi\over 2}\right)\nonumber \\
  &=& \rap V_\perp \cos^2{\varphi\over 2}-{\q e}_n\,V_\perp\,{\q e}_n\sin^2{\varphi\over 2}+({\q e}_n\,V_\perp -V_\perp \,{\q e}_n)\cos{\varphi\over 2}\sin{\varphi\over 2}\,.\label{eq55q}
\end{eqnarray}
Since $V_\perp$ is orthogonal to ${\q e}_n$, we have  ${\q e}_n\,V_\perp={\q e}_n\vec \times V_\perp=-V_\perp\,{\q e}_n$, and  $V_\perp={\q e}_n\,V_\perp\,{\q e}_n$ (see the corollary of \ref{appenD}).  If $W=\exp ({\q e}_n\varphi /2)\,V_\perp\exp (-\!{\q e}_n\varphi /2)$, Eq.\ (\ref{eq55q}) leads to
\begin{equation}
W=V_\perp\cos \varphi+ {\q e}_n\vec \times V_\perp\sin \varphi\,,\end{equation}
and we first conclude that $W$ is deduced from $V_\perp$ under the rotation of angle $\varphi$ in the plane ${\cal P}$, and then that $V'$, defined by Eq.\ (\ref{eq52q}), is such that $V'= \|V\|\,{\q e}_n\cos\alpha + W$, and is  deduced from $V$ under the rotation of angle $\varphi$ around ${\q e}_n$. The proof is complete. \qed

\subsubsection{Determinant}
Let $R$ be a rotation on ${\mathbb E}_3$ and let ${\cal R}$ be its matrix in the basis $\{{\qv e}_1,{\qv e}_2,{\qv e}_3\}$. The determinant of ${\cal R}$ (as a matrix) is also the determinant of $R$ (as a linear mapping on ${\mathbb E}_3$); since $R$ is a rotation, it preserves the orientation of the previous basis, so that 
$\det R=\det{\cal R}=1$.

%**********************************************
%FIN DE ROTE3
%**********************************************

\subsection{Minkowski space. Minquats}%**************************************************************************

The Minkowski  space is $({\mathbb R}^4,Q)$, where $Q$ is the {\bf Lorentz quadratic form} (or {\bf metric}), defined by
\begin{equation}
  Q(x_0,x_1,x_2,x_3)= (x_0)^2-(x_1)^2-(x_2)^2-(x_3)^2\,.\end{equation}
The Minkowski  space is thus isomorphic to $({\mathbb R}\oplus{\mathbb E}_3,Q)$.

The associated bilinear form (or polar form of $Q$) is the pseudo-Euclidean scalar product
\begin{equation}
  \langle \vec x,  \vec y\rangle =x_0y_0-x_1y_1-x_2y_2-x_3y_3\,,\end{equation}
where $\vec x=(x_0,x_1,x_2,x_3)$ and $\vec y=(y_0,y_1,y_2,y_3)$ belong to ${\mathbb R}^4$. 

A quaternion $x$ is a {\bf minquat} if it takes the form
\begin{equation}
x=x_0{\q e}_0+\I (x_1{\q e}_1+x_2{\q e}_2+x_3{\q e}_3)\,,\end{equation}
where the $x_\mu$'s are real numbers. The reason is that
\begin{equation} 
N(x)=(x_0)^2-(x_1)^2-(x_2)^2-(x_3)^2\,,\end{equation}
is equal to the Lorentz quadratic form associated with the real four-vector  $(x_0,x_1,x_2,x_3)$ of the Minkowski  space. (The word ``minquat'', due to Synge \cite{Syn}, is a contraction of ``Minkowski'', or ``Minkowskian'', and ``quaternion''. The definition of Synge's minquats is different from the  definition given in this section, see Remark \ref{rem22}.) 

Every minquat $x$ is such that $x-\overline{x}^*=0$.

Minquats form a 4--dimensional real vector-space, denoted ${\mathbb M}$, quadratically  isomorphic to the Minkow\-ski space, the correspondence being such that
\begin{equation}
 ({\mathbb R}\oplus {\mathbb E}_3,Q) \ni \vec x=(x_0,x_1,x_2,x_3)\longmapsto x=x_0{\q e}_0+\I (x_1{\q e}_1+x_2{\q e}_2+x_3{\q e}_3) \in {\mathbb M}\,,\end{equation}
with $Q(\vec x)= N(x)$.
  (Since it is a real vector-space, ${\mathbb M}$ is not a suspace of ${\mathbb H}_{\rm c}$, because  ${\mathbb H}_{\rm c}$ is a complex vector-space.)

For example, we are led  to represent four-dimensional vectors, as used in the special theory of relativity, by minquats.
Thus, if $x_1$, $x_2$ and $x_3$ are Cartesian coordinates related to a Galilean inertial frame and if $x_0={\rm c}t$ (${\rm c}$ denotes light velocity in vaccum and $t$ the time in the considered frame), the position quadrivector $(x_0,x_1,x_2,x_3)$ is represented by the minquat  $x_0{\q e}_0+\I (x_1{\q e}_1+x_2{\q e}_2+x_3{\q e}_3)$. 

In electromagnetism, if $\Phi$ denotes the scalar potential and $\vec a$ the (three-dimensional) vector potential, in the coordinate system $\{ x_\mu\}=\{ x_0,x_1,x_2,x_3\}$ (in a Galilean frame), the electromagnetic-potential four-vector is
$(\Phi /{\rm c}, a_1,a_2,a_3)$ and the associated quaternion (a minquat) is
\begin{equation}
A={\Phi\over {\rm c}}{\q e}_0+\I (a_1{\q e}_1+a_2{\q e}_2+a_3{\q e}_3)\,.\end{equation}

\begin{remark}\label{rem22} {\rm Synge \cite{Syn} represents the 4--vector $\vec x=(x_0,x_1,x_2,x_3)$ (the $x_\mu$'s are real numbers) by the quaternion
     $ x=\I x_0{\q e}_0+x_1{\q e}_1+x_2{\q e}_1+x_3{\q e}_3$,
    which he calls a minquat. 
    The space of Synge's minquats is then $(\I\,{\mathbb R}\oplus {\mathbb E}_3,Q')$, where the quadratic form $Q'$ is such that
    \begin{equation}
      Q'(\vec x)=-(x_0)^2+(x_1)^2+(x_2)^2+(x_3)^2=N(x)\,.\end{equation}
Every Synge's minquat $x$ is such that $x+\overline{x}^{\, *}=0$.
  }\end{remark}

\bigskip
\noindent{\bf Null, scalarlike and  spacelike minquats.} %*******************************
We adopt the following vocabulary:
\begin{itemize}
\item A minquat $x$ is a {\bf null minquat}  (or an {\bf isotropic minquat}) if $N(x)=0$. Example: $x_0{\q e}_0+\I\,x_1{\q e}_1$ is a null minquat if $|x_0|=|x_1|$. More generally, if ${\q e}_n$ is a real unit pure  quaternion, then  ${\q e}_0+ \I\,{\q e}_n$ and ${\q e}_0- \I\,{\q e}_n$   are null minquats.
\item A minquat $x$ is a {\bf scalarlike minquat} if $N(x)>0$. Example:  $x_0{\q e}_0+ \I\,x_1{\q e}_1$ is a scalarlike minquat if $|x_0|>|x_1|$.  Since minquats represent quadrivectors of the Minkowski  space, which is used in the special theory of relativity under the name of Minkowski  spacetime, a scalarlike minquat is also called a {\bf timelike} minquat.
   \item A minquat $x$ is a {\bf spacelike minquat} if $N(x)<0$. Example:  $x_0{\q e}_0+ \I\,x_1{\q e}_1$ is a spacelike minquat if $|x_0|<|x_1|$.
  \end{itemize}

%******************
%VECTORSPACE
%******************
\subsection{Vector spaces}

We sum up the vector spaces we deal with:
\begin{itemize}
\item ${\mathbb E}_3$ denotes the {\bf 3-dimensional Euclidean vector-space}, which is a real vector-space. It is ${\mathbb R}^3$ equipped with the Euclidean quadratic form: $\|(x_1,x_2,x_3)\|^2=(x_1)^2+(x_2)^2+(x_3)^2$. The associated scalar product (polar form) is defined for vectors $\vec x=(x_1,x_2,x_3)$ and $\vec y=(y_1,y_2,y_3)$ by $\langle \vec x,\vec y\rangle = x_1y_1+x_2y_2+x_3y_3$. We have   $2\langle \vec x,\vec y\rangle=\|\vec x+\vec y\|^2-\|\vec x\|^2-\|\vec y\|^2$.
\item $({\mathbb R}^4,Q)=({\mathbb R}\oplus{\mathbb E}_3,Q)$ is the {\bf Minkowski space}. It is a real vector-space. The quadratic form $Q$ is the Lorentz quadratic form: $Q(x_0,x_1,x_2,x_3)= (x_0)^2-(x_1)^2-(x_2)^2-(x_3)^2$.
  The associated scalar product is the pseudo-Euclidean scalar product defined for vectors $\vec x=(x_0,x_1,x_2,x_3)$ and $\vec y=(y_0,y_1,y_2,y_3)$ by   $\langle \vec x,\vec y\rangle =x_0y_0-x_1y_1-x_2y_2-x_3y_3$. We have
  $2\langle \vec x,\vec y \rangle =Q(\vec x+\vec y)-Q(\vec x)-Q(\vec y)$.
  \item ${\mathbb H}$ denotes the {\bf real algebra of Hamilton's quaternions}. If we consider only the addition of quaternions and  the multiplication of quaternions by  real numbers,  ${\mathbb H}$ is a real vector-space.
\item ${\mathbb H}_{\rm c}$ denotes the {\bf complex algebra of complex quaternions}. If we consider only the addition of quaternions and the  multiplication of quaternions by  complex numbers,  ${\mathbb H}_{\rm c}$ is a complex vector-space.
\item ${\mathbb M}$ denotes the {\bf real vector-space of minquats}. It is a quadratic space for the quaternionic norm $N$, such that $N(x)= x\,x^*=(x_0)^2-(x_1)^2-(x_2)^2-(x_3)^2$ for  the minquat $x=x_0{\q e}_0+\I\,x_1{\q e}_1+\I\,x_2{\q e}_2+\I\,x_3{\q e}_3$. Moreover, if $y=y_0{\q e}_0+\I\,y_1{\q e}_1+\I\,y_2{\q e}_2+\I\,y_3{\q e}_3$, the polar form of the norm is the dot product, defined by
  $x\vec\cdot y=x_0y_0-x_1y_1-x_2y_2-x_3y_3$, and such that
  $2\,x\vec\cdot y =N(x+y)-N(x)-N( y)$.
  The minquat space $({\mathbb M},N)$ is
  isomorphic to the Minkowski space $({\mathbb R}^4,Q)$. The isomorphism is quadratic: $N$ and $Q$
  are such that  $N(x)=N(x_0{\q e}_0+\I\,x_1{\q e}_1+\I\,x_2{\q e}_2+\I\,x_3{\q e}_3) =Q(x_0{\qv e}_0+x_1{\qv e}_1+x_2{\qv e}_2+x_3{\qv e}_3)=Q(\vec x)$.
\end{itemize}
%********************
%FIN DE VECTORSPACE
%********************

\subsection{Proper Lorentz rotations}\label{sect14}%******************************************************

A rotation on a quadratic space is an automorphism whose determinant is equal to $1$. 
A rotation $R_3$ on ${\mathbb E}_3$ can naturally be extended to the Minkowski  space. For that purpose we write an element of the Minkowski  space ${\mathbb R}\oplus {\mathbb E}_3$ in the form $(x_0,\vec V)$, where  $x_0$ is a real number and $\vec V$ a vector in ${\mathbb E}_3$.
Let $R$ be the mapping defined on ${\mathbb R}\oplus {\mathbb E}_3$  by  $R(1,\vec 0)=1$, and $R(0,\vec V)=R_3(\vec V)$, for every 3-dimensional vector $\vec V$ of ${\mathbb E}_3$. If ${\cal R}_3$ denotes the matrix of $R_3$ in the basis $\{{\qv e}_1,{\qv e}_2,{\qv e}_3\}$ (in ${\mathbb E}_3$),  the matrix of $R$ in the basis $\{ 1, {\qv e}_1,{\qv e}_2, {\qv e}_3\}$ is
\begin{equation}{\cal R}=\left(\begin{array}
    {c|ccc}1  & 0  & 0 & 0  \\    \hline  \\ %\cdashline{1-4} 0 \\
  0 & & {\cal R}_3 \\
  0 & \end{array}\right)\,,\label{eq65}\end{equation}
and $\det R=\det  {\cal R}=\det {\cal R}_3=1$, which means that $R$ is a rotation. Moreover  $R$ separately preserves the orientations of both ${\mathbb R}$ and ${\mathbb E}_3$, and is called a {\bf proper rotation}.

A transformation on the Minkowski  space that preserves the Lorentz quadratic form is called a {\bf Lorentz transformation}. If its determinant is equal to 1, the transformation is a rotation.
On the Minkowski  space, a rotation that {\bf separately} preserves  the orientation of ${\mathbb R}$ and that of ${\mathbb E}_3$  is  a {\bf proper Lorentz rotation}. 
(Consequently every proper Lorentz rotation  preserves the orientation of the Minkowski  space; the converse does not hold.)
In the special theory of relativity, the transformations that preserve the orientation of ${\mathbb R}$ are often called {\bf orthochronous}; proper Lorentz rotations are thus orthochronous Lorentz rotations.

The previous rotation $R$, as mentioned above, is a proper Lorentz rotation, but every proper Lorentz rotation is not necessarily written, in matrix form, as  in Eq.\ (\ref{eq65}), as we shall see.
Proper Lorentz rotations form a (non-commutative) group, denoted SO$_+(1,3)$ \cite{Deh}, generated by hyperbolic rotations
and  elliptic rotations (that will be introduced in the next sections).

\subsubsection{Elliptic rotations on the Minkowski space}\label{sect232}%**********************

On the Minkowski  space, the effect of an {\bf elliptic} (or pure, or spatial) rotation of angle $\varphi$ around the axis ${\qv e}_1$ is expressed, in matrix form, by
\begin{equation}
\begin{pmatrix}x'_0\\ x'_1 \\ x'_2 \\ x'_3 \\
\end{pmatrix}
=\begin{pmatrix}1&0&0&0 \\
0&1&0&0 \\0&0& \cos \varphi &-\sin \varphi  \\
0&0&\sin\varphi &\cos\varphi \\
\end{pmatrix}
\begin{pmatrix}x_0\\ x_1 \\ x_2 \\ x_3 \\
\end{pmatrix}\,.
\label{eq30}
\end{equation}
The 4--vector $\vec x=(x_0,x_1,x_2,x_3)$ is transformed into the 4--vector $\vec x'=(x'_0,x'_1,x'_2,x'_3)$.

With minquats we have
\begin{equation}
x'=\exp \left({\q e}_1{\varphi\over 2}\right)x\,\exp \left(-\,{\q e}_1{\varphi\over 2}\right)\,,\label{eq31}\end{equation}
that is
\begin{equation}
\left[\begin{matrix}
x'_0\\
\I x'_1\\
\I x'_2 \\
\I x'_3\\
\end{matrix}
\right]
=
\left[\begin{matrix}
\cos(\varphi / 2)\\
\sin (\varphi / 2)\\
0 \\
0\\
\end{matrix}
\right]
\left[\begin{matrix}
x_0\\
\I x_1\\
\I x_2 \\
\I x_3\\
\end{matrix}
\right]
\left[\begin{matrix}
\cos (\varphi /2)\\
-\sin (\varphi/ 2)\\
0 \\
0\\
\end{matrix}
\right]
\,.\label{eq32}\end{equation}
Displaying quaternions as in Eq.\ (\ref{eq32}) affords a technical way to actually check that Eq.\ (\ref{eq31}) is equivalent to Eq.\ (\ref{eq30}): quaternions products are performed according to Eq.\ (\ref{eq10}).

If the rotation is around the axis ${\q e}_n=n_1{\q e}_1+n_2{\q e}_2+n_3{\q e}_3$  (${\q e}_n$ is a real unit pure quaternion: ${n_1}^2+{n_2}^2+{n_3}^2=1$), then
\begin{equation}
x'=\exp \left({\q e}_n{\varphi\over 2}\right)x\,\exp \left(-\,{\q e}_n{\varphi\over 2}\right)\,,\label{eq33}\end{equation}
that is
\begin{equation}
\left[\begin{matrix}
x'_0\\
\I x'_1\\
\I x'_2 \\
\I x'_3\\
\end{matrix}
\right]
=
\left[\begin{matrix}
\cos (\varphi /2)\\
n_1\sin (\varphi/ 2)\\
n_2 \sin (\varphi/ 2)\\
n_3\sin (\varphi/ 2)\\
\end{matrix}
\right]
\left[\begin{matrix}
x_0\\
\I x_1\\
\I x_2 \\
\I x_3\\
\end{matrix}
\right]
\left[\begin{matrix}
\cos (\varphi / 2)\\
-n_1\sin (\varphi/ 2)\\
-n_2\sin (\varphi/ 2 ) \\
-n_3\sin (\varphi/ 2)\\
\end{matrix}
\right]
\,.\label{eq32b}\end{equation}

\begin{remark}\label{rem21} {\rm  Passing from Eq.\ (\ref{eq31}) to Eq.\ (\ref{eq33}) illustrates the efficiency of quaternions in writing a proper Lorentz rotation around an arbitrary axis. Moreover Eq.\ (\ref{eq32b}) shows that such a writing is not merely a theoretical notation but truly a practical one, allowing  us to actually manage derivations. Equation (\ref{eq33}) can also be used for intrinsic derivations (see Sect.\ \ref{sect215}).
}
\end{remark}

\subsubsection{Hyperbolic rotations on the Minkowski space}
 
The hyperbolic rotation of direction $x_1$ %(or ${\q e}_1$)
and parameter $\delta$ ($\delta \in {\mathbb R}$) is the following transformation, on the Minkowski  space
\begin{eqnarray}
x'_0\rap &=&\rap x_0\cosh \delta +  x_1\sinh\delta\,,\label{eq21b}\\
x'_1\rap &=&\rap x_0\sinh\delta +x_1\cosh\delta\,,\label{eq22b}\\
x'_2\rap &=&\rap x_2\,,\label{eq23b}\\
x'_3\rap &=&\rap x_3\,,\label{eq24b}\end{eqnarray}
and its  matrix expression is %, in the canonical basis, is
\begin{equation}
\begin{pmatrix}x'_0\\ x'_1 \\ x'_2 \\ x'_3 \\
\end{pmatrix}
=\begin{pmatrix}\cosh \delta &\sinh \delta & 0 & 0 \\
\sinh\delta &\cosh\delta & 0 & 0 \\
0 & 0 & 1 & 0 \\
0 & 0 & 0 & 1 \\
\end{pmatrix}
\begin{pmatrix}x_0\\ x_1 \\ x_2 \\ x_3 \\
\end{pmatrix}\,.
\label{eq26}
\end{equation}

With minquats  $x=x_0{\q e}_0+\I (x_1{\q e}_1+x_2{\q e}_2+x_3{\q e}_3)$ and  $x'=x'_0{\q e}_0+\I (x'_1{\q e}_1+x'_2{\q e}_2+x'_3{\q e}_3)$, the quaternionic form of Eqs.\ (\ref{eq21b}--\ref{eq24b}) is
\begin{equation}
x'=\exp \left(\I \,{\q e}_1{\delta\over 2}\right)x\, \exp \left(\I \,{\q e}_1{\delta\over 2}\right)\,,\label{eq27}\end{equation}
and also, if we display quaternions as in Eq.\ (\ref{eq10}),
\begin{equation}
\left[\begin{matrix}
x'_0\\
\I x'_1\\
\I x'_2 \\
\I x'_3\\
\end{matrix}
\right]
=
\left[\begin{matrix}
\cosh (\delta / 2)\\
\I\, \sinh (\delta / 2)\\
0 \\
0\\
\end{matrix}
\right]
\left[\begin{matrix}
x_0\\
\I x_1\\
\I x_2 \\
\I x_3\\
\end{matrix}
\right]
\left[\begin{matrix}
\cosh (\delta / 2)\\
\I\, \sinh (\delta / 2)\\
0 \\
0\\
\end{matrix}
\right]
\,.\label{eq28}\end{equation}

We also say that the direction of the previous hyperbolic rotation is along ${\q e}_1$, or by comparison with elliptic rotations (see Sect.\ \ref{sect232}), that ${\q e}_1$ is the {\bf axis} of the hyperbolic rotation.  If the direction is  along ${\q e}_n$, Eq.\ (\ref{eq27}) is replaced with
\begin{equation}
x'=\exp \left(\I \,{\q e}_n{\delta\over 2}\right)x\,\exp \left(\I \,{\q e}_n{\delta\over 2}\right)\,,\label{eq29}\end{equation}
and Eq.\ (\ref{eq28}) with
\begin{equation}
\left[\begin{matrix}
x'_0\\
\I x'_1\\
\I x'_2 \\
\I x'_3\\
\end{matrix}
\right]
=
\left[\begin{matrix}
\cosh (\delta / 2)\\
\I\, n_1\sinh (\delta /2)\\

\I\,n_2\sinh (\delta / 2) \\
\I \,n_3\sinh (\delta / 2)\\
\end{matrix}
\right]
\left[\begin{matrix}
x_0\\
\I x_1\\
\I x_2 \\
\I x_3\\
\end{matrix}
\right]
\left[\begin{matrix}
\cosh (\delta /2)\\
\I\, n_1\sinh (\delta / 2)\\
\I\, n_2\sinh (\delta / 2) \\
\I\, n_3\sinh (\delta / 2)\\
\end{matrix}
\right]
\,.\label{eq28b}\end{equation}
with ${\q e}_n=n_1{\q e}_1+n_2{\q e}_2+n_3{\q e}_3$.

Remark \ref{rem21} is also appropriate here.

\begin{remark} {\rm From a strict geometrical point of view, there is some abuse in calling ${\q e_n}$ the axis of the hyperbolic rotation, but the abuse makes sense in polarization optics by reference to the Poincar\'e sphere, on which ${\q e}_n$ defines an axis that intercepts the sphere at two points. Those two points  represent the (polarization) eigenstates of a dichroic device, exactly as for a birefringent device, represented by an elliptic rotation for which the notion of rotation axis makes sense (see Section \ref{sect5}). We also abusively call $\I\,\delta$ the angle of an hyperbolic rotation of parameter $\delta$.
  }
  \end{remark}

\subsubsection{Hyperbolic rotations and Lorentz boosts}\label{sect253}
In the special theory of relativity, the Lorentz boost in the direction $x_1$ and velocity $v\,{\q e}_1$ is the following transformation
\begin{eqnarray}
x'_0\rap &=&\rap \gamma (x_0- \beta  x_1)\,,\label{eq21}\\
x'_1\rap&=&\rap\gamma (x_1-\beta x_0)\,,\label{eq22}\\
x'_2\rap &=&\rap x_2\,,\label{eq23}\\
x'_3\rap &=&\rap x_3\,,\label{eq24}\end{eqnarray}
where $\beta =v/{\rm c}$  and $\gamma =(1-\beta^2)^{-1/2}$.
Equations (\ref{eq21}--\ref{eq24}) link coordinates of two Galilean systems of inertia ${\cal S}$ and ${\cal S}'$ in relative translatory motion, that is, coordinates $\{ x_\mu\}$ in ${\cal S}$ to coordinates $\{ x'_\mu\}$ in ${\cal S}'$. Spatial axes are chosen parallel ($x' _j\parallel x_j$, $j=1,2,3$). If $t$ and $t'$ denote the times in ${\cal S}$ in ${\cal S}'$ respectively,  time origins are chosen so that spatial origins coincide at $x_0=0=x'_0$ (with $x_0={\rm c}t$ and $x'_0={\rm c}t'$). The relative velocity of ${\cal S}$ with respect to ${\cal S}$ is $\vec v$ and is along the $x_1$--axis.

The link with a hyperbolic rotation is obtained by writing $\tanh \delta =-\beta$,
so that
\begin{equation}
\cosh\delta =\gamma\,,\hskip .5cm {\rm and}\hskip .5cm
\sinh\delta =-\gamma \beta\,,\label{eq29a}\end{equation}
and Eqs. (\ref{eq21}--\ref{eq24}) become  Eqs. (\ref{eq21b}--\ref{eq24b}). (The parameter $\delta$ is sometimes called {\bf rapidity} \cite{Gou,Pen}.)

Equation (\ref{eq29}) is the quaternionic representation of a Lorentz boost, in the form of a hyperbolic rotation. It holds if the considered Lorentz boost is applied to 4--vectors  of the Minkowski  space.
We may suppose, however, that physical quantities other than 4--vectors of the Minkowski  space are transformed when passing from a Galilean inertial frame to another. We also say that such   physical quantities  undergo a Lorentz boost, but the boost is not expressed by a hyperbolic rotation on the Minkowski  space. In electromagnetism, for example, the scalar potential and the vector potential (a 3--vector) form together a 4--vector, called 4--potential vector and element of an appropriate Minkowski  space (see  Sect.\ \ref{sect32}). In a boost,  a 4--potential vector transforms according to Eq.\ (\ref{eq29}). On the other hand, both the electric field $\vec E$ and the magnetic field $\vec B$ are  transformed under a Lorentz boost between two Galilean inertial frames. Since they are not 4--vectors, Eq.\ (\ref{eq29}) can be applied separately neither to $\vec E$ nor to $\vec B$.  The electric field $\vec E$  and the magnetic field $\vec B$ can be mixed to form a complex vector and we shall show in Section \ref{sect34} that the effect of a Lorentz boost on the quaternion representative of such a vector is an elliptic rotation that operates on ${\mathbb H}_{\rm c}$, and not a hyperbolic rotation (that should operate on ${\mathbb M}$).

In conclusion we are led to make a difference between Lorentz boosts and hyperbolic rotations. Hyperbolic rotations are proper Lorentz rotations on the Minkowski  space, operating on 4--vectors, whereas Lorentz boosts may also apply to  sets of physical quantities distincts from  Minkowski  4--vectors.

\subsubsection{Composition of an elliptic and a hyperbolic rotations with a common axis}%**********************************
A proper Lorentz rotation is not necessarily an elliptic rotation, nor a hyperbolic rotation, that is, its representative quaternion does not necessarily take the form $\exp ({\q e}_n\varphi /2)$ or $\exp (\I\,{\q e}_n\delta /2)$ as in Eqs.\ (\ref{eq33}) or (\ref{eq29}). Nevertheless the group of proper Lorentz rotations on the Minkowski  space, denoted SO$_+(1,3)$, is generated by elliptic  rotations and hyperbolic rotations: every proper Lorentz rotation can be shown to be the product (or composition) of elliptic and hyperbolic rotations \cite{PPF7}.
Before we give the most general forms of proper Lorentz rotations on the Minkowski  space (see Section \ref{sect256}), we  examine the composition of an elliptic rotation and a hyperbolic one, the two rotations sharing a common axis.

If $\psi$ is a complex number and ${\q e}_n$ a real unit pure  quaternion, the quaternion $u=\exp ({\q e}_n{\psi /2})$
represents an elliptic rotation if $\psi$ is real, a hyperbolic rotation if $\psi$ is purely imaginary. If $\psi=\varphi +\I\delta$, we obtain
\begin{equation}
  u=\exp \left({\q e}_n{\psi \over 2}\right)=\exp \left({\q e}_n{\varphi \over 2}\right)\exp \left(\I\,{\q e}_n{\delta \over 2}\right)\,,\label{eq86}
  \end{equation}
which shows that $u$ represents the product (or composition) of an elliptic rotation with a hyperbolic one, that is, a proper Lorentz rotation. (A proof is as follows: if $R_1$ and $R_2$ are proper Lorentz rotations, then $\det R_1=\det R_2=1$, so that  $\det R_1\circ R_2=\det R_1\det R_2=1$; and $R_1\circ R_2$ preserves the orientation of both ${\mathbb R}$ and ${\mathbb E}_3$, because both $R_1$ and $R_2$ do.)

The effect on a minquat of the proper Lorentz rotation represented by $u$ above,   is deduced from Eqs.\ (\ref{eq33}) and (\ref{eq29}) and is expressed by
\begin{equation}
  x'=u\,x\,\overline{u}^{\, *}\,.\label{eq34}\end{equation}

\begin{remark}\label{rem26}{\rm The representation of proper Lorentz  rotations by  complex unit quaternions is twofold since both $u$ and $-u$ represent the same rotation, as a consequence of
    \begin{equation}
      u\,x\,\overline{u}^{\, *}=(-u)\,x\,(-\overline{u}^{\, *})\,.\end{equation}
    }
\end{remark}

%*************
%INFINITESIMAL
%*************
\subsubsection{Infinitesimal rotations}\label{sect255}

The quaternion $u$ in Eq.\ (\ref{eq86}), with $\psi$ a complex number and ${\q e}_n$ a real unit pure quaternion, is not a representative of the most general proper Lorentz rotation, as we shall show in this section.

\bigskip
\noindent{\bf Complex unit pure quaternions.} Let ${\q e}_m$ and ${\q e}_n$ be two real unit pure quaternions and let $\varphi$ and $\delta$ be two real numbers; we assume $\varphi\delta\ne 0$ and denote $\alpha$ the angle between ${\q e}_m$ and ${\q e}_n$, so that ${\q e}_m\!\vec\cdot{\q e}_n=\cos\alpha$ . The norm of ${\q e}_m\varphi + \I\,{\q e}_n\delta$ is
\begin{equation}
  N({\q e}_m\varphi + \I\,{\q e}_n\delta)= -({\q e}_m\varphi + \I\,{\q e}_n\delta)^2=\varphi^2-\delta^2+2\I\varphi\delta \,{\q e}_m\!\vec\cdot{\q e}_n=\varphi^2-\delta^2+2\I\varphi\delta\cos\alpha\,.
\end{equation}
Generally, the norm $N({\q e}_m\varphi + \I\,{\q e}_n\delta)$ is a complex number. It is equal to $0$ only if $\delta =\pm \varphi$ and  $\alpha = \pi /2\mod \pi$. Let us assume $N({\q e}_m\varphi + \I\,{\q e}_n\delta)\ne 0$, and let us introduce a complex number $\psi$, such that $\psi^2= N({\q e}_m\varphi + \I\,{\q e}_n\delta)$. The number $\psi$ takes the form $\psi =|\psi|\exp \I\chi$ with
\begin{equation}
  |\psi |=\sqrt[4]{\varphi^4+2\varphi^2\delta^2\cos 2\alpha+\delta^4}\,,\label{eq91n}
\end{equation}
and
\begin{equation}
  \tan 2\chi ={2\varphi\delta\cos\alpha\over \varphi^2-\delta^2}\,.\label{eq92n}\end{equation}
(We remark that since $-1\le \cos2\alpha$, we have $0\le (\varphi^2-\delta^2)^2\le \varphi^4+2\varphi^2\delta^2\cos 2\alpha+\delta^4$, so that taking the fourth root in Eq.\ (\ref{eq91n}) makes sense.)

Since $N({\q e}_m\varphi + \I\,{\q e}_n\delta)\ne 0$, we have $\psi \ne 0$ and we define ${\q e}_u$ by
\begin{equation}
  {\q e}_u={1\over \psi} ({\q e}_m\varphi + \I\,{\q e}_n\delta)
  ={\E^{-\I\chi }\over |\psi |} ({\q e}_m\varphi + \I\,{\q e}_n\delta)\,.\label{eq93n}\end{equation}
The quaternion ${\q e}_u$ is a complex  unit pure  quaternion ($N({\q e}_u)=1$). It is such that ${{\q e}_u}^{\! \! 2}=-1$, and
\begin{equation}
  \exp ({\q e}_m\varphi + \I\,{\q e}_n\delta)=\exp {\q e}_u\psi =\cos \psi + {\q e}_u\sin\psi\,,\end{equation}
  so that $N[ \exp ({\q e}_m\varphi + \I\,{\q e}_n\delta)]=1$ (see Eqs.\ (\ref{eq387}) and (\ref{eq388}) of \ref{appenA}).

\bigskip
\noindent{\bf One-parameter subgroup.} Let $g={\q e}_u\xi$, where ${\q e}_u$ is a unit pure quaternion (it may be complex) and let $\xi$ be a complex number. Let $s$ be a real parameter belonging to an interval $[0, S]$. We have
\begin{equation}
  {\D\over \D s}\exp gs =g\,\exp gs\,,\end{equation}
and
\begin{equation}
  \left.{\D\over \D s}\exp gs\right|_{s =0} =g\,.\end{equation}
The mapping $s\longmapsto \exp gs$ is a one-parameter subgroup of the multiplicative group of quaternions and $g$ is called its {\bf infinitesimal generator}. We have $\exp (gs')\exp (gs'')=\exp [g(s'+s'')]$, and in particular
\begin{equation}
\exp (gs)\exp ( g\,\D s)=\exp [g(s+\D s)]\,.\label{eq97n}\end{equation}

\bigskip
\noindent{\bf Elliptic and hyperbolic infinitesimal rotations.} If ${\q e}_m$ is a real unit pure quaternion and $\eta$ a real number, the quaternion $v=\exp ({\q e}_m\eta \,\D s/2)$ represents an infinitesimal elliptic rotation (axis ${\q e}_m$;  angle $\eta \,\D s$), which operates on minquats according to $x'=v\,x\,\overline{v}^{\, *}$.
Similarly, if ${\q e}_n$ is a real unit pure quaternion and $\zeta$ a real number, the quaternion $w=\exp (\I\,{\q e}_n\zeta \D s/2)$ represents an infinitesimal hyperbolic rotation (axis ${\q e}_n$;  parameter $\zeta \D s$ or angle $\I \zeta \D s$), which operates on minquats according to $x'=w\,x\,\overline{w}^{\, *}$. Since infinitesimal rotations commute, the product of the two rotations is an infinitesimal proper Lorentz rotation (as  product of two proper Lorentz rotations), represented by
\begin{equation}
u=vw=\exp {{\q e}_m\eta\,\D s\over 2}\,\exp {\I\,{\q e}_n\zeta \D s\over 2}=\exp {({\q e}_m\eta +\I\,{\q e}_n\zeta)\D s\over 2}\,,\end{equation}
and operating on minquats according to $x'=v(w\,x\,\overline{w}^{\, *})\overline{v}^{\, *}
=vw\,x\,\overline{vw}^{\, *}
=u\,x\,\overline{u}^{\, *}$.

We assume that the norm of ${\q e}_m\eta+\I\,{\q e}_n\zeta$ is not zero, and we define $\xi =|\xi|\exp\I\chi$ and ${\q e}_u$ with respect to $\eta$ and $\zeta$ as $\psi$ and ${\q e}_u$  with respect to $\varphi$ and $\delta$ in Eqs.\ (\ref{eq91n}--\ref{eq93n}). If $\alpha$ is the angle between ${\q e}_m$ and ${\q e}_n$, we obtain
\begin{equation}
  |\xi |=\sqrt[4]{\eta^4+2\eta^2\zeta^2\cos 2\alpha+\zeta^4}\,,
 \hskip 1cm  \tan 2\chi ={2\eta\zeta\cos\alpha\over \eta^2-\zeta^2}\,.\label{eq92q}\end{equation}
and \begin{equation}
  {\q e}_u  ={\E^{-\I\chi }\over |\xi |} ({\q e}_m\eta + \I\,{\q e}_n\zeta)\,.\label{eq93q}\end{equation}
The unit quaternion $u=\exp ({\q e}_u \xi \,\D s/2)$ represents an infinitesimal proper Lorentz rotation whose axis is the complex unit pure quaternion ${\q e}_u$ and ``angle'' is $\xi\,\D s$.

The mapping $s\longmapsto \exp ({\q e}_u \xi s/2)$ is a one-parameter subgroup, whose generator is  ${\q e}_u\xi /2$.
As a consequence of Eq.\ (\ref{eq97n}), the product of such infinitesimal rotations for $s$ varying from $0$ to $S$  is given by
\begin{equation}
  \exp \left({\q e}_u{\xi\over 2}\int_0^S\D s\right)=\exp\left({\q e}_u{\xi S\over 2}\right)=\exp\left({\q e}_u{\psi\over 2}\right)\,,\label{eq101p}
\end{equation}
where $\psi =\xi S$.

The quaternion $u=\exp ({\q e}_u\psi /2)$ %in Eq.\ (\ref{eq101p})
represents a proper Lorentz rotation more general than that represented by $u$ in Eq.\ (\ref{eq86}), because both ${\q e}_u$ and $\psi$ are complex. The rotation operates on minquats according to  $x'=u\,x\,\overline{u}^{\, *}$\!, or  more explicitly
\begin{equation}
  x'=\exp \left({\q e}_u{\psi \over 2}\right)\,x\,\exp \left(-\overline{{\q e}_u}\,{\overline\psi \over 2}\right)\,.\label{eq101q}\end{equation}

%********************
%FIN DE INFINITESIMAL
%********************

\subsubsection{General form of proper Lorentz rotations}\label{sect256}

We write the representative quaternion of a proper Lorentz rotation in the general form
\begin{equation}
  u=\exp \left({\q e}_u{\psi \over 2}\right)\,,\label{eq100q}\end{equation}
where $\psi$ is a complex number and ${\q e}_u$ a complex pure quaternion. (We recall that $-u$ also represents the same rotation, see Remark \ref{rem26}). The rotation operates on minquats according to $x'=u\,x\,\overline{u}^{\,*}$ (see Eq.\ (\ref{eq101q}) for a more explicit expression).

In the previous section ${\q e}_u$ is a complex pure quaternion, but also a unit quaternion. If $N({\q e}_u)=0$, we shall see, however, that $u$, given by Eq.\ (\ref{eq100q}), still represents a proper Lorentz rotation. Hence we have to consider two kinds of proper Lorentz rotations:
\begin{itemize}
  \item {\bf Regular} proper Lorentz rotations, for which $N({\q e}_u)=1$. The quaternion $u$ can be written $u=\cos(\psi /2)+{\q e}_u\sin(\psi /2)$. The rotation is an elliptic rotation if both $\psi$ and ${\q e}_u$ are real; it is a hyperbolic rotation if $\psi$ is an imaginary number and if ${\q e}_u$ is real.
\item {\bf Singular} proper Lorentz rotations, for which $N({\q e}_u)=0$. (See below for examples and properties.)
\end{itemize}

\bigskip
\noindent {\bf Eigenvalues and eigenminquats (regular rotations).} We consider the linear mapping $q\longmapsto u\,q\,\overline{u}^{\, *}$, defined on ${\mathbb H}_{\rm c}$, where $u$ is given by Eq.\ (\ref{eq100q}). The following quaternions are eigenquaternions \cite{PPF3,PPF7} (an eigenquaternion $q$ is such that $u\,q\,\overline{u}^{\, *}=\Lambda q$) :
\begin{eqnarray}
  x'\rap &=&\rap ({\q e}_0+\I\,{\q e}_u)({\q e}_0+\I\,\overline{{\q e}_u})\,,\label{eq106}\\
  x''\rap &=&\rap ({\q e}_0-\I\,{\q e}_u)({\q e}_0-\I\,\overline{{\q e}_u})\,,\label{eq107}\\
  q'\rap &=&\rap ({\q e}_0-{\q e}_u)({\q e}_0-\overline{{\q e}_u})\,, \\
  q''\rap &=&\rap ({\q e}_0+{\q e}_u)({\q e}_0+\overline{{\q e}_u})\,. 
\end{eqnarray}
(See \ref{appenE} for a proof.)
Only $x'$ and $x''$ are minquats (and are denoted as such). They are eigenminquats of the rotation given by Eq.\ (\ref{eq101q}). The corresponding eigenvalues (for minquats $x'$ and $x''$) are
\begin{eqnarray}
  \Lambda '\rap &=&\rap (C-\I\,S)(\overline{C}+\I\,\overline{S})=C\,\overline{C}+S\,\overline{S}+\I\,C\,\overline{S}-\I\,\overline{C}\,S \,,\label{eq110}\\
   \Lambda ''\rap &=&\rap (C+\I\,S)(\overline{C}-\I\,\overline{S})=C\,\overline{C}+S\,\overline{S}-\I\,C\,\overline{S}+\I\,\overline{C}\,S \,,
\end{eqnarray}
where $C=\cos (\psi /2)$ and $S=\sin (\psi /2)$.

Here are some examples, with ${\q e}_n$ a real unit pure quaternion, $\varphi$ and $\delta$ real numbers:
\begin{itemize}
\item $u=\exp ({\q e}_n\varphi /2)=\cos (\varphi /2)+{\q e}_n\sin(\varphi /2)$. Null eigenminquats are $x'=1+\I\,{\q e}_n$ and $x''=1-\I\,{\q e}_n$. They define eigendirections in ${\mathbb M}$. The corresponding eigenvalues are $\Lambda '=\Lambda ''=1$.
   \item $u=\exp (\I\,{\q e}_n\delta /2)=\cosh (\delta /2)+\I\,{\q e}_n\,\sinh (\delta /2)$. Null eigenminquats are $x'=1+\I\,{\q e}_n$ and $x''=1-\I\,{\q e}_n$. The corresponding eigenvalues are $\Lambda '=\exp \delta$ and $\Lambda ''=\exp (-\delta)$.
\end{itemize}

\subsubsection{Singular proper Lorentz rotations}%****************************************

In Section \ref{sect255}, real unit  pure  quaternions ${\q e}_m$ and ${\q e}_n$ are assumed to be not orthogonal to each other. We now assume that they are: ${\q e}_m\vec\cdot{\q e}_n= 0$, so that
$N({\q e}_m \pm \I\,{\q e}_n)=0$.

Let ${\q e}_u={\q e}_m + \I\,{\q e}_n$, let $\eta$ be a real number, and let $s$ be a real parameter in the interval $[0,S]$. Both $\exp ({\q e}_m\eta\,\D s/2)$  and $\exp (\I\,{\q e}_n\eta\,\D s/2)$ represent infinitesimal proper Lorentz rotations and their commutative product  $\exp ({\q e}_u\eta\,\D s/ 2)=\exp ({\q e}_m\eta\,\D s/ 2)\exp (\I\,{\q e}_n\eta\,\D s/ 2)$ also does. The mapping $s\longmapsto  \exp ({\q e}_u\eta\, s/ 2)$ is a one-parameter subgroup whose generator is ${\q e}_u\eta /2$. If $\varphi =\eta S$, the pro\-duct of infinitesimal rotations, represented by $\exp ({\q e}_u\eta\,\D s/ 2)$, for $s$ varying from $0$ to $S$ is a proper Lorentz rotation, represented by
\begin{equation}
  u_{_+}=\exp {{\q e}_u\varphi\over 2} %\exp\left[({\q e}_m + \I\,{\q e}_n){\eta S\over 2}\right]
  =\exp\left[({\q e}_m + \I\,{\q e}_n){\varphi\over 2}\right]={\q e}_0+({\q e}_m +\I\,{\q e}_n){\varphi\over 2}\,.\label{eq109s}\end{equation}

The mapping $x\longmapsto u_{_+}x\,\overline{u_{_+}\!\!}^{\, *}$ is a proper Lorentz rotation; the minquat $x={\q e}_0+\I\,{\q e}_m{\q e}_n$ is an eigenminquat, the eigenvalue being equal to 1 (see \ref{appenF} for a proof). There is no other null-minquat eigendirection.

The result holds if we replace  ${\q e}_u={\q e}_m + \I\,{\q e}_n$ with  ${\q e}_u={\q e}_m - \I\,{\q e}_n$, and $u_{_+}$ with
\begin{equation}
  u_{_-}=\exp {{\q e}_u\varphi\over 2}
  ={\q e}_0+({\q e}_m - \I\,{\q e}_n){\varphi\over 2}\,,\end{equation}
the minquat $x={\q e}_0-\I\,{\q e}_m{\q e}_n$ providing the unique null-minquat eigendirection of the proper Lorentz rotation $x\longmapsto u_{_-}x\,\overline{u_{_-}\!\!}^{\, *}$.

Quaternions $u_{_+}$ and $u_{_-}$ represent {\bf singular} proper Lorentz rotations. Such a rotation has only one null-minquat eigendirection in ${\mathbb M}$.

\begin{remark} {\rm Here,  ``singular'' means that the considered Lorentz rotation has only one null-minquat eigendirection and is represented by $u$ as in Eq.\ (\ref{eq100q}), with $N({\q e}_u)=0$. As mentioned by Synge \cite{Syn}, it is not singular in the usual sense of possessing no inverse (the inverse of  $\exp ({\q e}_u{\psi / 2})=1+ ({\q e}_u{\psi / 2})$ is $\exp (-{\q e}_u{\psi / 2})=1- ({\q e}_u{\psi / 2})$, if ${{\q e}_u}^{\! 2}=0$). Singular proper Lorentz rotations are also called {\bf null} rotations \cite{Gou,Pen}. ``Regular'', as used above, means that ${\q e}_u$ in Eq.\ (\ref{eq100q}) is such that $N({\q e}_u)=1$. The corresponding rotation admits two  null-minquat eigendirections.
 }
  \end{remark}

%*********************************
%COMPOROT
%*********************************

\subsubsection{Some compositions of proper Lorentz rotations}\label{sect258}

\noindent{\bf Composition of two elliptic rotations.}
We consider the elliptic rotation of angle $\varphi$ around ${\q e}_m$ followed by the elliptic rotation of angle $\theta$ around the axis ${\q e}_n$. We assume $-\pi<\varphi <\pi$ and  $\varphi\ne 0$ ; also $-\pi<\theta <\pi$ and $\theta\ne 0$. The composition of the two rotations is an elliptic rotation represented by
\begin{equation}
u=\exp\left({\q e}_n{\theta\over 2}\right)\,\exp\left( {\q e}_m{\varphi\over 2}\right)\,.\end{equation}
We shall show that $u$ takes the form $\exp ({\q e}_q\psi /2)$, where $\psi$ is a real number and ${\q e}_q$ a real pure quaternion.

We first remark that if ${\q e}_m=\pm \,{\q e}_n$, we obtain $u=\exp [{\q e}_n(\theta\pm \varphi )/2]$, that is ${\q e}_q={\q e}_n$ and $\psi =\theta\pm\varphi$.

We then assume ${\q e}_m\ne\pm {\q e}_n$. We have
\begin{eqnarray}
u&\!\!\!\!=&\!\!\!\!\left(\cos{\theta\over 2}+{\q e}_n\sin{\theta\over 2}\right)\left(\cos{\varphi\over 2}+{\q e}_m\sin{\varphi\over 2}\right) \nonumber \\
&\!\!\!\! = &\!\!\!\! \cos{\theta\over 2}\cos{\varphi\over 2}-{\q e}_n\vec\cdot{\q e}_m\sin{\theta\over 2}\sin{\varphi\over 2}\nonumber \\
& & \hskip 2cm +\;{\q e}_n\cos{\theta\over 2}\sin{\varphi\over 2}+{\q e}_m\cos{\varphi\over 2}\sin{\theta\over 2}
+{\q e}_n\vec \times {\q e}_m\sin{\theta\over 2}\sin{\varphi\over 2}\nonumber \\
 &\rap =\rap & u_0+V\,.\label{eq64a}
\end{eqnarray}
where $u_0$ is a real number and $V$ a real pure  quaternion.

Since  $\exp ({\q e}_m\varphi /2)$ and  $\exp ({\q e}_n\theta /2)$ are unit quaternions, their product $u$ is also a unit quaternion. From $1=N(u)={u_0}^2+\|V\|^2$ we conclude that $|u_0|\le 1$,  and there is a number $\psi$ ($-2\pi \le \psi \le 2\pi$) such that %$\cos (\psi /2)=u_0$, that is
\begin{equation}
\cos{\psi\over 2}=u_0=\cos{\theta\over 2}\cos{\varphi\over 2}-{\q e}_n\vec\cdot{\q e}_m\sin{\theta\over 2}\sin{\varphi\over 2}\,.\label{eq66a}\end{equation}
(So far $\psi$ is defined up to sign.)

In \ref{appenB}, we prove that $V$ takes the form $V={\q e}_q\sin (\psi /2)$.
We eventually obtain that $u$ takes the form
\begin{equation}
u=\exp\left({\q e}_q{\psi\over 2}\right)=\cos{\psi\over 2}+{\q e_q}\sin{\psi\over 2}\,,\label{eq65a}
\end{equation}
with
\begin{equation}
{\q e}_q={1\over \sqrt{1-\cos^2\displaystyle{\psi\over 2}}}\left({\q e}_n\cos\displaystyle{\theta\over 2}\sin{\varphi\over 2}+{\q e}_m\cos{\varphi\over 2}\sin{\theta\over 2}
+{\q e}_n\vec \times {\q e}_m\sin{\theta\over 2}\sin{\varphi\over 2}\right)
\,.\label{eq67a}
\end{equation}

Equation (\ref{eq66a}) provides the angle of the rotation, up to sign; Eq.\ (\ref{eq67a}) provides the axis. In fact the sign of ${\q e}_q\sin(\psi /2)$ is perfectly defined by Eq.\ (\ref{eq64a}): once ${\q e}_q$ is defined  by Eq.\ (\ref{eq67a}), the sign of $\psi$ is perfectly defined, according to  Eq.\ (\ref{eq64a}). (We obtain the same rotation if we change $\psi$ into $-\psi$ and ${\q e}_q$ into $-{\q e}_q$,  the sign of ${\q e}_q\sin(\psi /2)$ being unchanged.)

The effect of the compound rotation on a minquat $x$ is given by
\begin{equation}
x'=\E^{{\q e}_q\psi /2}\,x\,\E^{{-\q e}_q\psi /2}=\E^{{\q e}_n\theta /2}\,\E^{{\q e}_m\varphi /2}\,x\,\E^{{-\q e}_m\varphi /2}\,\E^{-{\q e}_n\theta /2}\,.\end{equation}

\begin{remark}\label{rem25}{\rm 
The interest of quaternions in representing elliptic rotations on the Minkowski space is twofold:
\begin{enumerate}
\item The most representative (most ``natural'') parameters of a rotation
  are: its axis, given by the unit vector ${\qv e}_n$, and the angle of rotation $\varphi$ around ${\qv e}_n$. 
  The rotation is represented by $\exp ({\q e_n}\varphi /2)$, which explicitly includes the rotation angle as well as the axis. Moreover such a notation has an additional operational virtue: under the considered rotation an arbitrary minquat $x$ is transformed into
\begin{equation}
  x'=\E^{{\q e}_n\varphi /2}\,x\,\E^{{-\q e}_n\varphi /2}\,,\label{eq115}\end{equation}
and Eq.\ (\ref{eq115}) can actually be applied to derive $x'$ from $x$. 

\item The exponential representation of a rotation is preserved in the product of rotations. That is,
  the product of   quaternions $\exp ({\q e}_m\varphi /2)$ and  $\exp ({\q e}_n\theta /2)$ ``naturally'' takes the form
$\exp ({\q e}_q\psi /2)=\cos (\psi /2)+{\q e}_n\sin (\psi /2)$, which explicity highlights the angle and the axis of the compound rotation. 
\end{enumerate}

The two previous items hold for elliptic rotations on the 3-dimensional Euclidean vector-space ${\mathbb E}_3$ (see Sect.\ \ref{sect22}).
}
\end{remark}

\bigskip
\noindent{\bf Composition of two hyperbolic rotations.}
 The composition of the hyperbolic rotation  of parameter $\delta_m$ along ${\q e}_m$ followed by the hyperbolic rotation  of parameter $\delta_n$ along ${\q e}_n$ (we assume $\delta_m\delta_n\ne 0$) is represented by
\begin{equation}
u=\exp\left(\I {\q e}_n{\delta_n\over 2}\right)\,\exp\left(\I {\q e}_m{\delta_m\over 2}\right)\,.
\end{equation}
Then
\begin{eqnarray}
u&\!\!\!\!=&\!\!\!\!\left(\cosh{\delta_n\over 2}+\I{\q e}_n\sinh{\delta_n\over 2}\right)\left(\cosh{\delta_m\over 2}+\I{\q e}_m\sinh{\delta_m\over 2}\right) \label{eq70a} \\
&\!\!\!\!=&\!\!\!\! \cosh{\delta_n\over 2}\cosh{\delta_m\over 2}+{\q e}_n\vec\cdot {\q e}_m\sinh{\delta_n\over 2}\sinh{\delta_m\over 2}\nonumber \\
& & \hskip .7cm +\; \I {\q e}_n\sinh{\delta_n\over 2}\cosh{\delta_m\over 2}+
\I{\q e}_m\sinh{\delta_m\over 2} \cosh{\delta_n\over 2}
-{\q e}_n\vec\times{\q e}_m\sinh{\delta_n\over 2}\sinh{\delta_m\over 2}\,.\nonumber
\end{eqnarray}
(Equation (\ref{eq70a}) is no more than Eq.\ (\ref{eq64a}) in which $\varphi$ is replaced with $\I\delta_m$ and $\theta$ by $\I\delta_n$.)

If ${\q e}_m$ and ${\q e}_n$ are collinear  (${\q e}_m=\pm\,{\q e}_n$), then $u$ is a hyperbolic rotation along ${\q e}_m$. If 
${\q e}_m$ and ${\q e}_n$ are not collinear, then ${\q e}_n\vec\times {\q e}_m$ is not zero and $u$ does not represent a pure hyperbolic rotation anymore, because of the term ${\q e}_n\vec \times{\q e}_m\sinh(\delta_n/ 2)\sinh(\delta_m/ 2)$, which is not imaginary. There is an additional elliptic rotation (see Section \ref{sect62}).

\medskip
\noindent {\bf Application to Relativity.}
The composition of two hyperbolic rotations is used in the special theory of relativity to obtain the relativistic composition-law of velocities. Let us consider three Galilean frames ${\cal S}$, ${\cal S}'$ and ${\cal S}''$. The relative velocity of ${\cal S}'$ with respect ${\cal S}$ is $v_m\,{\q e}_m$ and that of ${\cal S}''$ with respect to ${\cal S}'$ is $v_n\,{\q e}_n$. The corresponding rapidities are $\delta_m$ and $\delta_n$ with $\tanh\delta_m=-\beta_m=-v_m/{\rm c}$ and  $\tanh\delta_n=-\beta_n=-v_n/{\rm c}$  (see Section \ref{sect253} for a definition).

If ${\q e}_n={\q e}_m$, we have $u=\exp [\I\,{\q e}_m(\delta_m+\delta_n)/2]=\exp (\I{\q e}_m\delta /2)$, with $\delta =\delta_m+\delta_n$, so that $u$ is also a hyperbolic rotation, corresponding to the relative velocity of ${\cal S}''$ with respect to ${\cal S}$ in the form $v\,{\q e}_m$, with $v={\rm c}\beta =-{\rm c}\tanh \delta$. The equality $\delta =\delta_m+\delta_n$ constitutes the {\bf addition law of rapidities} in that case.
From $\delta =\delta_m+\delta_n$,  we deduce
\begin{equation}
\beta =-\tanh\delta =-\tanh (\delta_m+\delta_n)=-{\tanh\delta_m+\tanh\delta_n\over 1+\tanh\delta_m\tanh\delta_n}
={\beta_m+\beta_n\over 1+\beta_m\beta_n}\,,\label{eq118v}\end{equation}
that is
\begin{equation}
v=
{v_m+v_n\over 1+\displaystyle{v_mv_n\over {\rm c}^2}}\,,\label{eq119v}\end{equation}
which is the relativistic composition-law for collinear velocities.

If ${\q e}_m$ and ${\q e}_n$ are not collinear, the product of the two hyperbolic rotations includes a pure rotation in addition to a boost and leads to the effect known as Thomas precession (see Section \ref{sect62}).

%***************
%FIN DE COMPOROT
%***************

\subsubsection{Superposition of finite rotations}\label{sect259}
The infinitesimal rotations represented by 
$\exp ({\q e}_v\eta \,\D s/2)$ and $\exp ({\q e}_w\zeta \,\D s/2)$ commute and
\begin{equation}
  \exp \left({\q e}_v{\eta\over 2} \,\D s\right) \exp \left({\q e}_w{\zeta\over 2} \,\D s\right) =\exp \left[{1\over 2}({\q e}_v \eta +{\q e}_w\zeta )\,\D s\right]\,.
\end{equation}
The succession of such rotations for $s$ varying from $0$ to $S$ is represented by
\begin{equation}
  u=\exp \left[({\q e}_v \eta +\I\,{\q e}_w\zeta ) {S\over 2}\right]\,,
\end{equation}
which takes the form $u=\exp ({\q e}_u\psi /2)$, where $\psi$ is a complex number and ${\q e}_u$ a complex pure quaternion.

We will say that $u$ represents the {\bf superposition} of finite rotations represented by quaternions $\exp ({\q e}_v\eta S/2)$ and $\exp ({\q e}_w\zeta S/2)$. In particular, %with the previous notations,
the superposition of finite proper Lorentz rotations represented by $\exp ({\q e}_m\varphi /2)$ and $\exp (\I\,{\q e}_n\delta /2)$ is represented by $u=\exp ({\q e}_u\psi /2)$, where ${\q e}_u$ is given by
Eq.\ (\ref{eq93n}) and $\psi$ is defined by Eqs.\ ({\ref{eq91n}) and (\ref{eq92n}); it is the pro\-duct of infinitesimal rotations of the form
$\exp [({\q e}_m\eta +\I\,{\q e}_n\zeta )\,\D s/2]$, with $\eta =\varphi /S$ and $\zeta =\delta / S$, that are generated by infinitesimal elliptic rotations $\exp ({\q e}_m\eta \,\D s /2)$ and infinitesimal hyperbolic rotations $\exp (\I\,{\q e}_n\zeta \,\D s /2)$.

The complex quaternion $u$ represents a proper Lorentz rotation and operates on minquats according to  Eq.\ (\ref{eq34}): $x'=u\,x\,\overline{u}^{\, *}$.

\begin{remark} {\rm The superposition of  the rotations represented by  $\exp ({\q e}_m\varphi /2)$ and $\exp (\I\,{\q e}_n\delta /2)$ is distinct from their composition, because for finite rotations, in general
    \begin{equation}
      \exp {{\q e}_m\varphi+\I\,{\q e}_n\delta \over 2}\ne \exp {{\q e}_m\varphi \over 2}\exp {\I\,{\q e}_n\delta \over 2}\,.\label{ineq126}\end{equation}
    The left member of Inequality (\ref{ineq126}) represents the superposition of the two rotations, and the right member their composition (or product).

    A physical example of a superposition of two rotations is given in Section \ref{sect544}.
    }
  \end{remark}

\subsection{The four-nabla and its transforms}%************************

\subsubsection{The four-nabla differential form}

We use the notation
\begin{equation}
\partial_\mu ={\partial \over\partial x_\mu}\,,\end{equation}
where the $x_\mu$'s are four coordinates in the Minkowski  space. (In Relativity they are spatio-temporal coordinates with $x_0={\rm c}t$.)
The quaternionic representation of the four-nabla differential operator is defined by
\begin{equation}
\nabla ={\q e}_0\partial_0+\I ({\q e}_1\partial_1 +{\q e}_2\partial_2+{\q e}_3\partial_3)\,.\end{equation}

\subsubsection{Transformation under a hyperbolic rotation}

We consider a hyperbolic rotation on the Minkowski  space, given by Eqs.\ (\ref{eq21b}--\ref{eq24b}), linking coordinates $\{x'_\mu\}$ to $\{x_\mu\}$. We denote
\begin{equation}
  \partial_\mu '={\partial \over \partial x'_\mu}\,.\end{equation}
We apply the ``chain rule''
\begin{equation}
 \partial_\mu= {\partial \over \partial x_\mu}=\sum_{\nu =0}^{\nu =3}{\partial \over \partial x'_\nu}\,{\partial x'_\nu\over \partial x_\mu}=\sum_{\nu =0}^{\nu =3}\partial_\mu x'_\nu \partial'_\nu\,,
  \end{equation}
and  use Eqs.\ (\ref{eq21b}--\ref{eq24b}) to obtain
\begin{equation}
\partial_0=\partial_0x'_0\partial_0'+\partial_0x'_1\partial_1'+
\partial_0x'_2\partial_2'+\partial_0x'_3\partial_3'=\cosh\delta\; \partial_0' +\sinh\delta\;\partial_1'\,.\end{equation}
We also obtain
\begin{equation}
\partial_1=\partial_1x'_0\partial_0'+\partial_1x'_1\partial_1'+
\partial_1x'_2\partial_2'+\partial_1x'_3\partial_3'=\cosh\delta\; \partial_1' +\sinh\delta\;\partial_0'\,,\end{equation}
and 
\begin{equation}
\partial_2=\partial_2'\,,\hskip 1cm \partial_3=\partial_3'\,.\end{equation}
 Then Eqs.\ (\ref{eq21b}--\ref{eq22b}) lead us to write
\begin{eqnarray}
\partial_0'\rap&=&\rap \cosh\delta\; \partial_0 -\sinh\delta\;\partial_1\,,\label{eq40}\\
\partial_1'\rap&=&\rap -\sinh\delta \;\partial_0 + \cosh\delta \;\partial_1\,,\label{eq41}\\
\partial_2'\rap&=&\rap \partial_2\,,\label{eq42}\\
\partial_3'\rap&=&\rap \partial_3\,.\label{eq43}
\end{eqnarray}
The comparison of Eqs.\ (\ref{eq40}--\ref{eq43}) and Eqs.\ (\ref{eq21b}--\ref{eq24b}) shows that partial derivatives transform according to the inverse hyperbolic rotation of the previous one ($\delta$ is replaced with $-\delta$). With quaternions, we write
\begin{equation}
  \nabla '=\exp \left(-\I \,{\q e}_1{\delta\over 2}\right)\nabla \exp \left(-\I \,{\q e}_1{\delta\over 2}\right)\,.\label{eq44}\end{equation}
If the hyperbolic rotation is along ${\q e}_n$ we have
\begin{equation}
  \nabla '=\exp \left(-\I \,{\q e}_n{\delta\over 2}\right)\nabla \exp \left(-\I\, {\q e}_n{\delta\over 2}\right)\,.\label{eq45}\end{equation}

\subsubsection{Transformation under an elliptic rotation}%********************
The same method is applied to the rotation of angle $\varphi$ around the axis ${\q e}_1$, given by Eq.\ (\ref{eq30}). We obtain
\begin{eqnarray}
\partial_0'\rap &=&\rap  \partial_0\,,\\
\partial_1'\rap&=&\rap \partial_1\,,\\
\partial_2'\rap&=&\rap \cos\varphi \,\partial_2-\sin\varphi\,\partial_3\,,\\
\partial_3'\rap&=&\rap \sin\varphi\,\partial_2+\cos\varphi\,\partial_3\,,\end{eqnarray}
that is
\begin{equation}
\nabla '=\exp \left({\q e}_1{\varphi\over 2}\right)\nabla \exp \left(- \,{\q e}_1{\varphi\over 2}\right)\,.\label{eq50a}\end{equation}
We conclude that under an elliptic rotation the four-nabla transforms as coordinates do.

If the rotation is around ${\q e}_n$, we have
\begin{equation}
\nabla '=\exp \left({\q e}_n{\varphi\over 2}\right)\nabla \exp \left(- \,{\q e}_n{\varphi\over 2}\right)\,.\label{eq50}\end{equation}

\subsubsection{General transformation}
Let $u$ be a proper Lorentz rotation on the Minkowski  space. From Eqs.\ (\ref{eq45}) and (\ref{eq50}) we deduce that under this rotation, the four-nabla is transformed according to
\begin{equation}
\nabla '=\overline u\,\nabla\, u^*\,.\label{eq51}\end{equation}
Equation (\ref{eq51}) includes both Eqs.\ (\ref{eq45}) and (\ref{eq50}).

\subsection{Elliptic rotations on ${\mathbb H}_{\rm c}$}%********************************************
An elliptic rotation on minquats is represented by a quaternion of the form $u=\exp ({\q e}_n\psi /2)$, where ${\q e}_n$ is a  real unit pure  quaternion (the rotation axis) and $\psi$ a real number (the rotation angle). A minquat $x$ is transformed into the minquat $x'$ such that 
\begin{equation}
  x'=u\,x\,u^*\,.\label{eq55}\end{equation}
The rotation can be extended to complex quaternions: we consider the mapping defined on ${\mathbb H}_{\rm c}$ by
\begin{equation}
  q\;\longmapsto\; q'=u\,q\,u^*\,,\label{eq56}\end{equation}
where $u$ is a unit quaternion, so that $N(q')=N(q)$.

If $u=\exp ({\q e}_n\psi /2)$, with $\psi$ a real number, we prove in \ref{appenC} that the mapping $R_{\rm r}$, defined by $R_{\rm r}(q)=u\,q\,u^*$, is a rotation on ${\mathbb H}_{\rm c}$, called a  {\bf real elliptic} rotation.

We may also consider imaginary values of $\psi$: if $\psi =\I\theta$, we have $u=\exp (\I \,{\q e}_n\theta /2)$, and we define a mapping $R_{\I}$ on ${\mathbb H}_{\rm c}$ by Eq.\ (\ref{eq56}) once more. We prove  in \ref{appenC} that $R_\I$ is a rotation, called an {\bf imaginary elliptic} rotation. A more general elliptic rotation is obtained for $\psi =\varphi+\I\theta$ and corresponds to the notion of a {\bf real-axis complex elliptic}
rotation, represented by
\begin{equation}
u=\exp\left({\q e}_n{\psi \over 2}\right)=\exp\left({\q e}_n{\varphi +\I\theta \over 2}\right)=\exp\left({\q e}_n{\varphi \over 2}\right)\,\exp\left(\I {\q e}_n{\theta \over 2}\right)\,,\end{equation} 
and operating according to Eq.\ (\ref{eq56}), that is
\begin{equation}
  q'= \exp\left({\q e}_n{\varphi +\I\theta \over 2}\right)\, q\,
  \exp\left(-\,{\q e}_n{\varphi +\I\theta \over 2}\right)\,.\end{equation}
Finally the most general elliptic rotation is represented by
\begin{equation}
  u=\exp\left({\q e}_u{\psi \over 2}\right)\,,\end{equation}
where ${\q e}_u$ is a complex unit pure quaternion and $\psi$ a complex number (see \ref{appenC}). We call it a {\bf complex-axis elliptic} rotation; it operates on quaternions according to
\begin{equation}
  q'= u\,q\,u^*=\exp\left({\q e}_u{\psi \over 2}\right)\, q\,
  \exp\left(-{\q e}_u{\psi \over 2}\right)\,.\end{equation}

  A property of the previous elliptic rotations is the following.
Let $q=q_0{\q e}_0+ V$ ($V$ is the vector part of $q$; it is a pure quaternion) and let  $q'=u\,q\,u^*=q'_0{\q e}_0+V'$. Since $uu^*=1$,  we have $uq_0{\q e}_0u^*=q_0{\q e}_0$. We also have $(u\,V\,u^*)^*=u\,{V}^{\! *}\,u^*=-u\,V\,u^*$, which means that $u\,V\,u^*$ is a pure quaternion. Then necessarily $V'=u\,V\,u^*$ and  $q'_0=q_0$. An  elliptic rotation on ${\mathbb H}_{\rm c}$ operates separately on scalar and vector parts of complex quaternions: the scalar part is preserved, and the vector part is transformed into the vector part of the image. (The same property holds for real elliptic rotations, so that it is legitimate to extend the notion of elliptic rotation to rotations with complex angles and to designate them as ``complex elliptic rotations''.)

\begin{remark} {\rm  Formally, a complex elliptic rotation on ${\mathbb H}_{\rm c}$ is no more than a usual elliptic rotation whose angle is a complex number, because its representative quaternion  can be written 
     \begin{equation}
      u=\exp{{\q e}_n\psi\over 2}=\cos{\psi\over 2}+{\q e}_n\sin{\psi\over 2}\,,\end{equation}
     where $\psi$ is a complex number.

     It can also be understood as a rotation on ${\mathbb C}^3$, and is formally similar to a usual elliptic rotation on ${\mathbb E}_3$, whose angle becomes a complex number. 
   For example, the matrix form of the rotation of angle $\psi$ and axis ${\qv e}_1$ on ${\mathbb C}^3$ is similar to the submatrix ${\cal R}_3$ in Eq.\ (\ref{eq65}) or Eq.\ (\ref{eq30}) and is written
\begin{equation}
\!\begin{pmatrix}z'_1 \\ z'_2 \\ z'_3 \\
\end{pmatrix}
=\begin{pmatrix}
1&0&0 \\ 0& \cos \psi &-\sin \psi  \\
0&\sin\psi &\cos\psi \\
\end{pmatrix}
\begin{pmatrix} z_1 \\ z_2 \\ z_3 \\
\end{pmatrix}\,,
\label{eq30b}
\end{equation}
where both $(z_1,z_2,z_2)$ and $(z'_1,z'_2,z'_3)$ belong to ${\mathbb C}^3$, and $\psi =\varphi +\I\theta$.

Finally, we indicate that $u$ may also be written
  \begin{eqnarray}
    u =\exp {\q e}_n{\psi\over 2}\rap &=&\rap\exp\left({\q e}_n{\varphi+\I\theta\over 2}\right)=\cos\left({\varphi+\I\theta\over 2}\right)+{\q e}_n\sin \left({\varphi+\I\theta\over 2}\right) \\
    &=&\rap \cos{\varphi\over 2}\cosh{\theta\over 2}-\I \sin{\varphi\over 2}\sinh{\theta\over 2}
    +{\q e}_n\!\left(\sin{\varphi\over 2}\cosh{\theta\over 2}+\I \cos{\varphi\over 2}\sinh{\theta\over 2}\right)\,.
    \nonumber
  \end{eqnarray}
}
\end{remark}

\subsection{Classification of rotations}\label{sect17}%*******************************

At this point, it is useful to recapitulate  all the rotations that have been introduced and to inspect whether they preserve the scalar product of pure quaternions. Those rotations are represented by unitary quaternions in the form
\begin{equation}
  u=\exp \left({\q e}_u \,{\psi \over 2}\right)\,,\end{equation}
where $\psi$ is a complex number and ${\q e}_u$ is generally a complex unit pure quaternion or possibly a zero-norm pure quaternion.

We point out, however, that knowing the representative quaternion of a rotation is not enough to characterize the rotation. The knowledge of the vector space on which the rotation is defined is also necessary and  induces the way the rotation operates. For example a quaternion of the form $\exp (\I\,{\q e}_n\theta /2)$ ($\theta$ a real number) may represent a hyperbolic rotation (a proper Lorentz
rotation) on ${\mathbb M}$ and it operates according to $x'=\exp (\I\,{\q e}_n\theta /2)\,x\,\exp (\I\,{\q e}_n\theta /2)$; it may also represent a complex elliptic rotation on ${\mathbb H}_{\rm c}$ operating according to
 $q'=\exp (\I\,{\q e}_n\theta /2)\,q\,\exp (-\I\,{\q e}_n\theta /2)$.

Of course all the mentioned rotations are represented by unit quaternions and preserve the dot products and the norms of quaternions. 
 
The rotations we are dealing with are the followings:
\begin{enumerate}
\item {\its Elliptic rotations on ${\mathbb H}_{\rm c}$:} by definition, they operate on complex quaternions, according to Eq.\ (\ref{eq56}), that is
\begin{equation}
  q'=u\,q\,u^*\,,\label{eq151a}\end{equation}
and they preserve  the dot products and the norms of complex quaternions.
 They preserve the scalar parts of quaternions and they transform a pure quaternion into a pure quaternion. They preserve the scalar product of real pure quaternions (that is, they preserve the scalar product in ${\mathbb E}_3$).

Particular elliptic rotations on ${\mathbb H}_{\rm c}$ are (${\q e}_n$ denotes a real unit pure quaternion):
\begin{enumerate}
\item {\its Real  (elliptic) rotations}, represented by $u=\exp ({\q e}_n\varphi  /2)$, with $\varphi$ a real number. They are extension of elliptic rotations (defined on ${\mathbb E}_3$) to ${\mathbb H}_c$.
\item {\its Imaginary (elliptic) rotations,} represented by $u=\exp ( \I\, {\q e}_n\theta /2)$, with $\theta$ a real number.
\end{enumerate}

\item {\its Proper Lorentz rotations:} they operate on minquats, that is, on quaternions of the form
\begin{equation}
x=x_0{\q e}_0+\I (x_1{\q e}_1+x_2{\q e}_2+x_3{\q e}_3)\,,\hskip .5cm x_\mu \in {\mathbb R}\,,\end{equation}  according to
\begin{equation}
x'=u\,x\,\overline{u}^{\, *}\,.\label{eq61a}\end{equation}

Particular proper Lorentz rotations are (${\q e}_n$ is a real unit pure quaternion):
\begin{enumerate}
  \item {\its Elliptic rotations}, represented by quaternions of the form $u=\exp ( {\q e}_n\varphi  /2)$, where $\varphi$ is a real number. Then $\overline{u}=u$ and Eq.\ (\ref{eq61a}) reduces to
\begin{equation}
  x'=u\, x\,\overline{u}^{\, *}=u\, x\,u^*=\exp \left({\q e}_n{\varphi \over 2}\right)\,x\,\exp \left(-\,{\q e}_n{\varphi \over 2}\right)
  \,.\end{equation}
Elliptic rotations  transform  vector parts of minquats into vector parts of their images (and scalar parts into scalar parts). They preserve both the dot product of minquats and the scalar product of their vector parts.
\item {\its Hyperbolic rotations},
  represented by $u=\exp (\I\, {\q e}_n\delta /2)$, where $\delta$ is a real number. Then $\overline{u}^{\, *}=u$ and Eq.\ (\ref{eq61a}) reduces to
\begin{equation}
x'=u\, x\,\overline{u}^{\, *}=u\,x\,u =\exp \left(\I \,{\q e}_n{\delta \over 2}\right)x\exp \left(\I\, {\q e}_n{\delta \over 2}\right) \,.\label{eq62a}\end{equation}
A hyperbolic rotation preserves the dot product of minquats, but not the scalar product of their vector parts. If $x=x_0{\q e}_0+ \I V$ is a minquat (with $x_0\ne 0$ and $V$ a pure real quaternion) and if $x'$ is its image by a hyperbolic rotation, with $\delta \ne 0$, then $x_0{\q e}_0$ is  not  transformed into the scalar part of $x'$, and $\I V$ is not necessarily transformed into the vector part of $x'$.
\item {\its Singular rotations}, represented by $u =\exp [({\q e}_m\pm \I\,{\q e}_n)\varphi /2]= {\q e}_0+({\q e}_m\pm \I\,{\q e}_n)\varphi /2$, where $\varphi $ is a real number and ${\q e}_m$ and ${\q e}_n$ two orthogonal real unit pure quaternions. They operate according to
  \begin{equation}
    x' =u\, x\,\overline{u}^{\, *}=\left[{\q e}_0+({\q e}_m\pm \I\,{\q e}_n){\varphi \over 2}\right]\,x\,
    \left[{\q e}_0-({\q e}_m\mp \I\,{\q e}_n){\varphi \over 2}\right]\,.\end{equation}

\end{enumerate}
\end{enumerate}

%**************************************
\section{Electromagnetism}\label{sect3}
%**************************************

\subsection{Classical expressions of the fields}%*******************************
Both the electric and magnetic fields are described as vector fields that are deduced from the scalar and vector potentials, respectively denoted $\Phi$ and $\vec a$. 
The three-dimensional nabla operator is denoted by $\vec \nabla$. If ${\rm c}$ denotes light velocity, the electric-field vector  is
\begin{equation}
\vec E=-\vec\nabla\Phi-{\rm c}\,\partial_0\vec a\,,\label{eq52}\end{equation}
and the magnetic-field vector (also called magnetic-induction vector, to distinguish it from the magnetic-field $\vec H$)
\begin{equation}
  \vec B=\vec\nabla\vec \times\vec a\,.\label{eq53}\end{equation}
At every point $(x_1,x_2,x_3)$ and time $t$,  by Eq.\ (\ref{eq52}) the electric field is
\begin{equation}
  \vec E(x_0,x_1,x_2,x_3)=-\vec\nabla\Phi (x_0,x_1,x_2,x_3)-{\rm c}\,\partial_0\vec a (x_0,x_1,x_2,x_3)\,,\label{eq52c}\end{equation}
where $x_0={\rm c}t$. A similar coordinate dependence  can be established for the magnetic field from Eq.\ (\ref{eq53}).

\subsection{Expression from the four-potential}\label{sect32}%*******************************

\noindent{\bf Four-potential.}
In the coordinate system $\{ x_\mu\}$, attached to a Galilean inertial frame ${\cal S}$, the 4-potentiel is
$\vec A=(\Phi /{\rm c}, \vec a)=(\Phi /{\rm c}, a_1,a_2,a_3)$, where $\Phi$ and the $a_j$'s are functions of the $x_\mu$'s. The 4--potential is represented by the minquat 
 $A=(\Phi /{\rm c})\,{\q e}_0+\I a$, where $a=a_1{\q e}_1+a_2{\q e}_2+a_3{\q e}_3$.

If $A'$ denotes the quaternionic representation of the 4--potential in the coordinate system $\{ x'_\mu\}$ of another Galilean inertial frame ${\cal S}'$ in uniform translatory motion with respect to ${\cal S}$, since $A$ and $A'$ are minquats, they are linked by a proper Lorentz rotation, represented by $u$, and $A'=u\,A\,\overline{u}^{\, *}$.
The proper rotation is  hyperbolic only if the spatial-coordinate axes are parallel to each other ($x_j\parallel x'_j$, $j=1,2,3$).  More generally, the proper rotation may be the compound of hyperbolic and elliptic rotations.

\medskip
\noindent{\bf Field complex-vector.}
The calculation of $\nabla \! A$ gives
\begin{equation}
\nabla\! A=\left[\begin{matrix} \partial_0\\ \I\partial_1\\ \I\partial_2 \\ \I\partial_3\end{matrix}\right]
\left[\begin{matrix} \Phi /{\rm c}\\ \I a_1\\ \I a_2 \\ \I a_3\end{matrix}\right]
=\left[\begin{matrix}
(1/ {\rm c})\partial_0\Phi +\partial_1a_1+\partial_2a_2+\partial_3a_3\\
\I\partial_0a_1+(\I / {\rm c})\partial_1\Phi -\partial_2a_3+\partial_3a_2\\
\I\partial_0a_2+(\I / {\rm c})\partial_2\Phi +\partial_1a_3-\partial_3a_1\\
\I\partial_0a_3+(\I / {\rm c})\partial_3\Phi -\partial_1a_2+\partial_2a_1
\end{matrix}\right]\,,
\end{equation}
that is
\begin{equation}
\nabla\! A=\left({1\over {\rm c}}\partial_0\Phi+\nabla\vec\cdot  a\right){\q e}_0
+\I\partial_0 a+{\I\over {\rm c}}\nabla\Phi-\nabla\vec \times  a\,.\end{equation}
We adopt the Lorentz gauge, according to which
\begin{equation}
{1\over {\rm c}}\partial_0\Phi+\nabla\vec\cdot a=0\,,\end{equation}
and we obtain
\begin{equation}
-\nabla \! A=B+{\I\over {\rm c}}\,E\,.\label{eq57}\end{equation}
Equation (\ref{eq57}) is the quaternionic form of Eqs.\ (\ref{eq52}) and (\ref{eq53}).
The magnetic field is opposite to the real part of $\nabla\! A$, and the electric field is the imaginary part multiplied by $\I {\rm c}$.

We remark that $\nabla \!A$ is not a minquat; it is a complex vector, sometimes called a bivector, an element of ${\mathbb C}^3$.

A more explicit form of Eq.\ (\ref{eq57}) is
\begin{equation}
  -\nabla\! A=B_1{\q e}_1+B_2{\q e}_2+B_3{\q e}_3+{\I\over {\rm c}}(E_1{\q e}_1+E_2{\q e}_2+E_3{\q e}_3)\,,\label{eq62}\end{equation}
where the $B_j$'s and the $E_j$'s ($j=1,2,3$) are functions of the $x_\mu$'s ($\mu =0,1,2,3$).

\subsection{Maxwell equations}%******************************************************
\subsubsection{Quaternionic expression of Maxwell equations}
Maxwell equations in vacuum are classically expressed as follows
\begin{equation}
\vec\nabla\vec \times\vec E+{\rm c}\,\partial_0\vec B=0\,,\label{eq58}\end{equation}
\begin{equation}
\vec\nabla\vec \times \vec B-{1\over {\rm c}}\partial_0\vec E=\mu_0\vec j\,,\label{eq59}\end{equation}
\begin{equation}
\langle \vec\nabla ,\vec E\rangle ={\rho\over\varepsilon_0}\,,\label{eq60}\end{equation}
\begin{equation}
\langle \vec\nabla ,\vec B\rangle =0\,,\label{eq61}\end{equation} 
where $\rho$ denotes the electric charge density and $\vec j$ the current density at the considered point.

We use Eq.\ (\ref{eq62}) and calculate
\begin{eqnarray}
-\nabla^*\,(\nabla A)\rap &=&\rap \left[\begin{matrix} \partial_0\\ -\I\partial_1\\ -\I\partial_2 \\ -\I\partial_3\end{matrix}\right]
\left[\begin{matrix} 0\\ B_1+(\I/ {\rm c}) E_1\\ B_2+(\I /{\rm c})E_2\\ B_3+(\I / {\rm c}) E_3\end{matrix}\right]\\
&=&\rap
\left[\begin{matrix}
-\displaystyle{1\over {\rm c}}(\partial_1E_1+\partial_2E_2+\partial_3E_3)+\I (\partial_1B_1+\partial_2B_2+\partial_3B_3)\nonumber \\
\nonumber \\
\partial_0B_1+ \displaystyle{\I \over {\rm c}}\partial_0E_1-\I\partial_2B_3+\displaystyle{1\over c}\partial_2E_3+\I\partial_3B_2-\displaystyle{1\over {\rm c}}\partial_3E_2\nonumber\\
\nonumber \\
\partial_0B_2+ \displaystyle{\I \over {\rm c}}\partial_0E_2+\I\partial_1B_3-\displaystyle{1\over c}\partial_1E_3-\I\partial_3B_1+\displaystyle{1\over {\rm c}}\partial_3E_1\nonumber \\
\nonumber \\
\partial_0B_3+ \displaystyle{\I \over {\rm c}}\partial_0E_3-\I\partial_1B_2+\displaystyle{1\over c}\partial_1E_2+\I\partial_2B_1-\displaystyle{1\over {\rm c}}\partial_2E_1
\nonumber
\end{matrix}
\right]\,,
\end{eqnarray}
that is
\begin{equation}
\nabla^*\,\nabla A=\left({1\over {\rm c}}\nabla\vec\cdot E -\I\nabla\vec\cdot  B\right){\q e}_0
-\partial_0 B-{1\over {\rm c}} \nabla\vec \times E-{\I\over {\rm c}}\partial_0 E+\I  \nabla\vec \times B\,.
\end{equation}

We introduce the  current density four-vector, represented by
\begin{equation}
J={\rm c}\,\rho\, {\q e}_0+\I j={\rm c}\,\rho \,{\q e}_0+\I (j_1{\q e}_1+j_2{\q e}_2+j_3{\q e}_3)\,.\end{equation}
Since $\varepsilon_0\mu_0{\rm c}^2=1$,  Maxwell equations---Eqs.\ (\ref{eq58}--\ref{eq61})---reduce to
\begin{equation}
\nabla^*\,\nabla A=\mu_0J\,.\label{eq66}\end{equation}
Equation (\ref{eq66}) is the quaternionic form of Maxwell equations. (We recall that $A$ is the quaternionic representation of the 4--potential; both $A$ and $J$ are minquats.)

\subsubsection{Relativistic invariance of Maxwell equations}%**********************************
Maxwell equations are invariant under  every proper Lorentz rotation as shown by the following derivation. Let $u$ be the representing quaternion of a proper Lorentz rotation that links two Galilean inertial frames ${\cal S}$ and ${\cal S}'$. Then, according to Eqs.\ (\ref{eq34}) and (\ref{eq51}), and since $u$ is unitary ($u\,u^*=1$), we have
\begin{equation}
\nabla'^*\,\nabla' A'=(\overline u\,\nabla \,u^{*})^{*}\,(\overline u\,\nabla \,u^{*})\,(u\,A\,\overline u^{\, *})=
u\,\nabla^*\,\overline u^{\, *}\,\overline u\,\nabla \,u^*\,u\,A\,\overline u^{\, *}=u\,\nabla^*\,\nabla A\,\overline u^{\, *}\,.
\end{equation}
Since $J$ is a 4-vector, it transforms according to 
\begin{equation}
J'=u\,J\,\overline u^*\,.\end{equation}
Then $\nabla'^*\,\nabla' A'=\mu_0J'$ is equivalent to $\nabla^*\,\nabla A=\mu_0J$, and we conclude  that Maxwell equations are invariant under every proper Lorentz rotation.

\subsubsection{Wave equation}%****************************************
Since $\nabla^*\,\nabla =\partial_0^2
-\partial_1^2-\partial_2^2-\partial_3^2$, Maxwell equations are equivalent to the set of the two equations
 \begin{subequations}
   \begin{numcases}{}
     (\partial_0^2
     -\partial_1^2-\partial_2^2-\partial_3^2)\,\Phi ={\rho\over\varepsilon_0}\,,\label{eq81}\\
     (\partial_0^2
-\partial_1^2-\partial_2^2-\partial_3^2)\,\vec a =\mu_0\,\vec j\,.\label{eq81b}
      \end{numcases}
 \end{subequations}
Equations (\ref{eq81}) and (\ref{eq81b}) are wave equations for the scalar potential $\Phi$ and for the vector potential $\vec a$. Hence the notion of an electromagnetic wave.

\subsection{Electromagnetic-field complex-vector and its transformation\\ under a boost}\label{sect34}%****************************

\subsubsection{Definition and effect of a boost}
To represent the electromagnetic field, we introduce the complex vector $\vec F$ such that
\begin{equation}
\vec F=\vec B+{\I\over {\rm c}}\vec E\,,\end{equation}
or in quaternionic form
\begin{equation}
F=B+{\I\over c} E=-\nabla A\,.\label{eq177n}\end{equation}
 
Let us consider a boost, represented by the quaternion $u$, connecting two Galilean inertial frames ${\cal S}$ and ${\cal S}'$. The 4--vector $A$ transforms according to $A'=u\,A\,\overline{u}^{\, *}$, and the 4--nabla transforms according to $\nabla '= \overline{u}\,\nabla\,u^*$, so that $\nabla A$ transforms according to
\begin{equation}
  \nabla '\!A'=(\overline u\,\nabla u^*)(u \,A\,\overline{u}^{\, *})=\overline u\,\nabla \!A\,\overline{u}^{\, *}\,,\end{equation}
because $u$ is unitary ($u^*\,u=1$).
Since $u$ is a boost, we have  $\overline{u}=u^*$, so that
\begin{equation}
  \nabla '\!A'=u^*\,\nabla \! A\,u\,,\label{eq84n}\end{equation}
and by Eq.\ (\ref{eq177n}) we conclude
\begin{equation}
F'=u^*F\,u\,,\label{eq99a}\end{equation}
which  can be written $F'=u^*F\,(u^*)^*$ and  which means that when passing from ${\cal S}$ to ${\cal S}'$, the field complex-vector $F$ is transformed under the elliptic rotation on ${\mathbb H}_{\rm c}$ represented by $u^*$\!, according to Eq.\ (\ref{eq151a}). More precisely, $u$ represents a boost and takes the form $u=\exp (\I\,{\q e}_n\delta /2)$, where both ${\q e}_n$ and $\delta$ are real, so that Eq.\ (\ref{eq99a}) correponds to the imaginary elliptic rotation represented by $u^*=\exp (-\I\,{\q e}_n\delta /2)$ (which operates on ${\mathbb H}_{\rm c}$).

\subsubsection{Effect of a boost on the electric and magnetic fields} We now consider a Lorentz boost and we shall obtain the usual expressions of the transformations of the electric and magnetic fields. According to Eq.\ (\ref{eq99a}) the boost operates on $F$ as an imaginary elliptic rotation. For explicit derivations, we use coordinates in ${\cal S}$, so that 
\begin{equation}
F=  B+{\I \over {\rm c}} E=\left[\begin{matrix}0 \\
B_1+(\I /{\rm c})E_1\\
B_2+(\I /{\rm c})E_2\\
B_3+(\I /{\rm c})E_3
\end{matrix}
\right]\,,\label{eq71a}
\end{equation}
and in ${\cal S}'$
\begin{equation}
F'=  B'+{\I \over {\rm c}}E'=\left[\begin{matrix}0 \\
B'_1+(\I /{\rm c})E'_1\\
B'_2+(\I /{\rm c})E'_2\\
B'_3+(\I /{\rm c})E'_3
\end{matrix}
\right]\,.\label{eq71}
\end{equation}

We consider a Lorentz boost, represented by $u=\exp (\I \,{\q e}_1\delta /2)$, and operating on ${\mathbb H}_{\rm c}$. Then
\begin{eqnarray}
u^*\,F\,u\rap &=&\rap \exp \left(-\I \,{\q e}_1{\delta\over 2}\right)\,F\,\exp \left(\I \,{\q e}_1{\delta\over 2}\right)\nonumber \\
&=&\rap
\left[\begin{matrix}
\cosh\displaystyle{(\delta / 2)}\\
-\I\sinh\displaystyle{(\delta/2)}\\
0\\
0\end{matrix}\right]
\left[\begin{matrix} 0\\  B_1+(\I/ {\rm c}) E_1\\  B_2+(\I /{\rm c})E_2\\   B_3+(\I / {\rm c}) E_3\end{matrix}\right]
\left[\begin{matrix}
\cosh\displaystyle{(\delta / 2)}\\
\I\,\sinh\,\displaystyle{(\delta / 2)}\\
0\\
0\end{matrix}\right]\nonumber\\
& &\nonumber \\
&=&\rap
\left[\begin{matrix}
\I\left[B_1+\displaystyle{(\I / {\rm c})}E_1\right]\sinh\displaystyle{(\delta / 2)}\\
\left[B_1+\displaystyle{(\I / {\rm c})}E_1\right]\cosh\displaystyle{(\delta / 2)}\\
\left[B_2+\displaystyle{(\I / {\rm c})}E_2\right]\cosh\displaystyle{(\delta/ 2)}+\I \left[B_3+\displaystyle{(\I / {\rm c})}E_3\right]\sinh\displaystyle{(\delta / 2)}\\
\left[B_3+\displaystyle{(\I / {\rm c})}E_3\right]\cosh\displaystyle{(\delta / 2)}-\I \left[B_2+\displaystyle{(\I / {\rm c})}E_2\right]\sinh\displaystyle{(\delta/ 2)}\\
\end{matrix}\right]
\left[\begin{matrix}
\cosh\displaystyle{(\delta / 2)}\\
\I\,\sinh\,\displaystyle{(\delta / 2)}\\
0\\
0\end{matrix}
\right]\nonumber \\
& &  \nonumber \\
&=&\rap\left[\begin{matrix} 0\\
B_1+\displaystyle{(\I / {\rm c})}E_1\\
\left[B_2+\displaystyle{(\I / {\rm c})}E_2\right]\cosh \delta +\I \left[B_3+\displaystyle{(\I / {\rm c})}E_3\right] \sinh \delta\\
\left[B_3+\displaystyle{(\I / {\rm c})}E_3\right]\cosh \delta-\I \left[B_2+\displaystyle{(\I / {\rm c})}E_2\right] \sinh \delta\\
\end{matrix}
\right]\,.\label{eq72}
\end{eqnarray}
The comparison of Eqs.\ (\ref{eq71}) and (\ref{eq72}) leads to
\begin{equation}
\left\{\begin{matrix}E'_1=E_1\,,\\
E'_2= E_2\cosh\delta +{\rm c}B_3\sinh\delta \,, \\
E'_3= E_3\cosh\delta - {\rm c} B_2\sinh\delta\,, \end{matrix}\right.
\hskip 1cm
\left\{
\begin{matrix}
 B'_1=B_1\,,\\
B'_2=B_2\cosh\delta-(1/{\rm c})E_3\sinh\delta \,,\\
 B'_3=B_3\cosh\delta+(1/ {\rm c})E_2\sinh\delta\,.
\end{matrix}\right.\label{eq76}
\end{equation}
Equations (\ref{eq76}) provide relations between electric and magnetic field components in two Galilean inertial frames in relative translatory motion. 
Under a boost, the complex pure vector $\vec F$ is transformed under an imaginary rotation \cite{Land}, as already mentioned.

For a boost along ${\q e}_n$ we have
\begin{equation}
F'=\exp \left(-\I {\q e}_n{\delta\over 2}\right)\,F\,\exp \left(\I {\q e}_n{\delta\over 2}\right)\,.\end{equation}

%*******************************************
\section{Electromagnetic waves}\label{sect4}
%*******************************************

\subsection{Some relativistic invariants}\label{sect41}%**************************************
We first estimate $(\nabla \!A)^2$ in the sense of quaternions: 
\begin{equation}
(\nabla\! A)^2=(\nabla \!A)(\nabla \!A)=\left( B+{\I\over {\rm c}} E\right)
\left( B+{\I\over {\rm c}} E\right)
={\|E\|^2\over {\rm c}^2}-\|B\|^2-2{\I\over {\rm c}}E\vec \cdot B\,,\label{eq91}\end{equation}
which means that $(\nabla \! A)^2$ is a scalar. (Hint for the above derivation: since both $E$ and $B$ are real pure quaternions, we have 
$-E^2=N(E)=\|E\|^2$,
and the same for $B$. We also have $E\,B=-E\vec \cdot B+E\times B$ and $B\,E=-E\vec\cdot B +B\times E$, so that $E\,B+B\,E=-2E\vec \cdot B$.)

Let us consider a proper Lorentz rotation, represented by $u$, that transforms $A$ into $A'$ and $\nabla$ into $\nabla '$. We apply Eq.\ (\ref{eq84n}) and write
\begin{equation}
(\nabla '\!A')^2=(\nabla '\!A')(\nabla '\!A')=(\overline u\,\nabla \!A\,\overline u^{\, *})(\overline u\,\nabla \!A\,\overline u^{\, *})\,,\end{equation}
and since $u$ is unitary (i.e.  $\overline u^{\, *}\;\overline u = \overline{u^*\,u}=1$), we obtain
\begin{equation}
(\nabla '\!A')^2=\overline u\,(\nabla \!A)^2\,\overline u^{\, *}\,.\label{eq78}\end{equation}
Since $(\nabla \!A)^2$ is scalar, the product on the right side of Eq.\ (\ref{eq78}) is  commutative, and we deduce 
\begin{equation}
(\nabla '\!A')^2=(\nabla \!A)^2\,,\label{eq79}\end{equation}
that is
\begin{equation}
{\| E'\|^2\over {\rm c}^2}-\| B'\|^2-2{\I\over {\rm c}} E'\vec \cdot B'
={\| E\|^2\over {\rm c}^2}-\| B\|^2-2{\I\over {\rm c}} E\vec \cdot B\,.\label{eq168}\end{equation}
Eventually, the isomorphism between 3--vectors in ${\mathbb E}_3$ and real pure quaternions allows us to write $\|E\|=\|\vec E\|$, $\|B\|=\|\vec B\|$ and $E\vec\cdot B=\langle\vec E, \vec B\rangle$, so that Eq.\ (\ref{eq168}) provides
 the two following relativistic invariants
\begin{eqnarray}
&&\hskip -.6cm{\|\vec E'\|^2\over {\rm c}^2}-\|\vec B'\|^2={\|\vec E\|^2\over {\rm c}^2}-\|\vec B\|^2\,,\label{eq82}\\
&&\hskip -.6cm\langle\vec E',\vec B'\rangle =\vec\langle\vec E , \vec B\vec\rangle\,.\label{eq83}\end{eqnarray}
Equations (\ref{eq82}) and (\ref{eq83}) hold true for the most general proper Lorentz rotation, and in particular for pure rotations as well as for hyperbolic rotations of the Minkowski  space.

\subsection{Application to electromagnetic waves}%*************************************
A progressive monochromatic electromagnetic plane wave, propagating in free isotropic space, possesses the following properties:
\begin{enumerate}
\item It is a transverse wave.
\item Field vectors $\vec E$ and $\vec B$ at a same point are such that
\begin{equation}
  \|\vec E\|={\rm c}\|\vec B\|\,.\label{eq84}\end{equation}
(Equation (\ref{eq84}) can be deduced from Maxwell equations).  Equation (\ref{eq82}) shows that this property holds true in every Galilean inertial frame.
\item Vectors $\vec E$ and $\vec B$ are orthogonal to each other, that is $\langle\vec E , \vec B\rangle =0$. According to Eq.~(\ref{eq83}), this holds true in every Galilean inertial frame. 
\end{enumerate}

Since $\nabla A$ is a pure quaternion, we have $N(\nabla A)=-(\nabla A)^2$, and we deduce from Eqs.\ (\ref{eq91}) and (\ref{eq84}) and from $E\vec \cdot B=0$ that $N(\nabla A)=0$, that is,  $F=-\nabla A$ is an isotropic (or null) vector.

We now prove that property of item 1 above holds true in every Galilean inertial frame.
We consider a Galilean frame ${\cal S}$. The energy-momentum of a photon, whose frequency is $\nu$ and direction of propagation is along the unit vector ${\qv e}_n$ (or pure quaternion ${\q e}_n$), is a 4--vector represented by the minquat
\begin{equation}
P={{\rm h}\nu\over {\rm c}} ({\q e}_0+\I\, {\q e}_n)\,,\label{eq112}\end{equation}
where ${\rm h}$ denotes the Planck constant. The transversality of vectors $\vec E$ and $\vec B$ in ${\cal S}$  is expressed by
${\q e}_n\vec\cdot E=\langle{\qv e_n}, \vec E\rangle =0$, and ${\q e}_n\vec\cdot B=\langle {\qv e_n}, \vec B \rangle =0$, so that
\begin{equation}
  0=\langle{\qv e}_n,  \vec B\rangle +{\I\over {\rm c}}\langle{\qv e}_n,  \vec E\rangle={\q e}_n\vec \cdot \left( B+{\I\over {\rm c}} E\right) ={\q e}_n\vec \cdot  F =-\,{\q e}_n\vec \cdot \nabla A\,.\end{equation}
If ${\cal S}'$ is another Galilean frame, in which the lightwave direction of propagation  is ${\q e}'_n$ and the elctromagnetic field is $F'$, we have to prove that ${\q e}'_n\vec\cdot F'=0$. It would be tempting to say that the dot product ${\q e}_n\vec\cdot\ F'$ is preserved under the boost. But we may not proceed this way, because $F'$ is transformed under an elliptic rotation on ${\mathbb H}_{\rm c}$, whereas ${\q e}_n$ is not. To obtain the result, we will proceed as follows.

We consider a boost, represented by the quaternion $u$, between the Galilean frames ${\cal S}$ (coordinates $\{x_\mu\}$)  and ${\cal S}'$ (coordinates $\{x'_\mu\}$). We assume $u=\exp (\I\,{\q e}_1\delta /2)$.

We first assume ${\q e}_n={\q e}_1$ (in ${\cal S}$, the lightwave propagates along ${\q e}_1$, or axis $x_1$). In ${\cal S}'$ the energy-momentum of the previous photon is represented by the minquat
\begin{equation}
  P'={{\rm h}\nu '\over {\rm c}}({\q e}_0+\I\, {\q e}_n')
  =u\,P\,\overline{u}^{\,*}
  ={{\rm h}\nu \over {\rm c}}\,\exp {\I\,{\q e}_1\delta \over 2}\, ({\q e}_0+\I\, {\q e}_1)\,\exp {\I\,{\q e}_1\delta \over 2}={{\rm h}\nu \over {\rm c}}\,\E^\delta\,({\q e}_0+\I\, {\q e}_1)\,,\label{eq198}\end{equation}
because ${\q e}_0+\I\,{\q e}_1$ is an eigenminquat of the hyperbolic rotation $x\longmapsto u\,x\,\overline{u}^{\,*}$ (with eigenvalue $\exp \delta$). From Eq.\ (\ref{eq198})  we conclude that ${\q e}'_n={\q e}_1$, which means that in ${\cal S}'$, the lightwave propagates along ${\q e}_1$ (axis $x'_1$, identical to $x_1$).
Since both the electric field $\vec E$ and the magnetic field $\vec B$ are transverse, their components on ${\q e}_1$ are $E_1=0$ and $B_1=0$. According to Eqs.\ (\ref{eq76}) we have $E'_1=E_1=0$, which means that $\vec E'$ is transverse. And $\vec B'$ is also transverse, because $B'_1=B_1=0$.

\begin{figure}%$$$$$$$$$$$$$$$$$$$$$$$$$$$$$$$$$$$$$$$$$$$$$$$$$$$$$$$$$$$$$
   \begin{center}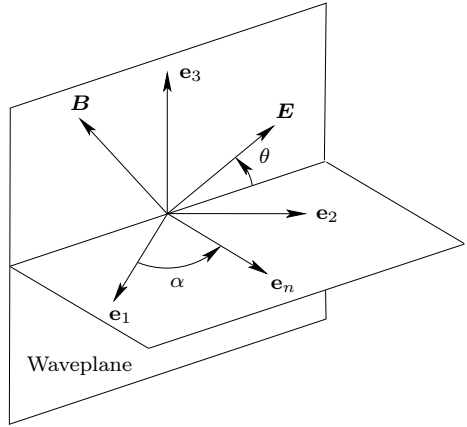
     \caption{\small Plane lightwave propagating along ${\qv e}_n$ in a Galilean reference frame $\{ {\qv e}_1, {\qv e}_2, {\qv e}_3\}$. The electric field $\vec E$ and the magnetic field $\vec B$ are transverse and orthogonal to each other.  
       \label{fig1ppbis}}   \end{center}
\end{figure}%$$$$$$$$$$$$$$$$$$$$$$$$$$$$$$$$$$$$$$$$$$$$$$$$$$$$$$$$$$$$$$$

We now assume that  ${\q e}_n$ belongs to the $({\q e}_1,{\q e}_2)$ plane and that ${\q e}_n\ne {\q e}_1$ (see Fig. \ref{fig1ppbis}). Then
${\q e}_n={\q e}_1\cos\alpha+\,{\q e}_2\sin\alpha$ ($\alpha \ne 0$).

In the coordinate system $\{ x'_\mu\}$ the  energy-momentum of the previous photon is represented by the minquat
\begin{equation}
  P'={{\rm h}\nu '\over {\rm c}}({\q e}_0+\I\, {\q e}_n')=u\,P\,\overline{u}^{\,*}
  ={{\rm h}\nu \over {\rm c}}\,\exp {\I\,{\q e}_1\delta \over 2}\, ({\q e}_0+\I\, {\q e}_n)\,\exp {\I\,{\q e}_1\delta \over 2} \,,\label{eq114}\end{equation}
that is
\begin{equation}
  P'={{\rm h}\nu \over {\rm c}}\left[\begin{matrix} \cosh \delta +\sinh \delta \cos\alpha \\
      \I\sinh\delta +\I\cosh\delta \cos\alpha \\
      \I\sin\alpha \\
      0\\
  \end{matrix} \right]\,.
\end{equation}
We note that ${\q e}'_n$ is proportional to the pure quaternion $N'= {\q e}_1(\sinh\delta +\cosh\delta \cos\alpha)
+\,{\q e}_2\sin\alpha$. To ligthen the notations we use Eq.\ (\ref{eq29a}) and  write $\cosh\delta =\gamma$ and $\sinh\delta =-\gamma \beta$, so that $N'=\gamma\,{\q e}_1 (\cos\alpha -\beta )+\,{\q e}_2\sin\alpha$.

If $\theta$ denotes the angle between the electric field $\vec E$ (represented by the pure quaternion $E$) and the plane $({\q e}_1,{\q e}_2)$, then (see Fig.\ \ref{fig1ppbis})
\begin{equation}
  E= \widetilde{E} (-\,{\q e}_1\cos\theta \sin\alpha +{\q e}_2\cos\theta \cos\alpha +{\q e}_3\sin\theta )\,,\end{equation}
where $\widetilde{E}$ denotes the electric field ampitude (it is a scalar complex-valued function depending on the $x_\mu$'s).

The magnetic field is represented by the pure quaternion
\begin{equation}
  B={\widetilde E\,\over {\rm c}} ({\q e}_1\sin\theta \sin\alpha -{\q e}_2\sin\theta \cos\alpha +{\q e}_3\cos\theta )\,.\end{equation}
(We may check that ${\q e}_n\vec\cdot E=0=\,{\q e}_n\vec\cdot B$ and $E\vec\cdot B=0$.)

According to Eqs.\  (\ref{eq76}) we have
\begin{eqnarray}
  E'\rap & = &\rap E'_1\,{\q e}_1+E'_2\,{\q e}_2+E'_3\,{\q e}_3\\
  \rap & = &\rap  E_1\,{\q e}_1+\gamma (E_2 -{\rm c}\beta B_3)\,{\q e}_2+\gamma (E_3+{\rm c}\beta B_2)\,{\q e}_3\nonumber \\
   \rap & = &\rap \widetilde{E} \bigl[-\,{\q e}_1\cos\theta\sin\alpha +{\q e}_2\gamma (\cos\theta\cos\alpha-\beta\cos\theta) +\,{\q e}_3\gamma (\sin\theta -\beta \sin\theta \cos\alpha )\bigr]
\end{eqnarray}
and we obtain
\begin{eqnarray}
  N'\vec\cdot E'
  \rap &=&\rap \widetilde{E}\bigl[-\gamma \cos\theta \sin\alpha\cos\alpha +\gamma \beta \cos\theta\sin\alpha
  +\gamma \cos\theta\cos\alpha\sin\alpha -\gamma\beta\cos\theta\sin\alpha\bigr]\nonumber \\
  \rap & =& \rap 0\,.  \end{eqnarray}
Similarly we have
\begin{eqnarray}
  {\rm c} B'\rap & = &\rap {\rm c}B'_1\,{\q e}_1+{\rm c}B'_2\,{\q e}_2+{\rm c}B'_3\,{\q e}_3\\
  \rap & = &\rap  {\rm c}B_1\,{\q e}_1+\gamma ({\rm c}B_2 + \beta E_3)\,{\q e}_2+\gamma ({\rm c}B_3-\beta E_2)\,{\q e}_3\nonumber \\
   \rap & = &\rap \widetilde{E} \bigl[{\q e}_1\sin\theta\sin\alpha +{\q e}_2\gamma (-\sin\theta\cos\alpha+\beta\sin\theta) +\,{\q e}_3\gamma (\cos\theta -\beta \cos\theta \cos\alpha )\bigr]
\end{eqnarray}
and eventually
\begin{eqnarray}
  N'\vec\cdot B'
  \rap &=&\rap (\widetilde{E}/{\rm c})\bigl[ \gamma \sin\theta \sin\alpha\cos\alpha -\gamma \beta \sin\theta\sin\alpha
  -\gamma \sin\theta\cos\alpha\sin\alpha +\gamma\beta\sin\theta\sin\alpha\bigr]\nonumber \\
  \rap & =& \rap 0\,.  \end{eqnarray}
Since $N'$ is collinear to ${\q e}'_n$, we conclude that ${\q e}'_n\vec\cdot F'=0$, so that both $\vec E'$ and $\vec B'$ are transverse in ${\cal S}'$.

To complete the proof, indeed, we eventually point out that the previous configuration is the most general one, because given two Galilean frames, we may choose the $x_1$ and $x'_1$--axes along the relative velocity, that is, ${\q e}_1$ collinear to the relative velocity,  and then, by an appropriate rotation around ${\q e}_1$,  make ${\q e}_2$ be coplanar with ${\q e}_1$ and ${\q e}_n$ (if ${\q e}_n\ne {\q e}_1$).

\subsection{Doppler effect and aberration of light  \cite{Mol,Gou,Rou}}%******************************************************

To show how quaternion algebra works, when applied to  electromagnetic waves, we consider the following problem. Two Galilean inertial frames ${\cal S}$ and ${\cal S}'$ are in relative motion with velocity $v{\q e}_m$ ($v$ is a real number and $|v|<{\rm c}$). A light ray is propagating  along ${\q e}_n$ in ${\cal S}$. What is its propagating direction in ${\cal S}'$? What is the relation between frequencies $\nu$ in ${\cal S}$ and $\nu '$ in ${\cal S}'$?

The relative motion of ${\cal S}'$ with respect to ${\cal S}$ corresponds to a hyperbolic rotation, represented by a unit quaternion $u$ such that
\begin{equation}
u=\exp{\I \delta\, {\q e}_m\over 2}=\cosh{\delta\over 2}+\I \, {\q e}_m\sinh{\delta\over 2}\,,\end{equation}
with the rapidity $\delta$ defined according to Eqs.\ (\ref{eq29a}).

The energy-momentum of a photon propagating along ${\qv e}_n$ in ${\cal S}$ is a 4--vector whose minquat expression is 
\begin{equation}
P={{\rm h}\nu\over {\rm c}}( {\q e}_0+\I \, {\q e}_n)\,,\end{equation}
and in ${\cal S}'$ it is
\begin{equation}
P'={{\rm h}\nu '\over {\rm c}}( {\q e}_0+\I \, {\q e}'_n)\,.\end{equation}
We have  (hyperbolic rotation)
\begin{equation}
P'=u\,P\,\overline{u}^{\,*}={{\rm h}\nu\over {\rm c}}\left(\cosh{\delta\over 2}+\I \, {\q e}_m\sinh{\delta\over 2}\right)
( {\q e}_0+\I \, {\q e}_n)\left(\cosh{\delta\over 2}+\I \, {\q e}_m\sinh{\delta\over 2}\right)\,.\end{equation}
By applying quaternion-algebra rules, we obtain
\begin{eqnarray}
P'\rap &=&\rap{{\rm h}\nu\over {\rm c}}\left(\cosh{\delta\over 2}+\I  \,{\q e}_m\sinh{\delta\over 2}
+\I \, {\q e}_n\cosh{\delta\over 2}- {\q e}_m {\q e}_n\sinh{\delta\over 2}
\right)
\left(\cosh{\delta\over 2}+\I \, {\q e}_m\sinh{\delta\over 2}\right)\nonumber \\
&=&\rap{{\rm h}\nu\over {\rm c}}\left(\cosh^2{\delta\over 2}+\sinh^2{\delta\over 2}
+2\I \, {\q e}_m\cosh{\delta\over 2}\sinh{\delta\over 2}
+\I \, {\q e}_n\cosh^2{\delta\over 2}
- {\q e}_n {\q e}_m\cosh{\delta\over 2}\sinh{\delta\over 2}\right.\nonumber \\
& &\hskip 1cm \left.
-\, {\q e}_m {\q e}_n\cosh{\delta\over 2}\sinh{\delta\over 2} -\I \, {\q e}_m {\q e}_n {\q e}_m\sinh^2{\delta\over 2}\right)
\nonumber \\
&=& \rap {{\rm h}\nu\over {\rm c}}\left(\cosh \delta + {\q e}_m\vec\cdot  {\q e}_n\sinh\delta +\I \, {\q e}_m\sinh \delta + \I \, {\q e}_n\cosh^2{\delta\over 2}
-\I \, {\q e}_m {\q e}_n {\q e}_m\sinh^2{\delta\over 2}\right)
\,.\label{eq205n}\end{eqnarray}
From $( {\q e}_m {\q e}_n {\q e}_m)^*= {\q e}_m^* {\q e}_n^* {\q e}_m^*=- \,{\q e}_m {\q e}_n {\q e}_m$, we conclude that $ {\q e}_m {\q e}_n {\q e}_m$ is a pure quaternion, so that the scalar part of $P'$ is
\begin{equation}
{{\rm h}\nu '\over {\rm c}}={{\rm h}\nu\over {\rm c}}(\cosh \delta + {\q e}_m\vec\cdot  {\q e}_n\sinh\delta )\,.\label{eq125a}\end{equation}
Equation (\ref{eq125a}) expresses the Doppler effect. Since $\cosh\delta =\gamma$ and $\sinh \delta =-\gamma \beta$, Eq. (\ref{eq125a}) also writes
\begin{equation}
{\nu '\over \nu}=\gamma (1-\beta \, {\q e}_m\vec\cdot  {\q e}_n)={1-\displaystyle{v\over {\rm c}}\; {\q e}_m\vec\cdot  {\q e}_n\over \sqrt{1-\displaystyle{v^2\over {\rm c}^2}}}\,.\label{eq215n}\end{equation}

The longitudinal Doppler effect is obtained when $ {\q e}_m=\pm  \,{\q e}_n$ ; the transverse Dopler effect (a relativistic effect) when $ {\q e}_m\vec\cdot  {\q e}_n=0$ ($ {\q e}_m$ and $ {\q e}_n$ are orthogonal to each other).

\begin{remark} {\rm The Doppler effect is usually referred to an observer at rest in ${\cal S}$ who is receiving light from a moving light source, attached to ${\cal S}'$. In Eq.\ (\ref{eq215n}) the frequency $\nu '$ is the frequency measured in the frame attached to the moving light source (proper frequency of the source) and $\nu$ is the frequency measured by the observer at rest. If the source moves radially away from the observer, then $v>0$. Moreover since light travels towards the observer, we have ${\q e}_n=-\,{\q e}_m$ and we obtain
  \begin{equation}
    {\nu '\over \nu}={1+\displaystyle{v\over {\rm c}} \over \sqrt{1-\displaystyle{v^2\over {\rm c}^2}}}
    =\sqrt{1+\displaystyle{v\over {\rm c}} \over 1-\displaystyle{v\over {\rm c}}}>1\,.\label{eq215p}\end{equation}
  There is a ``red shift'': the frequency in ${\cal S}$ is measured smaller than in ${\cal S}'$. If the light source approaches the observer radially ($v<0$), then $\nu >\nu '$, and there is a ``blue shift.''
  }
  \end{remark}

\medskip
\noindent{\bf Aberration of light (stellar aberration).}
According to Eq.\ (\ref{eq205n}), the light ray direction in ${\cal S}'$ is given by $ {\q e}'_n$ such that
\begin{equation}
 {\q e}'_n={ {\q e}_m\sinh \delta +  {\q e}_n\cosh^2\displaystyle{\delta\over 2}
- {\q e}_m {\q e}_n {\q e}_m\sinh^2\displaystyle{\delta\over 2}\over \cosh\delta + {\q e}_m\vec\cdot  {\q e}_n\sinh\delta}\,.\label{eq127a}
\end{equation}
Equation (\ref{eq127a}) expresses the ``aberration of light''.

If $ {\q e}_m=\pm \, {\q e}_n$, then $ {\q e}_m {\q e}_n {\q e}_m=- \,{\q e}_n$, so that $ {\q e}'_n= {\q e}_n$ (no aberration). If $ {\q e}_m\vec\cdot  {\q e}_n=0$, then $ {\q e}_m {\q e}_n {\q e}_m= {\q e}_n$ and
\begin{equation}
 {\q e}'_n={ {\q e}_m\sinh\delta + {\q e}_n\over \cosh\delta}\,.\label{eq209}
\end{equation}

In general the unit pure quaternion ${\q e}_n$ is chosen oriented from the star to the observer in ${\cal S}$, and ${\q e}'_n$ from the star to the observer in ${\cal S}'$. 
Let us assume that ${\q e}_m\ne \pm \,{\q e}_n$. Then---in the subspace of real pure quaternions, isomorphic to ${\mathbb E}_3$---the unit pure quaternion $-\,{\q e}_m{\q e}_n{\q e}_m$ is deduced from ${\q e}_n$ under the rotation of angle $\pi$ around ${\q e}_m$ and it belongs to the plane $({\q e}_m,{\q e}_n)$.  Eq.\  (\ref{eq127a}) shows that ${\q e}'_n$ also belongs to that plane. Moreover we have ${\q e}_m\vec\cdot {\q e}_n= -\,{\q e}_m\vec\cdot ({\q e}_m{\q e}_n{\q e}_m)$ (see Fig.~\ref{figC} in \ref{appenD} and replace ${\q e}_p$ with ${\q e}_m$). Let $\theta$ be the angle between ${\q e}_m$ and $-\,{\q e}_n$,  taken from ${\q e}_m$ to $-\,{\q e}_n$ (since ${\q e}_n$ is oriented from the star to the observer, $\theta$ is the angle under which the observer in ${\cal S}$ see the light coming from the star). Let $\theta '$ be the equivalent of $\theta$ for the observer in ${\cal S}'$. We have $\cos\theta =-\,{\q e}_m\vec\cdot {\q e}_n= {\q e}_m\vec\cdot ({\q e}_m{\q e}_n{\q e}_m)$ and $\cos\theta '=-\,{\q e}_m\vec\cdot {\q e}'_n$. Since $\cosh\delta =\gamma$ and $\sinh\delta =-v /{\rm c}$, Eq.\ (\ref{eq127a}) becomes
\begin{equation}
  \cos\theta '=-\,{\q e}_m\vec\cdot {\q e}'_n={\displaystyle{v\over {\rm c}}+\cos\theta \over 1+\displaystyle{v\over {\rm c}}\cos\theta}\,,\label{eq210}
\end{equation}
from which we deduce
\begin{equation}
  \tan^2{\theta '\over 2}={1-\cos\theta '\over 1+\cos\theta '}={1-\displaystyle{v\over {\rm c}}\over 1+\displaystyle{v\over {\rm c}}}\,\cdot\, {1-\cos\theta \over 1+\cos\theta }=
  {1-\displaystyle{v\over {\rm c}}\over 1+\displaystyle{v\over {\rm c}}}\,\tan^2{\theta\over 2}\,.
\end{equation}
Since $\theta$ and $\theta '$ belong to $[0,\pi]$ and since $|v|<{\rm c}$, we eventually obtain the general light aberration in the form \cite{Gou,Pen}
\begin{equation}
  \tan{\theta '\over 2}=
  \sqrt{{1-\displaystyle{v\over {\rm c}}\over 1+\displaystyle{v\over {\rm c}}}}\,\tan{\theta\over 2}\,.
\end{equation}

If ${\q e}_m$ and ${\q e}_n$ are orthogonal, then $\theta =\pi/2$; and we obtain
  \begin{equation}
  \tan{\theta '\over 2}=
  \sqrt{{1-\displaystyle{v\over {\rm c}}\over 1+\displaystyle{v\over {\rm c}}}}\,,\end{equation}
  that is, $\cos\theta '=v/{\rm c}$, which can also be deduced from Eq.\ (\ref{eq209}).

\subsection{Light polarization as a relativistic invariant}\label{sect43}%*********************************

We consider a Lorentz boost along $ {\q e}_1$ once more, corresponding to two Galilean frames ${\cal S}$ and ${\cal S}'$: light is propagating along $ {\q e}_n$ in ${\cal S}$, and $ {\q e}_n'$ in ${\cal S}'$. 
For the sake of simplicity, we  choose $ {\q e}_n$ in the $({\q e}_1, {\q e}_2)$ plane (see Fig.\ \ref{fig1ppbis}), that is
\begin{equation}
 {\q e}_n= {\q e}_1\cos\alpha + {\q e}_2\sin\alpha\,.\end{equation}
(We may assume $-\pi/2 \le \alpha \le \pi/2 $.)

The energy-momentum 4--vector of a photon is represented by $P$ in ${\cal S}$ and by $P'$ in ${\cal S}'$ according to Eqs.\ (\ref{eq112}) and (\ref{eq114}). The general form of the quaternion representing the electric-field vector  associated with a rectilinearly polarized plane wave is ($\widetilde{E}$ denotes the field complex-amplitude)
\begin{equation}
E= \widetilde{E}(- \,{\q e}_1\cos\theta \sin\alpha + {\q e}_2\cos\theta \cos\alpha + {\q e}_3\sin\theta )\,,\end{equation}
where $\theta$ is the angle of $E$ with the  $( {\q e}_1, {\q e}_2)$ plane (see Fig.\ \ref{fig1ppbis}).
The corresponding magnetic-field vector is
\begin{equation}
B= {\widetilde{E}\over {\rm c}}( {\q e}_1\sin\theta \sin\alpha - {\q e}_2\sin\theta \cos\alpha + {\q e}_3\cos\theta )\,.\end{equation}

We use Eqs.\ (\ref{eq76}) and obtain, in ${\cal S}'$,
\begin{eqnarray}
E'\rap&=&\rap\widetilde{E}\bigl[- \,{\q e}_1\cos\theta\sin\alpha + {\q e}_2(\cos\theta \cos\alpha\cosh\delta +\cos\theta\sinh\delta )\nonumber \\
& & \hskip 5cm+
  \,{\q e}_3(\sin\theta \cosh\delta +\sin\theta\cos\alpha \sinh\delta )\bigr]\,.\label{eq120}\end{eqnarray}

We first assume  that the considered lightwave is horizontally  polarized in ${\cal S}$: we may set $\theta =0$, in Eq.\ (\ref{eq120}), that is
\begin{equation}
  E'= \widetilde{E} \,\bigl[-\,{\q e}_1\sin\alpha  +{\q e}_2(\cos\alpha \cosh\delta +\sinh\delta )\bigr]\,.\label{eq226}
\end{equation}
We conclude that the polarization is also horizontal in ${\cal S}'$.

Moreover, from Eq.\ (\ref{eq226}) we deduce (we use $\| E\| =|\widetilde{E}|$)
\begin{equation}
\|E'\|= \|E\|(\sinh\delta\cos\alpha +\cosh\delta )\,.\label{eq122}\end{equation}

We now assume that the lightwave is vertically polarized in ${\cal S}'$: $\theta =\pi /2$. Then Eq.\ (\ref{eq120}) gives 
\begin{equation}
E'=\widetilde{E}\,{\q e}_3 (\cosh\delta +\cos\alpha \sinh\delta )\,,\end{equation}
and 
the polarization is also vertical in ${\cal S}'$; Equation (\ref{eq122}) still holds true in the vertical case.

We eventually conclude as follows.
\begin{enumerate}
\item A rectilinear polarization with an arbitrary orientation  (in the waveplane) is the superposition (in phase) of horizontal and vertical polarizations (linear combination). Since both horizontal and vertical polarizations are preserved with homothetic moduli when passing from ${\cal S}$ to ${\cal S}'$, as shown by Eq.\ (\ref{eq122}), which holds for both polarizations, the resulting polarization direction is also preserved: the polarization is the same whether observed in ${\cal S}$ or in ${\cal  S}'$.
\item If the considered lightwave is elliptically polarized, the electric field is the superposition of horizontal and vertical polarized fields with homothetic moduli and with a phase delay between them. Phase delays  in ${\cal S}$ and ${\cal S}'$ are equal, and the the polarization ellipse in ${\cal S}'$ is homothetic to the ellipse in ${\cal S}$. (A proof of the invariance of the phase of a plane wave is given in the book of M\o ller \cite{Mol}, see §3, p.\ 6--7.)
\end{enumerate}
The wave polarization is the same in ${\cal S}$ and in ${\cal S}'$ (this property of polarization is mentioned by Misner {\its et al.} in their book {\its Gravitation} \cite{Mis}, see Exercice 22.12 p. 577 and 581}). Only the lightwave power is different, due to the Doppler effect. 

\begin{remark} {\rm Eq.\ (\ref{eq122}) is no more than the relativistic longitudinal Doppler effect (see Eq.~(\ref{eq125a}) for ${\q e}_m$ and ${\q e}_n$ collinear). }
\end{remark}

\begin{remark} {\rm Light polarization is a relativistic invariant, but polarization states are not, because the polarization state of a lightwave includes the polarization but also the power of the lightwave (we anticipate  Sections \ref{sect45} and \ref{sect51}). Two observers in relative motion agree about the polarization of a lightwave in the sense that they mesure homothetic oriented-countour ellipses, but the areas of the ellipses are not equal because of the Doppler effect.  }
  \end{remark}

\subsection{Stokes parameters, degree of polarization}\label{sect45}%*******************************************************
\subsubsection{Stokes parameters \cite{Bor,Cla,Gol,Ram,Shu,Sim}}
We consider a polarized plane-lightwave with  narrow spectrum around the frequency $\widetilde \nu$. If $E_x(M,t)$ and $E_y(M,t)$ are the orthogonal components of the electric field in a plane wavefront (that is, an equiphase plane) at point $M$ and time $t$,
the {\bf Stokes parameters} of the polarized wave are defined by 
\begin{eqnarray}
X_0\rap&=&\rap \langle E_x\overline{E}_x+E_y\overline{E}_y\rangle_t\,,\\
X_1\rap&=&\rap \langle E_x\overline{E}_x-E_y\overline{E}_y\rangle_t\,,\\
X_2\rap&=&\rap \langle E_x\overline{E}_y+E_y\overline{E}_x\rangle_t\,,\\
X_3\rap&=&\rap \I \langle E_x\overline{E}_y-E_y\overline{E}_x)\rangle_t\,,\end{eqnarray}
where the bracket $\langle\;\rangle_t$ denotes a temporal average (taken on a time length much more longer than  $1/\widetilde \nu$).

Stokes parameters are observables: they are homogeneous to irradiances (up to a dimensional factor related to the propagation-medium impedance). Stokes parameters of a given lightwave are obtained by measuring light irradiances  after the wave has passed through polarizers (also called analyzers in that case) with appropriate orientations (including circular polarizers).
According to Section \ref{sect43}, two observers related to Galilean frames ${\cal S}$ and ${\cal S}'$ in relative motion  measure  Stokes parameters of a given lightwave that are proportionnal to each other  (Stokes parameters are subject to Doppler effect). With Stokes parameters $X_\mu$ in ${\cal S}$ ($X_0\ne 0$) and $X'_\mu$ in ${\cal S}'$, for $X_jX'_j\ne 0$ we have
\begin{equation}
  {X'_0\over X_0}={X'_j\over X_j}\,,\hskip .5cm j=1, 2, 3\,.\label{eq225}
  \end{equation}

\subsubsection{Stokes vectors}%*******************************************************

The four Stokes parameters $X_\mu$ $(\mu =0,1,2,3)$ form a 4-dimensional real vector belonging to ${\mathbb R}^4$ and written in matrix form as $\vec X=\,^t(X_0,X_1,X_2,X_3)$, where upperscript $t$ means ``transpose'', and called a {\bf Stokes vector}. A vector $\vec X$ in ${\mathbb R}^4$ is a Stokes vector, associated with a lightwave, only if $(X_0)^2\ge (X_1)^2+ (X_2)^2+ (X_3)^2$, and $X_0\ge 0$. The set of Stokes vector is thus a half cone ${\cal C}$ included in ${\mathbb R}^4$: if $\vec X\in {\cal C}$, then  $\alpha \vec X\in {\cal C}$, for every $\alpha \ge 0$.

If  $(X_0)^2= (X_1)^2+ (X_2)^2+ (X_3)^2$, the Stokes vector corresponds to a {\bf completely polarized} lightwave. The lightwave is {\bf partially polarized} if  $(X_0)^2> (X_1)^2+ (X_2)^2+ (X_3)^2>0$; it is {\bf unpolarized} (or  natural) if $X_0>0$ and $X_1=X_2=X_3=0$. The Stokes vector of a completely polarized lightwave lies on the surface  of ${\cal C}$. The interior of ${\cal C}$ corresponds to partially polarized lightwaves.

In the previous meaning, Stokes vectors do not form a vector space (for example if $\vec X$ is a Stokes vector, associated with a physical lightwave, the mathematical opposite $-\vec X$ is not a Stokes vector, in the sense that it does not correspond to a physical lightwave). Nevertheless the sum of two Stokes vectors makes sense. If $\vec X$ and $\vec Y$ are two Stokes vectors associated with lightwaves ${\cal X}$ and ${\cal Y}$, then the Stokes vector $\vec X+\vec Y$ represents the polarization of the {\bf incoherent superposition} of ${\cal X}$ and ${\cal Y}$ \cite{Ram}.  More generally, for every $\vec X$ and $\vec Y$ in ${\cal C}$ and for every positive number $\alpha$ and $\beta$,  a linear combination such as $\alpha \vec X+\beta \vec Y$ makes sense and actually represents a (partially) polarized lightwave. As a consequence, the half cone ${\cal C}$ is convex. (A complete analysis of partially coherent superpositions of partially polarized ligtwaves is given by Ramachandran and Ramaseshan \cite{Ram}.)

\bigskip
\noindent{\bf Example.} Let us consider two completely polarized lightwaves, whose electric fields (in the respective planewaves) are $\vec E_1=
(E_0,0)$ and  $\vec E_2=(0,E_0)$. The two lightwave polarizations are orthogonal to each other and their respective Stokes vectors are 
$\vec X_1= \,^t(|E_0|^2, |E_0|^2,0,0)$ and $\vec X_2= \,^t(|E_0|^2, -|E_0|^2,0,0)$. The Stokes vector of  the incoherent superposition of the two lightwaves  is $\vec X=\vec X_1+\vec X_2 = \,^t(2|E_0|^2, 0,0,0)$ and corresponds to natural light.

\subsubsection{Degree of polarization}%***************************************
The {\bf degree of polarization} of a partially-polarized wave is defined by
\begin{equation}
  \rho ={\sqrt{{X_1}^2+{X_2}^2+{X_3}^2}\over X_0}\,,\label{eq178}\end{equation}
where the $X_\mu$'s are the Stokes parameters of the lightwave ($X_0>0$).

If $\rho =1$,  the considered lightwave is completely polarized, and if $\rho =0$, the  lightwave is unpolarized (natural light).
As a consequence of Eq.\ (\ref{eq225}), the degree of polarization of a given wave is a relativistic invariant: two observers in two Galilean inertial frames receiving a light beam measure identical degrees of polarizations (that is $\rho =\rho '$).

%************************************************************************
\section{Quaternionic representation of polarization optics}\label{sect5}
%************************************************************************

\subsection{Poincaré sphere and Stokes parameters}\label{sect51}%***********************************

\begin{figure}[h]%$$$$$$$$$$$$$$$$$$$$$$$$$$$$$$$$$$$$$$$$$$
  \begin{center}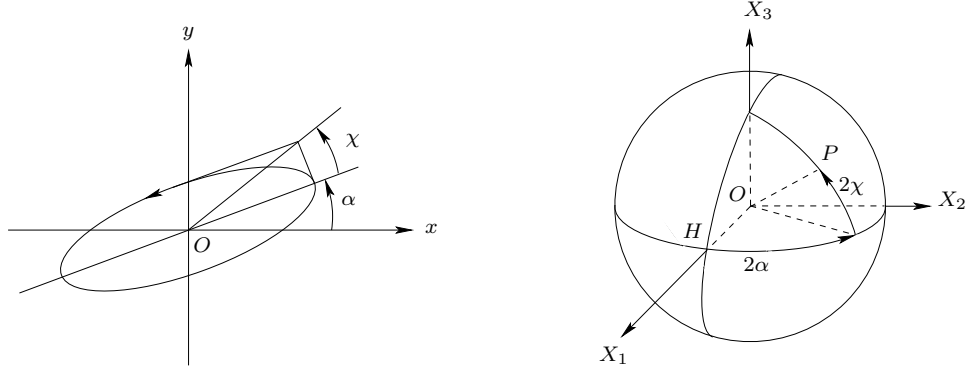
    \caption{\small On the left: in a waveplane, the polarization ellipse is characterized by its orientation (given by $\alpha$) with respect to the ho\-ri\-zontal axis, and by its ellipticity (given by $\chi$). On the right: on the Poincaré sphere, the corresponding polarization state in represented by the point $P$ whose longitude is $2\alpha$ and latitude is $2\chi$. The latitude is positive for levogyre states (the ellipse on the left diagram is left-handed). Point $H$ on the $X_1$--axis represents a horizontal rectilinear polarization state.\label{fig1}}
  \end{center}
\end{figure}

A harmonic plane wave generates an electromagnetic field at a given point $O$ in the physical space. Over time, the end point of the electric field at $O$ describes an ellipse in the plane tangent to the wave surface at $O$:  the wave is said to be {\bf elliptically polarized}. The ellipse %(and also the corresponding polarization state)
is {\bf left-handed}  if the electric field rotates in the trigonometric sense (anticlockwise) for an observer looking towards the source of light
(Fig.\ \ref{fig1}); it is {\bf right-handed} if the electric field rotates in the opposite sense. A {\bf oriented-contour ellipse} defines the {\bf polarization state} of the harmonic wave.  We say that the polarization state is left-handed (resp. right-handed) or levorotatory (resp. dextrorotatory) if the ellipse is left-handed (resp. right-handed). Particular polarization states are {\bf circularly polarized} states and {\bf rectilinearly polarized} states.

%$$$$$$$$$$$$$$$$$$$$$$$$$$$$$$$$$$$$$$$$$$$$$$$$

An ellipse is characterized by its orientation, given by the angle $\alpha$ its major axis makes  with an arbitrary predetermined direction in the plane wave (which is generally the ``horizontal'' direction) and by its ellipticity, related to the angle $\chi$ (see Fig.\ \ref{fig1}).  By abuse, we shall call $\chi$ the {\bf ellipticity} of the ellipse, and also the ellipticity of the corresponding polarization state.  We will adopt the following rule: $\chi >0$, for left-handed ellipses, and $\chi <0$, for  right-handed ellipses; we have $|\chi |\le \pi /4$. The power of a polarization state is proportional to $a^2+b^2$, where $a$ and $b$ are the lengths of the major and minor axes of the corresponding oriented-contour ellipse.  Two oriented-contour ellipses with the same $\alpha$ and the same $\chi$ correspond to a same polarization state, up to a power (or energy)  factor. We  shall abusively say ``the ellipse $(\alpha ,\chi )$'', assuming the  power of the polarization state being equal to 1 (in S. I. units for example).

A polarization state is also represented by a point on the {\bf Poincaré sphere} \cite{Bor,Ram}, which is a sphere of unit radius, referred to orthogonal axes $X_1$, $X_2$ and $X_3$ (Fig.\ \ref{fig1}).  The {\bf equator} is the intersection of the sphere with the plane $X_1$--$X_2$, so that the $X_3$--axis intercepts the sphere at the North and South poles. The {\bf longitude} is taken along the equator, with origin in the plane $X_1$--$X_3$, and is positive if taken from $X_1$ to $X_2$. The {\bf latitude} is taken from the equator, along a great circle passing through the poles; it is positive for $X_3\ge 0$. The ellipse $(\alpha ,\chi)$ is represented on the sphere by the point $P$ of longitude $2\alpha$ and latitude $2\chi$ (Fig.\ \ref{fig1}); we write $P(2\alpha ,2\chi)$. The North pole corresponds to {\bf left-circular} polarizations, the South pole to {\bf right-circular} polarizations.

\begin{figure}%[b]%$$$$$$$$$$$$$$$$$$$$$$$$$$$$$$$$$$$$$$$$$$
  \begin{center}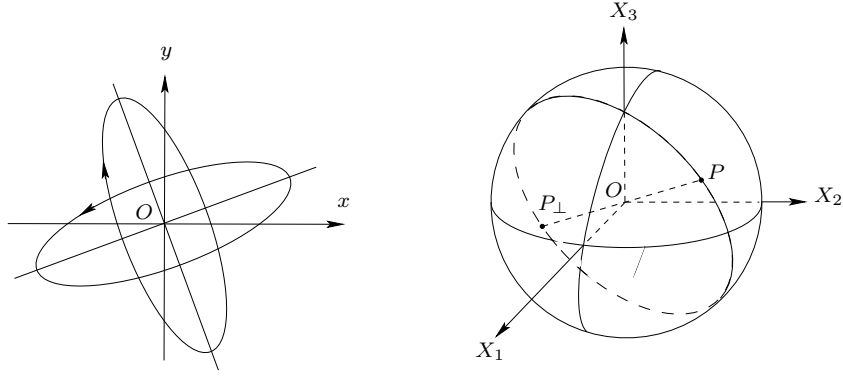
    \caption{\small Two (unitary) orthogonal oriented-contour ellipses (on the left) are represented by two diametrically opposite points ($P$ and $P_\perp$) on the Poincar\'e sphere (on the right).\label{fig2bis}}
  \end{center}
\end{figure}

Two oriented-contour ellipses are said to be {\bf orthogonal} if their major  axes are orthogonal to each other and if they have opposite ellipticities (with equal absolute values): one ellipse  is left-handed, the other right-handed. (Two orthogonal ellipses are similar: their major and minor axes are in the same ratio. They have the same shape and may differ by a scaling factor.) The corresponding polarization states are also said to be orthogonal. Orthogonality of elliptically-polarized states constitutes a generalization of the notion of orthogonal rectilinear-polarizations.  Right-circular polarization states are orthogonal to left-circular polarization states. Let $(\alpha ,\chi )$ be an oriented-contour ellipse, represented on the Poincar\'e sphere by the point $P(2\alpha ,2\chi )$. The orthogonal ellipse is $(\alpha_\perp,\chi_\perp)$ such that $(\alpha_\perp,\chi_\perp)=(\alpha +\pi /2, -\chi )$ and is represented on the Poincar\'e sphere by the point $P_\perp (2\alpha +\pi, -2\chi)$. Points $P$ and $P_\perp$ are diametrically opposite and they define an axis of the sphere (see Fig.\ \ref{fig2bis}).

The Poincaré sphere  may also be seen like a geometrical representation of Stokes vectors (defined in  Section \ref{sect45}) \cite{Bor,Ram}. Since the Stokes parameter $X_0$ of a lightwave is an irradiance, it is proportional to the lightwave power and the Poincar\'e sphere axes may be calibrated in irradiances. The Poincaré-sphere radius is generally taken equal to unity and 
a completely polarized lightwave with unitary irradiance $X_0$ is represented on the sphere by the point whose coordinates are $(X_1, X_2, X_3)$, where the $X_\mu$'s are the Stokes parameters of the polarized wave. Partially polarized waves with $X_0<1$  are represented by points inside the sphere.

%$$$$$$$$$$$$$$$$$$$$$$$$$$$$$$$$$$$$$$$$$$$$$$$$

The $X_1$--axis is associated with horizontal and vertical rectilinear-polarizations. More precisely the horizontal rectilinear-polarization state is represented by the point $(1,0,0)$ and the vertical polarization by $(-1,0,0)$. 

For a completely polarized  lightwave, we have
\begin{equation}
  {X_0}^2={X_1}^2+{X_2}^2+{X_3}^2\,.\end{equation}
If the lightwave is partially polarized, then
\begin{equation}
  {X_0}^2>{X_1}^2+{X_2}^2+{X_3}^2\,,\end{equation}
and the corresponding degree of polarization is given by Eq.\ (\ref{eq178}).

\subsection{Polarization states. Birefringence and dichroism. Polarizers}%***********************************************
Basic properties of polarizers, birefringent and dichroic devices are assumed to be known, as well as their associated Mueller matrices \cite{Cla,Gol,Ram,Shu,Sim}.

\subsubsection{Polarization states and minquats}
With the   Stokes vector $\vec X=\,^t(X_0, X_1,X_2,X_3)$, we associate the minquat \cite{PPF2,PPF5,PPF6,PPF7}
\begin{equation}
  X=X_0 {\q e}_0+\I (X_1 {\q e}_1+X_2 {\q e}_2+X_3 {\q e}_3)\,.\end{equation}
The norm of $X$ is
\begin{equation}
  N(X)= {X_0}^2-{X_1}^2-{X_2}^2-{X_3}^2\,,\end{equation}
so that  completely polarized waves are represented by  isotropic (or null) minquats (whose norms are zero), of the form $X_0(1+\I\,{\q e}_p)$, where ${\q e}_p$ is a real unit pure quaternion. If we refer to the Poincar\'e sphere (whose radius is equal to $1$), we have ${\qv e}_p=\vec {OP}$, see Fig.\ \ref{fig2bis}. By abuse we say that the minquat
$X_0(1+\I\,{\q e}_p)$ is polarized along ${\q e}_p$.

A partially polarized wave, with polarization degree $\rho$, is represented by a minquat of the form $X_0(1+\I\, \rho \,{\q e}_p$).

Two orthogonal unitary polarization states are represented by opposite points, say $P$ and $P_\perp$ on the Poincar\'e sphere:  $\vec{OP_\perp}=-\vec{OP}$. If $\vec{OP}={\qv e}_p$, the representative minquats are $1+\I\,{\q e}_p$ and $1-\I\,{\q e}_p$. More generally, minquats $X_0(1+\I\,{\q e}_p)$ and $Y_0(1-\I\,{\q e}_p)$ represent orthogonal states of polarization.

\begin{proposition}\label{prop51} {\its Two null minquats with equal (positive) scalar parts represent two orthogonal completely-polarized states if, and only if, their vector parts are opposite to each other.}
  \end{proposition}

\subsubsection{Polarized components and unpolarized components}

A minquat of the form $X=X_0(1+\I \,{\q e}_p)$ represents a completely-polarized state to which  $X_\perp=X_0(1-\I {\q e}_p)$ is orthogonally polarized.

The (completely) {\bf polarized component} of $X=X_0(1+\I\rho \, {\q e}_p)$---where $\rho$ is the degree of polarization---is $X_{\rm c}= \rho X_0(1+\I\,{\q e}_p)$; its power is $\rho X_0$. The {\bf unpolarized component} of $X$ is $X_{\rm u}=(1-\rho )X_0$. We have $X=X_{\rm c}+X_{\rm u}$.

If $X=X_0(1+\I\rho \,{\q e}_p)$ and $X'=X_0(1-\I\rho \,{\q e}_p)$, then $X'_{\rm c}=\rho X_0(1-\I\,{\q e}_p)=X_{{\rm c}\perp}$, and $X'_{\rm u}=\rho X_0=X_{\rm u}$ : the polarized components of $X$ and $X'$ are orthogonal to each other, whereas their unpolarized components are equal.

\subsubsection{Dichroic devices}\label{sect523}%*************************************************************
A dichroic device absorbs different polarized waves by different amounts. It owns two orthogonal unitary polarization-eigenstates that are represented by two opposite points on the Poincar\'e sphere. Those two points define an axis on the sphere, oriented from the more absorbed state to the less absorbed one.

If we replace $x_\mu$ with $X_\mu$ in Eq.\ (\ref{eq26}), we obtain the matrix expression of a pure dichroic device. More precisely, let $\vec X=\,^t(X_0,X_1,X_2,X_3)$ be the Stokes vector associated with a lightwave incident on a pure dichroic device whose eigenstates are horizontal and vertical polarization states, and whose dichroism is $\delta$. The corresponding emerging lightwave polarization  is represented by the vector $\vec X'=\,^t(X'_0,X'_1,X'_2,X'_3)$ such that
\begin{equation}
  \begin{pmatrix} X'_0\\X'_1 \\ X'_2\\ X'_3\end{pmatrix}
    =\begin{pmatrix} \cosh \delta & \sinh\delta & 0 & 0 \\
    \sinh\delta & \cosh \delta & 0 & 0 \\ 0 & 0& 1 & 0 \\ 0 & 0 & 0 & 1\end{pmatrix}
    \begin{pmatrix} X_0 \\ X_1 \\ X_2 \\ X_3\end{pmatrix}\,,\label{eq149}\end{equation}
where the square matrix is the Mueller matrix of the dichroic device.

According to Eqs.\ (\ref{eq26}) and (\ref{eq27}), the quaternionic form of Eq.\ (\ref{eq149}) is
\begin{equation}
  X'=\exp \left(\I \, {\q e}_1{\delta\over 2}\right)\,X\,\exp\left(\I\,  {\q e}_1{\delta\over 2}\right)\,,\label{eq205}\end{equation}
where ${\q e}_1$ is the axis of the dichroic.
The unit eingenstates of polarization are represented by $1+\I\,{\q e}_1$ and $1-\I\,{\q e}_1$ (we take $X_0=1$).

A more general pure-dichroic device has two orthogonal (unit) eigenstates represented by  $1+\I\,{\q e}_n$ and $1-\I\,{\q e}_n$,  and ${\q e}_n$ is its axis on the Poincaré sphere.
The dichroic device is represented by $u=\exp (\I \, {\q e}_n\delta /2)$ and operates on minquats according to
\begin{equation}
  X'=\exp \left(\I \, {\q e}_n{\delta\over 2}\right)\,X\,\exp\left(\I \, {\q e}_n{\delta\over 2}\right)\,.\label{eq206}\end{equation}
Equation (\ref{eq206}) shows that pure dichroics operate on polarization minquats as hyperbolic rotations.

\bigskip
\noindent{\bf Pure dichroics and physical dichroics.} If $\delta >0$ and $X_1>0$, Eq.\ (\ref{eq149}) shows that $X'_0>X_0$, a result that cannot be obtained with a passive device. To represent a physical dichroic device we have to introduce an {\bf isotropic absorption factor} $\kappa$, and we write
\begin{equation}
  \begin{pmatrix} X'_0\\X'_1 \\ X'_2\\ X'_3\end{pmatrix}
    =\kappa \begin{pmatrix} \cosh \delta & \sinh\delta & 0 & 0 \\
    \sinh\delta & \cosh \delta & 0 & 0 \\ 0 & 0& 1 & 0 \\ 0 & 0 & 0 & 1\end{pmatrix}
    \begin{pmatrix} X_0 \\ X_1 \\ X_2 \\ X_3\end{pmatrix}\,.\label{eq149b}\end{equation}
The isotropic factor $\kappa$ is a positive number. Since the power of $\vec X'$ should be less than or equal to  the power of $\vec X$, we have, for every $X_0$ ($X_0\ge 0$) and $X_1$ ($|X_1|\le X_0$)
\begin{equation}
  X'_0=\kappa (X_0\cosh\delta +X_1\sinh\delta )\le X_0\,.\end{equation}
Since $\max (X_0\cosh\delta +X_1\sinh\delta )=  X_0(\cosh\delta +\sinh |\delta|)=X_0\exp |\delta| $, we conclude that $\kappa$ is such that
\begin{equation}
  0\le \kappa\,\exp |\delta|\le 1\,.\end{equation}

With quaternions, Eq.\ (\ref{eq149b}) becomes
\begin{equation}
  X'=\kappa\,\exp \left(\I \, {\q e}_n{\delta\over 2}\right)\,X\,\exp\left(\I \, {\q e}_n{\delta\over 2}\right)\,,\label{eq206a}\end{equation}
and since $\kappa >0$, the dichroic device is represented by the quaternion $\sqrt{\kappa}\,\exp(\I\,{\q e}_n\delta /2)$.

If $X_+=X_0(1+\I\,{\q e}_n)$, we obtain $X_+'=\kappa \exp(\delta) X_+ $, and if
 $X_-=X_0(1-\I\,{\q e}_n)$, we obtain $X_-'=\kappa \exp(-\delta) X_-$. The polarization state $X_+$ is less absorbed than the state $X_-$ ($\delta >0$). On the Poincar\'e sphere the axis of the dichroic is ${\q e}_n=\vec{OP}$ and $P$ represents the less absorbed state.

A pure dichroic ($\kappa =1)$ can be  obtained if the device is active: the polarization state $X_0(1+\I\,{\q e}_n)$ is amplified, whereas the state $X_0(1-\I\,{\q e}_n)$ is mitigated  (for $\delta >0$). Such a result may be achieved in a laser cavity. More generally, we may have $\kappa >1$ in a laser cavity in which an eigenstate  is more amplified that the orthogonal eigenstate.

\subsubsection{Birefringent devices}%************************************************************
A birefringent device has two orthogonal unitary polarization-eigenstates that are represented by two opposite points on the Poincar\'e sphere, which define an axis; the device introduces a phase  shift (or phase delay)
$\varphi$---also called birefringence, with some abuse---between the eigenstates. If the fast axis corresponds to the horizontal polarization state and the slow axis to the vertical polarization, the Mueller matrix of the birefringent is as in Eq.\ (\ref{eq30}), that is
\begin{equation}
  \begin{pmatrix} X'_0\\ X'_1 \\ X'_2\\ X'_3\end{pmatrix}
    =\begin{pmatrix} 1 & 0 & 0 & 0 \\
    0 & 1 & 0 & 0 \\ 0 & 0& \cos\varphi  & -\sin\varphi \\ 0 & 0 & \sin\varphi & \cos\varphi\end{pmatrix}
    \, \begin{pmatrix} X_0\\ X_1 \\ X_2\\ X_3\end{pmatrix}\,.\end{equation}
The quaternionic representation of the birefringent is $u=\exp ( {\q e}_1\varphi /2)$ and according to Eq.\ (\ref{eq31}) we have
\begin{equation}
  X'=\exp \left( {\q e}_1{\varphi\over 2}\right)\,X\,\exp\left(- \,{\q e}_1{\varphi\over 2}\right)\,.\end{equation}
On the Poincar\'e sphere, the birefringent is represented  by the rotation of angle $\varphi$ around $ {\q e}_1$. For a general birefringence device, the rotation is around an axis defined by a unit vector $\vec  {\q e}_n$  and associated with the orthogonal eigenstates of the device; the axis (that is $\vec {\q e}_n$) is oriented from the slow-axis representative point to the fast one. The birefringent device is represented by $\exp ( {\q e}_n\varphi /2)$ and operates according to
\begin{equation}
  X'=\exp \left( {\q e}_n{\varphi\over 2}\right)\,X\,\exp\left(-\, {\q e}_n{\varphi\over 2}\right)\,.\end{equation}
Birefringents operate on polarization minquats as elliptic rotations.

\subsubsection{Polarizers}%**************************************************
A polarizer also admits two orthogonal unitary polarization-eigenstates that define an axis  on the Poincaré sphere (one eigenstate is transmitted, the other is totally absorbed or reflected). If the axis is along $ {\q e}_p$ (corres\-ponding to the transmitted eigenstate)  on the Poincar\'e sphere, the polarizer is represented by \begin{equation}
u={1\over 2}(1+\I  {\q e}_p)\,,\label{eq215}\end{equation} 
and it operates on minquats according to $x'=u\,x\,\overline{u}^{\, *}$. %Eq.\ (\ref{eq115}).
Explicitly, if the incident state is represented by  $X=X_0(1+\I  {\q e}_n)$, the emerging state is
\begin{equation}
  X'={X_0\over 4}(1+\I \, {\q e}_p)(1+\I \, {\q e}_n)(1+ \I\,  {\q e}_p)\,.\label{eq186}\end{equation}

If ${\q e}_n={\q e}_p$, from $(1+\I \,{\q e}_p)^2=2(1+\I\,{\q e}_p)$, we deduce $X'=X$; if  ${\q e}_n=-{\q e}_p$, we obtain $X'=0$, because $(1+\I\, {\q e}_p)(1-\I\, {\q e}_p)=0$.

\medskip
\noindent{\bf Malus Law.}
If $ {\q e}_p= {\q e}_1$ (polarizer transmitting the horizontal polarization state) and if the incident wave is polarized along  ${\q e}_n=n_1 {\q e}_1+n_2 {\q e}_2+n_3 {\q e}_3$ (with $(n_1)^2+(n_2)^2+(n_3)^2=1$), then
\begin{equation}
  X'={X_0\over 4}\left[\begin{matrix}1 \\ \I \\ 0 \\ 0\end{matrix}\right]
  \left[\begin{matrix}1 \\ \I  n_1 \\ \I n_2 \\ \I n_3\end{matrix}\right]
  \left[\begin{matrix} 1 \\ \I \\ 0 \\ 0\end{matrix}\right]
  ={X_0\over 4}\left[\begin{matrix}1 \\ \I \\ 0 \\ 0\end{matrix}\right]
  \left[\begin{matrix}1+n_1 \\ \I +\I n_1 \\ \I n_2-n_3\\ \I n_3+n_2\end{matrix}\right]
  ={X_0\over 2}\left[\begin{matrix}1+n_1 \\ \I (1+n_1)\\ 0 \\ 0\end{matrix}\right]\,,\label{eq187}
\end{equation}
that is,
\begin{equation}
  X'={(1+n_1)X_0\over 2}\,(1+\I\, {\q e}_1)\,.\label{eq216}\end{equation}
The output state is completely polarized along $ {\q e}_1$ (horizontal polarization), if $n_1\ne -1$. The transmitted power is proportional to $(1+n_1)/2$ (Malus law).

\medskip
\noindent{\bf Pure polarizers and real polarizers.} Equations (\ref{eq215}) and (\ref{eq216}) hold for pure polarizers. Physical (real) polarizers generally exhibit an isotropic absorption, as dichroic device do.  We introduce an isotropic absorption factor $\kappa$ ($0<\kappa \le 1$) and represent  a real polarizer by $(\sqrt{\kappa }/2)\,(1+\I\,{\q e}_p)$.  A pure polarizer corresponds to $\kappa =1$.

\begin{remark}[Polarizer as limit of a dichroic.]  {\rm A polarizer may be thought of as the limit of a dichroic device. We consider a dichroic device represented by
    \begin{equation} 
      u=\sqrt{\kappa}\,\exp{\I\,{\q e}_p\,\delta\over 2}=\sqrt{\kappa}\left(\cosh{\delta\over 2}+\I\,{\q e}_p\sinh{\delta\over 2}\right)\,.
    \end{equation}
    Let us make $\delta$ tend to infinity and $\kappa$ to $0$, under the constraint  \begin{equation}
      \doublelimite{\kappa\rightarrow 0}{\delta\rightarrow +\infty}\sqrt{\kappa}\,\exp {\delta \over 2}= \sqrt{\kappa_0}\le 1\,,
      \end{equation}
    Then
    \begin{equation}
      \doublelimite{\kappa\rightarrow 0}{\delta\rightarrow +\infty}\sqrt{\kappa}\,\exp{\I\,{\q e}_p\,\delta\over 2}={\sqrt{\kappa_0}\over 2}\, (1+\I\,{\q e}_p)\,.
        \end{equation}
    If $\kappa_0=1$, the limit is a pure polarizer.

    Many polarizers used in experiments are ``Polaroid'' also called sheet polarizers, initially and commercially developed by the Polaroid Corporation (created by Edwin Land, who invented those components) \cite{Cla,Shu,Gol}. Strictly speaking, polaroids are not polarizers but dichroic devices (i.e. partial polarizers) with very large $\delta$. Typical values are $\kappa_0 \sim 40 \%$ and $\exp\delta\sim 10^5$.
}\end{remark}

\begin{remark}[Analyzer] {\rm A polarizer, as a physical device, may be used to produce a polarization state---it acts as a polarizer in the strict sense---but also to analyze a polarization state---it acts then as an analyzer.  Most often we will use the word ``polarizer'' for both functions.
  }
  \end{remark}

\subsubsection{Pure dephasors}\label{sect526}%*****************************

Birefringent devices and pure dichroic devices are special cases of what we call {\bf pure dephasors}, represented by unit quaternions in the form $u=\exp ({\q e}_u\psi /2)$, with ${\q e}_u$ a complex pure quaternion and $\psi$ a complex number, and operating on polarization minquats according to
\begin{equation}
  X'=u\,X'\,\overline{u}^{\, *}\,,\end{equation}
where $X$ is the quaternionic representation of the polarization state incident on the pure dephasor, and $X'$ represents the emerging state. The complex quaternion ${\q e}_u$ may be unitary or a null quaternion (an example is provided in Section \ref{sect633}).

Since $u$ is a unit quaternion, we have $N(X')=N(X)$.  A pure dephasor preserves the norm of quaternions associated with polarization states: the quantity $(X_0)^2-(X_1)^2-(X_2)^2-(X_3)^2$ is invariant under a pure dephasor.  The degree of polarization is a relativistic invariant; it is generally not invariant under a pure dephasor (it is invariant under a birefringent device, but not under a dichroic device).

Pure dephasors form a group isomorphic to SO$_+(1,3)$ and generated by birefringent and dichroic devices \cite{PPF3,PPF7}. Pure dephasors operate on polarization minquats as proper Lorentz rotations.
As  already mentioned, dichroic devices are equivalent in crystal optics to Lorentz boosts
in Relativity; birefringents are equivalent to elliptic (or pure) rotations.

\bigskip
\noindent{\bf Polarization eigenstates.} 
We consider $u=\exp ({\q e}_u\psi /2)$ as above, with ${\q e}_u$ a complex pure quaternion that may be unitary or a null quaternion.

If ${\q e}_u$ is unitary, eigenminquats $x'$  and $x''$ of the mapping $q\longmapsto u\,q\,\overline{u}^{\, *}$ are given by Eqs.\ (\ref{eq106}) and (\ref{eq107}).
We begin with real ${\q e}_u$. Thus  $\overline{{\q e}_u}={\q e}_u$,  and $x'=(1+\I\,{\q e}_u)^2=2(1+\I\,{\q e}_u)$. Similarly $x''=(1-\I\,{\q e}_u)^2=2(1-\I\,{\q e}_u)$. According to Proposition \ref{prop51}, minquats $x'$ and $x''$ represent orthogonal polarization states (they are not unitary).

 We now assume that ${\q e}_u$ is a complex unit pure quaternion: ${\q e}_u={\q e}_m +\I\, {\q e}_n$, where ${\q e}_m$ and ${\q e}_n$ are real unit pure quaternions. We have ${\q e}_u+\overline{{\q e}_u}=2\,{\q e}_m$, and
\begin{equation}
  {\q e}_u\,\overline{{\q e}_u}=({\q e}_m +\I\, {\q e}_n)({\q e}_m -\I\, {\q e}_n)
  =2\I\,{\q e}_m\vec\times {\q e}_n\,,
\end{equation}
so that  eigenminquats with equal scalar parts are
\begin{equation}
  x'=(1+\I\,{\q e}_u)(1+\I\,\overline{{\q e}_u})={\q e}_0+ 2\I\,{\q e}_m +2\I\,{\q e}_m\vec\times {\q e}_n\,,\end{equation}
and
\begin{equation}
   x''=(1-\I\,{\q e}_u)(1-\I\,\overline{{\q e}_u})={\q e}_0- 2\I\,{\q e}_m +2\I\,{\q e}_m\vec\times {\q e}_n\,.
\end{equation}
Minquats $x'$ and $x''$ represent polarization states that are not orthogonal to each other, because the vector parts of $x'$ and $x''$ are not opposite to each other (see Proposition \ref{prop51}).

If ${\q e}_u$ is a complex null pure quaternion the mapping $x\longmapsto u\,x\,\overline{u}^{\, *}$ is a singular rotation on the minquat space.  As proved in \ref{appenF}, there is only one null-minquat eigendirection, the corresponding eigenvalue being iqual to 1,  that is, there is only one unitary polarization eigenstate (see Section \ref{sect633} for an exemple).

\begin{remark}[Linear and rectilinear operators] {\rm An operator ${\cal L}$ on a complex vector-space is {\bf linear} if for all vectors $\ket{a}$ and  $\ket{b}$, and all  complex numbers $\alpha$ and $\beta$, we have
 \begin{equation}{\cal L}(\alpha\ket{a}+\beta\ket{b})=\alpha {\cal L}\ket{a}+\beta {\cal L}\ket{b}\,.\end{equation}

 All birefringent and dichroic devices, as well as polarizers and their compounds, operating on polarization states, are linear operators in the previous  mathematical meaning.  On the other hand, a polarization operator is {\bf rectilinear} if its eingenstates are rectilinear polarization states (the electric field oscillates along a straight line). For example an optically active medium or a circular polarizer are linear operators but not rectilinear ones.
Thus  a birefringent, which is a linear operator, may be a rectilinear, elliptic or circular birefringent.
  }
\end{remark}

\begin{remark} {\rm The quaternionic representations of polarization states are introduced here from the usual definitions of Stokes vectors and the representations of birefringent and dichroic devices from their Mueller matrices. Doing so affords the advantage to refer to very known formalisms and to promote---if possible!---a rapid assimilation  of the proposed calculus. Another advantage is to link polarization optics to conventional expressions of proper Lorentz transformations (in their matrix forms) as used in the special theory of relativity.  We point out, however, that another way of introducing  quaternions in polarization optics can be established from fundamental geometrical properties of polarized light, without referring to usual Stokes vectors nor to Mueller matrices \cite{PPF3,PPF4,PPF6,PPF7}. That can be achieved in two steps: (i) Basic geometrical properties of pola\-ri\-zed light lead us to regard polarization operators as proper Lorentz rotations, independently of any actual representation \cite{PPF3,PPF4,PPF6}; (ii) A quaternionic representation of  proper Lorentz rotations  can directly be deduced from abstract considerations, as shown by Synge \cite{Syn}.  Linking  the two items leads us to intrinsically formulate the quaternionic representation of polarized light \cite{PPF3,PPF4,PPF6,PPF7}.
}
\end{remark}

\subsection{Consistency of representations}%*************************************************************
In the proposed representations  of polarized light, some conventions are chosen: definitions of left and right-handed polarizations; left-handed states located on the northern hemisphere of the Poincar\'e sphere;  birefringent axis on the Poincar\'e sphere orientated from the slow eigenvibration
\begin{figure}[h]%$$$$$$$$$$$$$$$$$$$$$$$$$$$$$$$$$$$$$$$$$$
  \begin{center}
  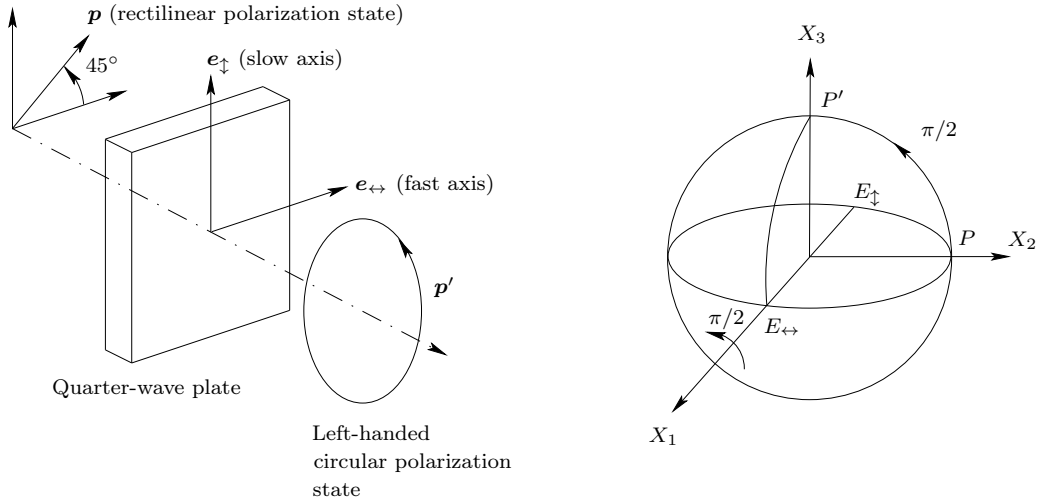
    \caption{\small Checking the consistency of the representations of polarized light. If the incident state of polarization $\vec p$ is rectilinear at $45^\circ$ from the fast axis of the quarter-wave plate ($\vec e_\leftrightarrow$), the emerging state $\vec p'$ is left-handed circular. On the Poincaré sphere, the quarter-wave plate axis is oriented from the slow-vibration representative-point $E_\updownarrow$ to the fast vibration $E_\leftrightarrow$ and is along $X_1$. The incident vibration is at point $P$, which is transformed into $P'$ (North pole) under the rotation of angle $\pi /2$ around the $X_1$--axis.\label{fig4}}
  \end{center}
\end{figure}%$$$$$$$$$$$$$$$$$$$$$$$$$$$$$$$$$$$$$$$$$$$$$$$
 representative point to the fast one; positive rotation angle  for an observer to which the birefringent axis is pointing.  Although arbitrary, those conventions have to be consistent with physical reality.

We check the consistency of our conventions with the following example, illustrated on Fig.~\ref{fig4}. A completely polarized lightwave is assumed to be incident on a quarter-wave plate whose fast axis is horizontal and slow axis vertical, in the waveplane (the quarter-wave plate is a birefringent device).
The lightwave is polarized at $45^\circ$ from the horizontal direction.
Physically, the emerging lightwave is circularly polarized and left-handed \cite{Bru}.

On the Poincaré sphere, the birefringent axis is the $X_1$--axis, and the effect of the waveplate is a $\pi /2$ rotation around $X_1$,   the sense of rotation being the trigonometric one (anticlockwise) for an observer to which  $X_1$ is pointing.  The input state is along $X_2$, which goes to $X_3$ under the previous rotation. The emerging polarization state is thus circularly polarized and left-handed, in accordance with physical reality.

With quaternions, the imput state is $X=X_0(1+\I\,{\q e}_2)$, and since the quarter-wave plate is represented
 by $u=\exp ({\q e}_1\pi /4)=(\sqrt{2}/2)(1+{\q e}_1)$, the emerging state is
\begin{eqnarray}
  X'=u\,X\,\overline{u}^{\, *}={X_0\over 2}(1 + {\q e}_1)\,(1+\I\,{\q e}_2)\, (1 -{\q e}_1)
  \rap &=&\rap  {X_0\over 2}(1+{\q e}_1 + \I\,{\q e}_2+\I\,{\q e}_3)(1 -{\q e}_1)\nonumber \\
  &=&\rap  X_0(1+\I\,{\q e}_3)\,.
\end{eqnarray}
The emerging state is circularly polarized and left-handed, in accordance with the previous results.

\begin{figure}[h]%$$$$$$$$$$$$$$$$$$$$$$$$$$$$$$$$$$$$$$$$$$
  \vskip .3cm
  \begin{center}
  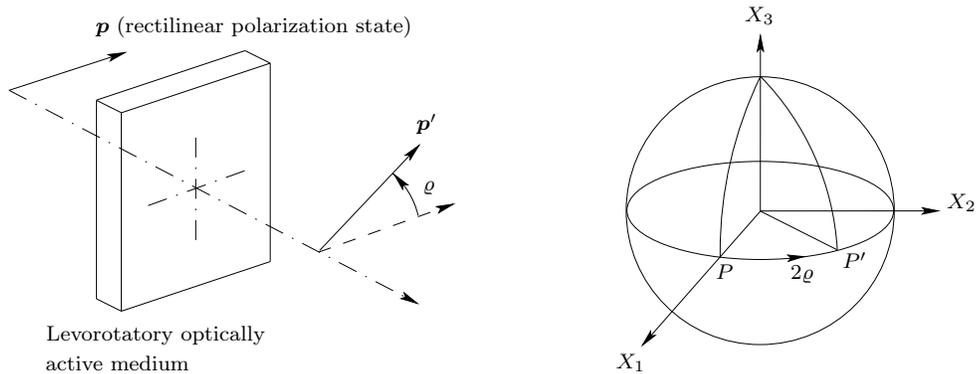
  \caption{\small After passing through the optically active medium, the incident rectilinear state of polarization $\vec p$ has turned an angle $\varrho$, which is positive for a levorotatory medium. On the Poincar\'e sphere, the representative point $P$ has turned an angle $2\varrho$, around the $X_3$--axis.
\label{fig5}}
  \end{center}
\end{figure}

Another feature to be checked is related to optical activity, or rotatory power, as shown on Fig.\ \ref{fig5}. A propagation medium is optically active if the polarization state of a wave has been rotated after propagation through the medium. The rotation angle is the rotatory power, say $\varrho$. Fresnel explained the phenomenon by regarding the medium as a circular birefringent medium whose eigenstates are circularly polarized. Those states may be regarded as fast and low vibrations of the medium.  The phase shift between left-handed and right-handed circular states is twice the rotatory power, namely $\theta =2\varrho$.  If the medium is left-handed, the fast vibration is the left-handed one, and the right-handed vibration is the slow one, so that $\varrho >0$.

On the Poincar\'e sphere, the medium is a circular birefringent whose axis is oriented from the slow axis (southern pole) to the fast one (northern pole), and is along $X_3$.  The effect of the medium is a rotation of angle $\theta=2\varrho >0$ around the $X_3$--axis, the positive sense being anticlockwise for an observer to which the axis $X_3$ is pointing. On the Poincar\'e sphere, the point $P$ is transformed into $P'$ under the rotation of angle $\theta =2\varrho$ around the $X_3$--axis (Fig.\ \ref{fig5}). 
%$$$$$$$$$$$$$$$$$$$$$$$$$$$$$$$$$$$$$$$$$$$$$$
 Since the longitude on the Poincar\'e sphere is twice the angle between the horizontal and the polarization direction (in the waveplane), the angle between ${\vec p}$ and ${\vec p}'$ is $\varrho$.

With quaternions the input state is $X=X_0(1+\I\,{\q e}_1)$ and the medium is represented by $u=\exp ({\q e}_3\theta /2)=\exp {\q e}_3\varrho $. The emerging state is obtained as follows
\begin{eqnarray}
  X'=u\,X\,\overline{u}^{\, *}\rap&=&\rap X_0(\cos\varrho +{\q e}_3\sin\varrho )(1+\I\,{\q e}_1)(\cos\varrho -{\q e}_3\sin\rho ) \nonumber \\
  \rap&=&\rap X_0(\cos\varrho +\I\,{\q e}_1\cos\varrho +\I {\q e}_2\sin\varrho +{\q e}_3\sin\varrho )(\cos\varrho -{\q e}_3\sin\varrho ) \nonumber \\
  \rap&=&\rap X_0[\cos^2\varrho +\sin^2\varrho +\I\,{\q e}_1(\cos^2\varrho -\sin^2\varrho)+2\I\,{\q e}_2 \cos\varrho\sin\rho]
   \nonumber \\
   \rap&=&\rap X_0 (1+\I\,{\q e}_1\cos 2\rho+\I\,{\q e}_2\sin 2\rho)\,.
\end{eqnarray}

Then for $X_0=1$, we obtain $X'_1=X_0\cos 2\varrho =\cos 2\varrho$,  $X'_2=X_0\sin2\varrho =\sin2\varrho$, and $X'_3=0$, so that on the Poincar\'e sphere $P'(\cos 2\varrho,\sin 2\varrho,0)$ is deduced from $P(1,0,0)$ under the rotation of angle $\theta =2\varrho$ around $X_3$, which is consistent with the previous results.

\subsection{Intrinsic derivations}%****************************************************************
The derivation  in Eq.\ (\ref{eq187}) is analytic in the sense that we explicitly use coordinates. Intrinsic derivations---that do not refer to coordinates---may be performed as illustrated by the following.

\subsubsection{Composition of two half-wave plates}%**********************************
    Since its birefringence (considered as  a phase shift) is equal to $\pi$, a half-wave plate is represented by $\exp ({\q e}_n\pi /2)={\q e}_n$, where ${\q e}_n$ is the plate axis (on the Poincar\'e sphere). The succession of a half-wave plate of axis ${\q e}_m$ (first plate to be crossed by light) and a half-wave plate of axis ${\q e}_n$ (${\q e}_n\ne \pm {\q e}_m$) is represented by the product ${\q e}_n\,{\q e}_m$, that is, by
    \begin{equation}
      u={\q e}_n\,{\q e}_m=-{\q e}_n\vec\cdot {\q e}_m+ {\q e}_n\vec \times {\q e}_m=-\cos\alpha -{\q e}_q\sin\alpha =
      -\exp {\q e}_q\alpha \,,
      \end{equation}
    where $\alpha$ is the angle between ${\q e}_m$ and ${\q e}_n$ (on the Poincar\'e sphere), oriented from ${\q e}_m$ to ${\q e}_n$ ($\alpha \ne 0$) and where ${\q e_q}$, defined by ${\q e_q}\sin\alpha ={\q e}_m\times {\q e}_n$, is orthogonal to both ${\q e}_m$ and ${\q e}_n$. Since $u$ and $-u$ represent the same birefringent, we may represent the composition of the two half-wave plates by $u=\exp {\q e}_q\alpha$. 
    Generally $u$ represents an elliptic birefringent whose birefringence is equal to $2\alpha$.

    If both wave plates are rectilinear, then ${\q e}_q={\q e_3}$,
    and $u$ represents a rotatory power equal to $\alpha$ (a well-known result), that is,  a circular-birefringence phase-shift equal to $2\alpha$. (We note that in the ``physical'' space, the angle between the respective fast axes of the two half-wave plates is $\alpha /2$.)

\subsubsection{Composition of two quater-wave plates}%**********************************
A quater-wave plate whose axis is ${\q e}_n$ is represented by
\begin{equation}
 \exp \left({\q e}_n{\pi\over 4}\right)={\sqrt{2}\over 2}\,(1+{\q e}_n)\,,\end{equation}
because the birefringence is $\varphi =\pi /2$.

The succession of a quater-wave plate, whose axis is ${\q e}_m$, and a quater-wave plate, whose axis is $ {\q e}_n$, is thus represented by
\begin{eqnarray}
  u=  \exp \left({\q e}_q{\psi\over 2}\right)=\exp \left({\q e}_n{\pi\over 4}\right)\,\exp \left({\q e}_m{\pi\over 4}\right)\rap&=&\rap{1\over 2}\,(1+{\q e}_n)\,(1+{\q e}_m)\nonumber \\
 \rap &=&\rap{1\over 2}(1-{\q e}_n\vec\cdot {\q e}_m+{\q e}_n+{\q e}_m+{\q e}_n\vec \times{\q e}_m)\,.\end{eqnarray}
In general $u$ represents an elliptic birefringent. The birefringence $\psi$ is given by
\begin{equation}
  \cos{\psi\over 2}={1\over 2}(1-{\q e}_n\vec\cdot {\q e}_m) ={1\over 2}(1-\cos\alpha)\,,\end{equation}
where $\alpha$ is the angle between ${\q e}_m$ and ${\q e}_n$ (on the Poincar\'e sphere). The axis is given by
\begin{equation}
  {\q e}_q \sin {\psi\over  2}={1\over  2} ( {\q e}_n+{\q e}_m+{\q e}_n\vec \times{\q e}_m)\,.
\end{equation}

If both wave plates are rectilinear birefringents, then ${\q e}_n\vec \times {\q e}_m=-\,{\q e}_3\sin\alpha$, where the angle $\alpha$ is taken from ${\q e}_m$ to ${\q e}_n$. Let ${\q e}_s$ be defined by
\begin{equation}
  {\q e}_s={{\q e}_n+{\q e}_m\over 2\cos (\alpha /2)}\,.\end{equation}
The quaternion ${\q e}_s$ is a real unit pure quaternion orthogonal to ${\q e}_3$.
We eventually obtain
    \begin{equation}
      {\q e}_q={1\over \sqrt{4\cos^2(\alpha /2)+\sin^2\alpha}}\,\left(2\,{\q e}_s\cos{\alpha\over 2}-{\q e}_3\sin\alpha\right)\,.\end{equation}
Both $\psi$ and ${\q e}_q$ only depends on the relative positions of ${\q e}_m$ and ${\q e}_n$ and not on their coordinates.

    \subsubsection{Composition of rectilinear birefringence and optical activity}%******************

    Let ${\cal B}$ be a birefringent of axis ${\q e}_n$ and birefringence $\varphi$ ($-\pi <\varphi \le \pi$), assumed to be rectilinear, which means that ${\q e}_n$ takes the form ${\q e}_n=n_1{\q e}_1+n_2{\q e}_2$. The birefringent ${\cal B}$ is followed by an optically active device ${\cal R}$.  Since the work of Fresnel, we know that the eigenstates of an optically active medium are the circularly polarized states, so that the axis of ${\cal R}$ on the Poincar\'e sphere is ${\q e}_3$. We also know that the phase shift $\theta$ between levorotatory and detrorotatory circular polarizations is  twice the rotatory power.  We assume $-2\pi <\theta \le 2\pi$. If $\theta >0$, the rotatory power is levorotatory.
    If $\theta <0$, the rotatory power is dextrorotatory, the axis still being ${\q e}_3$. (Equivalently we may change both the axis ${\q e}_3$ into $-\,{\q e}_3$ and $\theta$ into $-\theta$, the optically active device does not change.)

    The birefringent ${\cal B}$ is represented by $\exp ({\q e}_n\varphi /2)$ and ${\cal R}$ by $\exp ({\q e}_3\theta /2)$. If $\varphi\, \theta=0$  the composition of ${\cal B}$ and ${\cal R}$ is trivial: it reduces to ${\cal B}$, if $\theta =0$; to ${\cal R}$, if $\varphi =0$; to indentity, if $\theta =\varphi =0$. If $\varphi\,\theta \ne 0$, the composition ${\cal R}\circ{\cal B}$ is represented by
    \begin{eqnarray}
      u\rap &=&\rap\exp {{\q e}_3\theta \over 2}\,\exp {{\q e}_n\varphi \over 2}=
      \left(\cos{\theta\over 2}+{\q e}_3\sin{\theta\over 2}\right)\left(\cos{\varphi\over 2}+{\q e}_n\sin{\varphi\over 2}\right)\nonumber \\
      &=&\rap \cos{\theta\over 2}\cos{\varphi\over 2}+{\q e}_n\cos{\theta\over 2}\sin{\varphi\over 2}+
      {\q e}_q\sin{\theta\over 2}\sin{\varphi\over 2}+{\q e}_3\sin{\theta\over 2}\cos{\varphi\over 2}\,,
    \end{eqnarray}
    where ${\q e}_q={\q e}_3\,{\q e}_n={\q e}_3\times{\q e}_n$, because ${\q e}_3$ and ${\q e}_n$ are orthogonal. We remark that ${\q e}_q=-n_2{\q e}_1+n_1{\q e}_2$.

    The quaternion $u$ takes the form $\exp ({\q e}_p\psi /2)$ and it represents an elliptic birefringent whose birefringence is $\psi$ such that
    \begin{equation}
      \cos{\psi\over 2}= \cos{\theta\over 2}\cos{\varphi\over 2}\,.\end{equation}
    The birefringent axis is ${\q e}_p$ given by
    \begin{equation}
      {\q e}_p={1\over \sqrt{1- \cos^2\displaystyle{\theta\over 2}\cos^2\displaystyle{\varphi\over 2}}}
      \left({\q e}_n\cos{\theta\over 2}\sin{\varphi\over 2}+
      {\q e}_q\sin{\theta\over 2}\sin{\varphi\over 2}+{\q e}_3\sin{\theta\over 2}\cos{\varphi\over 2}\right)\,.
    \end{equation}

    \subsubsection{Superposition of rectilinear birefringence and optical activity}\label{sect544}%*************
    There are crystals that mix birefringence and optical activity at the same time: every infinite\-simal layer of a crystal sample exhibits rectilinear birefringence and optical activity. Such are photorefractive BSO or BGO crystals (Bi$_{12}$SiO$_{20}$ or Bi$_{12}$GeO$_{20}$) \cite{PPF0,Her,Hui,PPF13,Pet}, when an externally electric field is applied to them. If no electric field is applied to it, a BSO-crystal sample exhibits optical activity, which depends on the crystal thickness (for example the rotatory power is about $22^\circ\!/{\rm mm}$ at $\lambda =633\,{\rm nm}$). When an uniform electric field is applied, the crystal also exhibits rectilinear birefringence, due to the electro-optic effect (Pockels effect).  If $n_0$ is the refractive index at wavelength $\lambda$, if $r_{41}$ is the electro-optic coefficient of BSO-crystals and  $E$  the applied electric-field amplitude, the birefringence is
    $\Delta n=(1/2)\,n_0^{\, 3}\,r_{41} E$, for a tranverse-field,  perpendicular to  [100] crystallographic planes (light rays perpendicular to [0$\overline{1}$1] planes) \cite{PPF13}.

    If $\ell$ is the optical thickness of a BSO-crystal sample, $L$ the distance between the electrodes and $V$ the applied voltage ($V=EL$), the birefringence $\Delta n$ results in a phase shift (also abusively called birefringence) which is, at wavelength $\lambda$ \cite{PPF13}
     \begin{equation}
      \varphi= {2\pi \ell \Delta n\over \lambda}={\pi n_0^{\, 3}\,r_{41} V\over \lambda L}\,\ell=\eta \ell\,.
     \end{equation}
 We call $\varphi$ the {\bf intrinsinc rectilinear birefringence} of the crystal sample. It is the birefringence (phase shift) the crystal would exhibits under an applied electric-field if there were no optical activity.

 If $\varrho$ denotes the rotatory power per unit length of BSO-crystals, the optical activity of the crystal sample is $\varrho \ell$, and in phase, it is $\theta = 2\varrho \ell$. We call $\theta$ the {\bf intrinsic circular-birefringence} of the crystal ($\varrho \ell$ would be the intrinsic optical activity).
    
Each infinitesimal layer of the crystal, whose thickness is $\D \ell$, exhibits both intrinsic rectilinear birefringence $\eta \,\D \ell$ and intrinsic circular birefringence $2\varrho\,\D \ell$ (or optical activity $\varrho \,\D \ell$). The whole crystal is the superposition of intrinsic rectilinear birefringence $\varphi =\eta \ell$ and intrinsic circular birefringence $\theta =2\varrho\ell$, in the sense of Section \ref{sect259}. The rectilinear birefringence axis can be taken along ${\q e}_1$, whereas the circular birefringence axis is  ${\q e}_3$. The cristal (elliptic) birefringence is  represented by
     \begin{equation}
       u=\exp {(\eta\, {\q e}_1+2\varrho \,{\q e}_3)\ell\over 2}=\exp {\varphi\, {\q e}_1+\theta \,{\q e}_3\over 2}\,,\label{eq270}\end{equation}
     and is the superposition of intrinsic rectilinear birefringence, represented by $\exp ({\q e}_1\varphi /2)$, and intrinsic circular birefringence, represented by $\exp ({\q e}_3\theta /2)$.

     The quaternion $u$ in Eq.\ (\ref{eq270}) takes the form $\exp ({\q e}_n\psi /2)$, where $\psi$ is a real number and ${\q e}_n$  a real unit pure quaternion. We have
     \begin{equation}
       \psi=\sqrt{N({\q e}_n\psi )}=\sqrt{\varphi^2+\theta^2}\,,
     \end{equation}
     and
     \begin{equation}
       {\q e}_n={1\over \sqrt{\varphi^2+\theta^2}} (\varphi\, {\q e}_1+\theta \,{\q e}_3)
       ={\q e}_1\cos\chi +{\q e}_3\sin\chi\,,\hskip 1cm\tan\chi={\theta\over \varphi}\,.\label{eq272}
     \end{equation}
     (Only the sign of ${\q e}_n\psi$ is defined, so that ${\q e}_n$ and $\psi$ may be replaced with $-{\q e}_n$ and $-\psi$.)

     On the Poincar\'e sphere the vector ${\q e}_n$ is represented by the point of longitude $0$ and latitude $\chi$ with $\tan\chi =\theta /\varphi$, i.e. the point $E(0,\chi)$ (see Section \ref{sect51}). It represents an elliptic polarization; the oriented-contour ellipse is $(0,\chi /2)$. The crystal sample is an elliptic birefringent.

\bigskip
\noindent{\bf Equivalent parameters.}
     The previous ellipticl birefringent may be thought of as the product (or composition) of a rectilinear birefringent, represented by $\exp ({\q e}_m\varphi '/2)$, and an optically active device, represented by  $\exp ({\q e}_3\theta '/2)$, where $\varphi '$ and $\theta '$ are called {\bf equivalent rectilinear birefringence} and {\bf equivalent circular birefringence} (the equivalent optical activity being $\varrho '=\theta '/2$). The axis ${\q e}_m$ of the equivalent rectilinear birefringent corresponds to a rectilinear polarization, so that ${\q e}_m={\q e}_1\cos\alpha '+{\q e}_2\sin\alpha '$. The equivalent parameters  are such that
     \begin{equation}
       u=\exp {{\q e}_m\varphi '\over 2}\exp {{\q e}_3\theta '\over 2}\,,\end{equation}
     which means that the equivalent optical activity operates first. (Of course the choice of equi\-va\-lent birefringence operating first is possible.)

     In \ref{appenG} we prove the following relations between intrinsic parameters and equivalent ones:
     \begin{subequations}
  \begin{numcases}{}
    \sin{\varphi '\over 2}=\cos\chi\sin{\psi\over 2}={\varphi \over\sqrt{\varphi^2+\theta^2}}\sin{\sqrt{\varphi^2+\theta^2}\over 2}\,, \\
    \tan{\theta '\over 2}=\sin\chi\tan{\psi\over 2}={\theta \over\sqrt{\varphi^2+\theta^2}}\tan{\sqrt{\varphi^2+\theta^2}\over 2}\,,
    \\
    \alpha ' ={\theta '\over2}\,.
     \end{numcases}\label{s274}
     \end{subequations}

From an experimental point of view, it is interesting to reverse those equations and to express intrinsic parameters as functions of  equivalent ones. We first remark that the intrinsic optical activity can be obtained by measuring the rotation of a rectilinear polarization state after propagation across the crystal sample, when no electric field is applied, that is, when the crystal is free of rectilinear birefringence.  But the intrinsic birefringence, which appears when an electric field is applied to the crystal sample, is not directly accessible to experiment, because it is always mixed with the intrinsic optical activity. On the other hand, equivalent parameters can be measured according to ellipsometric methods. They thus offer us an indirect means of measuring the intrinsic birefringence. The method has been applied to the determination of the electro-optic coefficient of BSO crystals \cite{PPF0,PPF13}.

     \bigskip
     \noindent{\bf Application to an elliptic birefringent.} The method of the previous section may be applied to express every elliptic birefringent as the product of an equivalent optical activity and an equivalent rectilinear birefringent. The only difference is that the elliptic-birefringent axis is not necessarily in the $X_1$--$X_3$ plane (on the Poincar\'e sphere) but in a plane of longitude  $\alpha$~: if the elliptical birefringent is represented by $u=\exp ({\q e}_n\psi /2)$, then Eq.\ (\ref{eq272}) is replaced with
     ${\q e}_n={\q e}_1 \cos\alpha \cos\chi +{\q e}_2\sin\alpha\cos\chi+{\q e}_3 \sin\chi$. The equivalent parameters $\varphi '$, $\theta '$ and $\alpha '$ are given by
          \begin{subequations}
  \begin{numcases}{}
    \sin{\varphi '\over 2}=\cos\chi\sin{\psi\over 2}\,, \\
    \tan{\theta '\over 2}=\sin\chi\tan{\psi\over 2}\,,
    \\
    \alpha ' -\alpha ={\theta '\over2}\,.
     \end{numcases}\label{s275}
     \end{subequations}

\subsubsection{General Malus law}%*****************************************

Equation (\ref{eq186}) can also be treated intrinsically. Let $X=X_0(1+\I\,{\q e}_n)$ be incident on a polarizer whose axis is ${\q e}_p$. The outgoing state is represented by
\begin{eqnarray}
  X' =
  {X_0\over 4}(1+\I\,{\q e}_p)(1+\I\,{\q e}_n)(1+\I\,{\q e}_p)
  \rap  &=& \rap
  {X_0\over 4}(1+\I\,{\q e}_n+\I\,{\q e_p}-{\q e}_p{\q e}_n)(1+\I\,{\q e}_p)\nonumber  \\
 \rap  &=& \rap {X_0\over 4}(2+2\,{\q e}_n\vec\cdot {\q e}_p+2\,\I\,{\q e}_p+\I\,{\q e}_n-\I\, {\q e}_p{\q e}_n{\q e}_p)\,.\label{eq195}
\end{eqnarray}
We apply Lemma 2 of \ref{appenD} and obtain from
Eq.\ (\ref{eq195})  
\begin{equation}
  X'={X_0\over 2}(1+{\q e}_n\vec\cdot{\q e}_p)(1+\I\,{\q e}_p)={X_0\over 2}(1+\cos\alpha)(1+\I\,{\q e}_p)\,,
  \label{eq196}
\end{equation}
where $\alpha$ is the angle between ${\q e}_n$ and ${\q e}_p$, on the Poincar\'e sphere.

The Malus law takes the form
\begin{equation}
  X'_0={X_0\over 2}(1+{\q e}_n\vec\cdot{\q e}_p)\,.\label{eq197}\end{equation}

If $X=X_0(1+\I\rho \,{\q e}_n)$---where $\rho$ denotes the degree of polarization---we obtain
\begin{equation}
  X'={X_0\over 2}(1+\rho\,{\q e}_n\vec\cdot{\q e}_p)(1+\I\,{\q e}_p)={X_0\over 2}(1+\rho \cos\alpha)(1+\I\,{\q e}_p)\,,
  \label{eq196b}
\end{equation}
and \begin{equation}
  X'_0={X_0\over 2}(1+\rho\,{\q e}_n\vec\cdot{\q e}_p)\,,\label{eq197b}\end{equation}
which is a general form of Malus law.

Equations (\ref{eq196}), (\ref{eq197}), (\ref{eq196b}) and  (\ref{eq197b}) are intrinsic: only the relative positions of ${\q e}_n$ and ${\q e}_p$ matter.

%*********************
%BIREF
%*********************
\subsection{Transforming the axis of a rectilinear birefringent}%***********************
\begin{figure}[h]%$$$$$$$$$$$$$$$$$$$$$$$$$$$$$$$$$$$$$$$$$$
  \begin{center}
  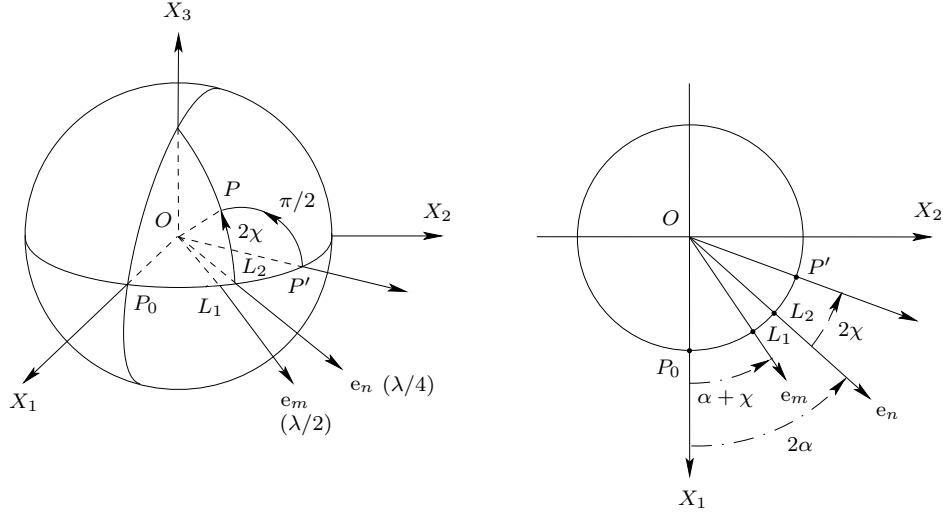
    \caption{\small The point $P_0$ is transformed into $P'$ under the rotation of angle $\pi$ around the axis ${\q e}_m$: the incident polarization state becomes the state $P'$, with the help of a half-wave plate whose axis is ${\q e}_m$. Then $P'$ is transformed into $P$ under the rotation of angle $\pi /2$ around the axis ${\q e}_n$: the polarization $P'$ is transformed into $P$, with the help of a quarter-wave plate.\label{fig6}}
  \end{center}
  \vskip .5cm
\end{figure}

We examine the following issue: how to transform a given  rectilinear birefringent ${\cal B}_1$, of axis ${\q e}_1$,  into a birefringent ${\cal B}_p$ of pre-defined axis ${\q e}_p$, with the help of rectilinear phase-plates, without changing the birefringence (phase delay)? The fast vibration of ${\cal B}_1$ is represented by $P_0$, which is on the $X_1$--axis, on the Poincar\'e sphere. The axis ${\q e}_p$ is along $OP$, where $P$ should represent the  fast vibration of the transformed birefringent ${\cal B}_p$. The longitude of $P$ is $2\alpha$ and its latitude is $2\chi$, so that (Fig.\ \ref{fig6})
\begin{equation}{\q e}_p={\q e}_1\cos 2\alpha\cos 2\chi +{\q e}_2\sin2\alpha \cos 2\chi +{\q e}_3\sin 2\chi\,.\end{equation}

We first solve the following problem: how to transform the polarization state $P_0$ into $P$?
A classical solution (non-unique) is obtained by using a half-wave plate, say ${\cal L}_1$, and a quarter-wave plate ${\cal L}_2$, whose axes are $OL_1$ and $OL_2$ respectively, on the Poincar\'e sphere.  The half-wave plate ${\cal L}_1$ transforms first the state $P_0$  into $P'$, such that $\angle L_2OP'=2\chi$, and  the quater-wave plate ${\cal L}_2$ transforms $P'$ into $P$, which is deduced from $P'$ in the rotation of angle $\pi /2$ around the axis $OL_2$. The longitude of $L_2$ is $2\alpha$, and since $\angle P_0OP'=2\alpha +2\chi$, the longitude of $L_1$ is $\alpha +\chi$ (Fig.\ \ref{fig6}).

The axis  ${\q e}_m$ of ${\cal L}_1$  (in the sense of quaternions) is then given by
\begin{equation}
{\q e}_m={\q e}_1\cos(\alpha +\chi)+{\q e}_2\sin(\alpha +\chi )\,,\end{equation}
and the axis ${\q e}_n$ of ${\cal L}_2$  by
\begin{equation}
  {\q e}_n={\q e}_1\cos2\alpha +{\q e}_2\sin 2\alpha\,.\end{equation}

We now use the quaternion calculus to check that $P$ is actually deduced from $P_0$ in the product of the rotation of angle $\pi$ around ${\q e}_m$, followed by the  rotation of angle $\pi /2$ around ${\q e}_n$.

We denote ${\q e}'_m$ the unit quaternion along $OP'$, deduced from ${\q e}_1$ 
%$$$$$$$$$$$$$$$$$$$$$$$$$$$$$$$$$$$$$$$$$$$$$$$
in the rotation of angle $\pi$ around ${\q e}_m$. We have
\begin{eqnarray}
  {\q e}'_m \rap &=&\rap\E^{{\q e}_m\pi /2}\,{\q e}_1\,\E^{-{\q e}_m\pi /2} = -\,{\q e}_m\,{\q e}_1\,{\q e}_m\nonumber \\
\rap &=&\rap-\left[{\q e}_1\cos(\alpha +\chi)+{\q e}_2\sin(\alpha +\chi )\right]{\q e}_1 \left[{\q e}_1\cos(\alpha +\chi)+{\q e}_2\sin(\alpha +\chi )\right]\nonumber\\
 \rap &=&\rap \left[\cos(\alpha +\chi)+{\q e}_3\sin(\alpha +\chi )\right] \left[{\q e}_1\cos(\alpha +\chi)+{\q e}_2\sin(\alpha +\chi )\right]\nonumber \\
 \rap &=&\rap {\q e}_1\cos(2\alpha +2\chi)+{\q e}_2\sin (2\alpha +2\chi )\,.
\end{eqnarray}
The rotation of angle $\pi /2$ around ${\q e}_n$ is expressed according to
\begin{eqnarray}
\E^{{\q e}_n\pi / 4}\,{\q e}'_m \,\E^{-{\q e}_n\pi/ 4}\rap&=&\rap{1\over 2}(1+{\q e}_n)\, \bigl[{\q e}_1\cos(2\alpha +2\chi)+{\q e}_2\sin (2\alpha +2\chi )\bigr]\, (1-{\q e}_n)\nonumber \\
&=& \rap {1\over 2}\begin{bmatrix} 1 \\ \cos2\alpha \\ \sin 2\alpha \\ 0 
\end{bmatrix}
\begin{bmatrix} 0\\  \cos(2\alpha +2\chi) \\ \sin (2\alpha +2\chi ) \\ 0
\end{bmatrix} 
\begin{bmatrix} 1 \\ -\cos2\alpha \\ -\sin 2\alpha \\ 0 
\end{bmatrix}\nonumber \\
& &\nonumber \\
\rap &=& \rap{1\over 2}
\begin{bmatrix}-\cos 2\alpha \cos(2\alpha +2\chi) -\sin 2\alpha\sin (2\alpha +2\chi )\\
 \cos(2\alpha +2\chi) \\
  \sin(2\alpha +2\chi) \\
   \cos 2\alpha \sin(2\alpha +2\chi) -\sin 2\alpha\cos (2\alpha +2\chi )
\end{bmatrix}
\begin{bmatrix} 1 \\ -\cos2\alpha \\ -\sin 2\alpha \\ 0 
\end{bmatrix}\nonumber \\
& &\nonumber \\
\rap &=& \rap{1\over 2}
\begin{bmatrix}-\cos 2\chi\\
 \cos(2\alpha +2\chi) \\
  \sin(2\alpha +2\chi) \\
   \sin 2\chi
\end{bmatrix}
\begin{bmatrix} 1 \\ -\cos2\alpha \\ -\sin 2\alpha \\ 0 
\end{bmatrix}
=
\begin{bmatrix} 0\\ \cos2\alpha \cos 2\chi\\
\sin 2\alpha \cos 2\chi \\
\sin 2\chi
\end{bmatrix}  ={\q e}_p\,.
\end{eqnarray}
In conclusion we have
\begin{equation}
{\q e}_p=\E^{{\q e}_n\pi /4}\,\E^{{\q e}_m\pi /2}\,{\q e}_1\,\E^{-{\q e}_m\pi /2}\,\E^{-{\q e}_n\pi /4}\,,\end{equation}
and then
\begin{equation}
1\pm\I\,{\q e}_p=
\E^{{\q e}_n\pi /4}\,\E^{{\q e}_m\pi /2}\,(1\pm\I\,{\q e}_1)\,\E^{-{\q e}_m\pi /2}\,\E^{-{\q e}_n\pi /4}
=u\,(1\pm\I{\q e}_1)\,u^*\,,\label{eq231a}\end{equation}
where
\begin{equation}
u=\exp {\q e}_n{\pi\over 4} \,\exp{\q e}_m{\pi\over 2}={\sqrt{2}\over 2}(1+{\q e_n})\,{\q e}_m\,.\end{equation}
The quaternion $u$ is the representation of a polarization operator, say ${\cal U}$, which operates on polarization states. 
Since ${\cal U}$ is the product of a half-wave plate ${\cal L}_1$ followed by a quater-wave plate ${\cal L}_2$, we write ${\cal U}={\cal L}_2\circ {\cal L}_1$, where $\circ$ denotes the composition law of operators.

\begin{figure}[t]%$$$$$$$$$$$$$$$$$$$$$$$$$$$$$$$$$$$$$$$$$$
  \begin{center}
  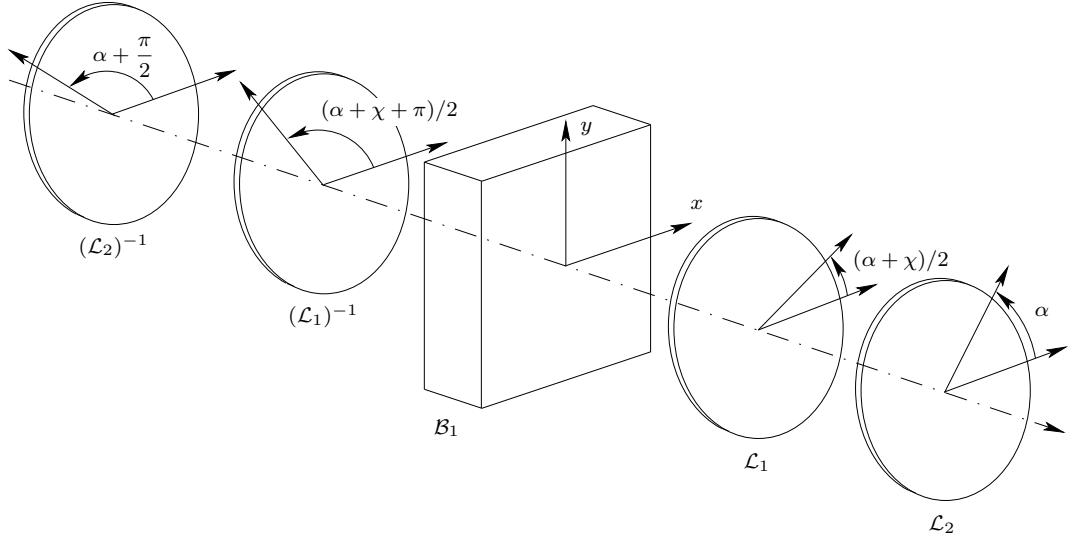
    \caption{\small Optical setup to transform the rectilinear birefringent ${\cal B}_1$ into a birefringent  ${\cal B}_p$ with a predefined axis. In general the eigenstates of ${\cal B}_p$ are elliptically polarized. If the fast vibration of ${\cal B}_1$ is along the $x$--axis (horizontal polarization), the major axis of the ellipse corresponding to the fast vibration of ${\cal B}_p$ makes the angle $\alpha$ with the $x$--axis, and the ellipticity is equal to  $\chi$. Waveplates ${\cal L}_1$ and ${\cal L}_1^{-1}$ are half-wave plates, and ${\cal L}_2$ and ${\cal L}_2^{-1}$ are quarter-wave plates. Only the fast vibration of each wave-plate is drawn, and its angle with the $x$--axis is given. Light propagates from left to right.\label{fig7}}
  \end{center}
  \end{figure}%$$$$$$$$$$$$$$$$$$$$$$$$$$$$$$$$$$$$$$$$$$$$$$

Let us go back now to the initial issue. To say that ${\q e}_p$ is the axis of ${\cal B}_p$ means that $1+\I\,{\q e}_p$ and $1-\I\,{\q e}_p$ represent the  eingenstates  of ${\cal B}_p$ (they respectively represent the fast and the slow vibrations). If a lightwave whose polarization state is $1+\I\,{\q e}_p$ is incident on ${\cal B}_p$, the polarization of the emerging lightwave should also be $1+\I\,{\q e}_p$. Such a result is obtained in three steps that are as follows. 
\begin{enumerate}
\item The incident state $1+\I\,{\q e}_p$ is transformed into the state $1+\I\,{\q e}_1$. According to  Eq.\ (\ref{eq231a}), since $u^{-1}=u^*$, we have
$1+\I{\q e}_1=u^*(1+\I\,{\q e}_p)u$. That means that the incident lightwave first goes across the inverse operator ${\cal U}^{-1}=({\cal L}_2\circ{\cal L}_1)^{-1}={\cal L}_1^{-1}\circ{\cal L}_2^{-1}$.
\item The lightwave then goes across ${\cal B}_1$. The birefringence is $\varphi$. Since $1+\I\,{\q e}_1$ is an eigenstate, we obtain
\begin{equation}
\E^{{\q e}_1\varphi /2} \,(1+\I\,{\q e}_1) \,\E^{-{\q e}_1\varphi /2}= 1+\I\,{\q e}_1\,.\end{equation}
\item Finally, we have to transform the state $1+\I\,{\q e}_1$ into  $1+\I\,{\q e}_p$. The light wave emerging from ${\cal B}_1$ goes across ${\cal U}$, 
according to Eq.\ (\ref{eq231a}).
\end{enumerate}

We eventually obtain ${\cal B}_p={\cal U}\circ {\cal B}_1\circ{\cal U}^{-1}$ and, since ${\cal U}={\cal L}_2\circ{\cal L}_1$, we have
\begin{equation}
  {\cal B}_p={\cal L}_2\circ{\cal L}_1\circ {\cal B}_1\circ{\cal L}_1^{-1}\circ{\cal L}_2^{-1}\,.\label{eq234b}\end{equation}
The result is illustrated by Fig.\ \ref{fig7}.

\begin{remark} {\rm The result obtained for a birefringent device can be applied to transform a rectilinear dichroic device or a rectilinear polarizer into a dichroic device or polarizer with arbitrary (orthogonal) eigenstates. Only ${\cal B}_1$ has to be remplaced by the appropriate rectilinear dichroic or rectilinear polarizer, whose axes are ${\q e}_1$.
}
\end{remark}

%*******************
%FIN DE BIREF
%*******************

%***************
%DEPOL
%***************
\subsection{Depolarizers \cite{PPF12}}

In this section we use greek letters $\beta$ and $\gamma$ with a distinct meaning of that of Section \ref{sect253} (special theory of  relativity).

\subsubsection{Perfect depolarizers}
A perfect depolarizer is such that for every incident state of polarization, the emerging state is unpolarized \cite{Ram,PPF12}. If $X=X_0(1+\I \,\rho \,{\q e}_n)$ represents the incident state, the emerging state should takes the form $X'=X'_0$, where $X'_0$ is a positive number.

Let $\Psi$ be the linear mapping defined on quaternions by
\begin{equation}
  \Psi (q)= {1\over 4}(q-{\q e}_1\,q\,{\q e}_1- {\q e}_2\,q\,{\q e}_2-{\q e}_3\,q\,{\q e}_3)\,.
\end{equation}
We have $\Psi(1)=1$ and $\Psi({\q e}_1)=\Psi({\q e}_2)=\Psi({\q e}_3)=0$, so that for every  minquat $X=X_0(1+\I \,\rho \,{\q e}_n)$, we obtain  $\Psi (X)=X_0$. We conclude that the restriction of $\Psi$ to minquats, denoted by $D$, represents a perfect depolarizer. It is such that
\begin{equation}
 D\;:\;X\longmapsto D (X)= {1\over 4}(X-{\q e}_1\,X\,{\q e}_1- {\q e}_2\,X\,{\q e}_2-{\q e}_3\,X\,{\q e}_3)\,.
\label{eq257b}\end{equation}

More generally let $\{{\q e}_u,{\q e}_v,{\q e}_w\}$ be a direct orthonormal basis of the subspace of pure  quaternions. The mapping $D'$ defined on minquats by
\begin{equation}
  D'(X)= {1\over 4}(X-{\q e}_u\,X\,{\q e}_u- {\q e}_v\,X\,{\q e}_v-{\q e}_w\,X\,{\q e}_w)\,,\label{eq298}
\end{equation}
is such that $D'(X)=X_0$ for every minquat $X=X_0(1+\I \,\rho \,{\q e}_n)$; thus $D'$ also represents a perfect depolarizer.

\begin{remark} {\rm For $X=X_0(1+\I\rho\,{\q e}_n)$ we obtain $-\,{\q e}_1\,X\,{\q e}_1-\, {\q e}_2\,X\,{\q e}_2-\,{\q e}_3\,X\,{\q e}_3=X_0(3-\I\rho\,{\q e}_n)$, whose polarized component  $\rho X_0(1-\I\,{\q e}_n)$ is orthogonal to the polarized component of $X$, with the same power. According to  Eq.\ (\ref{eq257b}) the depolarizing effect is obtained by {\bf incoherently} adding two lightwaves whose polarized components are orthogonally polarized (and have the same power).
  }\end{remark}
\subsubsection{Partial depolarizers}

A partial depolarizer depolarizes by a factor $1-\gamma$ (with $0\le \gamma \le 1$),  if the outgoing polarization state  corresponding to the incident state $X=X_0(1+\I \,\rho \,{\q e}_n)$ is
$X'=X'_0(1+ \I\,\gamma \,\rho \,{\q e}_n)$. In other words, the degree of polarization $\rho$ is changed into $\gamma\,\rho$.

According to Eq.\ (\ref{eq298}) the vectorial part of the polarized component of $X$ is balanced by the vectorial part of  $-\,{\q e}_u\,X\,{\q e}_u-\, {\q e}_v\,X\,{\q e}_v-\,{\q e}_w\,X\,{\q e}_w$. For the superposition to provide a partially polarized lightwave, not necessarily an unpolarized wave, we reduce the weight of $-\,{\q e}_u\,X\,{\q e}_u-\, {\q e}_v\,X\,{\q e}_v-\,{\q e}_w\,X\,{\q e}_w$ by introducing a factor $\beta$ ($0\le \beta\le 1$) and we consider the mapping
\begin{equation}
  D_{\rm p}\;:\,X\longmapsto D_{\rm p}(X)= {1\over 1+3\beta}\bigl[X-\beta ({\q e}_u\,X\,{\q e}_u- {\q e}_v\,X\,{\q e}_v-{\q e}_w\,X\,{\q e}_w)\bigr]\,,\label{eq257}
\end{equation}
where $0\le \beta \le 1$. We have
\begin{equation}
  D_{\rm p}(1+\I\,\rho\,{\q e}_n)=1+{1-\beta\over 1+3\beta}\,\I\,\rho\,{\q e}_n\,.
  \end{equation}
Then $D_{\rm p}(1+\I\,\rho\,{\q e}_n)$ is a minquat and its degree of polarization is $(1-\beta)\rho /(1+3\beta)$. To reduce the degree of polarization from $\rho$ to $\gamma \,\rho$, we choose $\beta$ such that
\begin{equation}\gamma ={1-\beta\over 1+3\beta}\,,\end{equation}
that is \begin{equation}
  \beta ={1-\gamma \over 1+3\gamma}\,,\end{equation}
and Eq.\ (\ref{eq257}) becomes
\begin{equation}
  D_{\rm p}(X)= {1+3\gamma\over 4}X-{1-\gamma \over 4}({\q e}_u\,X\,{\q e}_u- {\q e}_v\,X\,{\q e}_v-{\q e}_w\,X\,{\q e}_w)\,.\label{eq259}
\end{equation}
Equation (\ref{eq259}) is the general expression of a partial depolarizer that depolarizes by a factor $1-\gamma$.

\subsubsection{Actual design of a depolarizer}
We   deal with $\{{\q e}_u,{\q e}_v, {\q e}_w\}=\{ {\q e}_1,{\q e}_2, {\q e}_3\}$ and begin with a perfect depolarizer. We notice that $-{\q e}_1\,X\,{\q e}_1=\exp({\q e}_1{\pi / 2})\,X\,\exp (-{\q e}_1{\pi/2})$, so that the  mapping $X\longmapsto -{\q e}_1X{\q e}_1$ expresses the effect of  a half-wave plate on the polarization state represented by the minquat $X$.

\begin{figure}[h]%$$$$$$$$$$$$$$$$$$$$$$$$$$$$$$$$$$$$$$$$$$$$$$$$$$$$$$$$$$$$$$$
  \begin{center}
    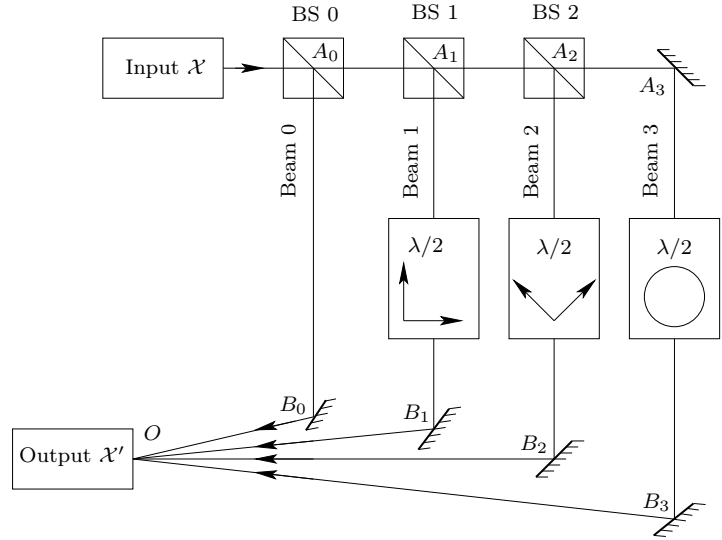
    \caption{\small Experimental setup for depolarizing a polarized lightwave ${\cal X}$, represented by a minquat $X=X_0(1+\I\,\rho\,{\q e}_n)$. Beam splitter BS 0 transmission factor is $T_0=0.75$, and its reflection factor is $R_0=0.25$. For BS 1, we have $T_1=2/3\approx 0.67$ and $R_1=1/3\approx 0.33$; for  BS 2, $T_2=R_2=0.5$. The four beams 0, 1, 2, 3 thus have equal powers (equal to $X_0/4$). %, where $X_0$ is the power of state $X$).
      Beam~1 passes through a half-wave plate whose eigenvibrations are horizontal and vertical; beam 2 passes through a half-wave plate whose eigenvibrations are at $\pm 45^\circ$ from the horizontal; beam 3 passes through an optically active medium whose rotatory power is $\pi /2$ (which is a half-wave plate whose eigenvibrations correspond to circularly polarized states). To make  the wave superposition at $O$ incoherent, the optical path differences $[A_2A_3B_3O]-[A_2B_2O]$, $[A_1A_2B_2O]-[A_1B_1O]$ and  $[A_0A_1B_1O]-[A_0B_0O]$ must be larger than the coherence length of the light source. For every input polarized wave ${\cal X}$, the ouput is an unpolarized lightwave ${\cal X}'$, represented by a minquat of the form $X'=X_0$ (scalar quaternion). \label{fig9}}
  \end{center}
\end{figure}%$$$$$$$$$$$$$$$$$$$$$$$$$$$$$$$$$$$$$$$$$$$$$$$$$$$$$$$$$$$$$$$$$$$$

 The  axis of the plate is $X_1$ on the Poincar\'e sphere, which means that the fast eigenvibration of the plate is horizontal, whereas its slow eigenvibration is vertical.

Similarly the  mapping $X\longmapsto -{\q e}_2X{\q e}_2$ expresses the effect of  a half-wave plate, whose eigenvibrations are at $\pm 45^\circ$ of the horizontal ($X_2$--axis on the Poincar\'e sphere); 
 and the  mapping $X\longmapsto -{\q e}_3X{\q e}_3$ expresses the effect of  a half-wave plate, whose eigenvibrations are circularly polarized ($X_3$--axis on the Poincar\'e sphere), that is, a rotaroy power equal to $\pi/2$.

A setup is shown on Fig.\ \ref{fig9}. Explanations are given in the caption. To obtain a partial depolarizer we replace the beam splitter 0 with a beam splitter whose transmission factor is $T_0= 3(1-\gamma )/4$ and reflection factor is $R_0=(1+3\gamma)/4$. For example, if we want $\gamma =0.5$, then we have to set $T_0=0.375$ and  $R_0=0.625$.

%*****************
%FIN DE DEPOL
%*****************

%***************
%RELAPOL
%***************
\section{Some equivalences between special theory  of relativity and polarization optics}\label{sect6}

\subsection{Effect of a hyperbolic rotation}\label{sect61}

\subsubsection{Analysis in the Minkowski space}\label{sect611}

We consider a hyperbolic rotation of axis ${\q e}_n$ and parameter $\delta$, represented by $\exp (\I\,{\q e}_n\delta /2)$, and a minquat $x=x_0({\q e}_0+\I a\,{\q e}_m)$, with $x_0>0$ and
$|a|\le 1$. Under the rotation, $x$ is changed into $x'$ such that
\begin{equation}
  x'=\exp \left(\I\,{\q e}_n{\delta \over 2}\right)\,x\,\exp \left(\I\,{\q e}_n{\delta \over 2}\right)\,.\end{equation}
Then
\begin{eqnarray}
  {x'\over x_0}\rap &=&\rap  \left({\q e}_0\cosh{\delta\over 2}+\I\,{\q e}_n\sinh{\delta\over 2}\right)
  ({\q e}_0+\I a\,{\q e}_m)
  \left({\q e}_0\cosh{\delta\over 2}+\I\,{\q e}_n\sinh{\delta\over 2}\right)\nonumber\\
  \rap &=&\rap   \left({\q e}_0\cosh{\delta\over 2}+\I\,{\q e}_n\sinh{\delta\over 2}
  ++\I a\,{\q e}_m\cosh{\delta\over 2}-a\,{\q e_n}{\q e}_m\sinh{\delta\over 2}\right)
  \left({\q e}_0\cosh{\delta\over 2}+\I\,{\q e}_n\sinh{\delta\over 2}\right)\nonumber\\
  \rap &=&\rap  {\q e}_0\left(\cosh^2{\delta\over 2}+\sinh^2{\delta \over 2}\right)
  +2\I\,{\q e}_n\cosh{\delta\over 2}\sinh{\delta\over 2}-a\,({\q e}_n{\q e}_m+{\q e}_m{\q e}_n)\cosh{\delta\over 2}\sinh{\delta\over 2}\nonumber \\
  & &\hskip 2cm +\I a\,{\q e}_m\cosh^2{\delta\over 2}-\I a\,{\q e}_n{\q e}_m{\q e}_n\sinh^2{\delta\over 2} \nonumber \\
  \rap &=&\rap {\q e}_0(\cosh\delta +a\,{\q e}_m\vec\cdot {\q e}_n \sinh\delta)
  +\I\,{\q e}_n\sinh\delta +\I a\,{\q e}_m\cosh^2{\delta\over 2}-\I a\,{\q e}_n{\q e}_m{\q e}_n\sinh^2{\delta\over 2}\,.\label{eq6.202}
\end{eqnarray}
We examine two particular cases:

\begin{enumerate}
\item We assume ${\q e}_n={\q e}_m$. Then ${\q e}_n{\q e}_m{\q e}_n=-{\q e}_m$ and Eq.\ (\ref{eq6.202}) becomes
  \begin{equation}
    x'=x_0{\q e}_0(\cosh\delta +a\sinh\delta ) +\I x_0{\q e}_m (\sinh\delta +a\cosh\delta )\,.\label{eq6.203}
  \end{equation}
\item We assume ${\q e}_n\vec\cdot{\q e}_m =0$. According to the corollary of \ref{appenD} we have  ${\q e}_n{\q e}_m{\q e}_n={\q e}_m$ and Eq.\ (\ref{eq6.202}) becomes
  \begin{equation}
    x'=x_0{\q e}_0\cosh\delta +\I x_0{\q e}_n\sinh\delta +\I x_0a\,{\q e}_m\,.\label{eq6.204}
  \end{equation}
\end{enumerate}

\subsubsection{Application to the special theory of relativity}
The result of Section \ref{sect611} is applied to the change of the components of a 4-vector when passing from a Galilean frame to another. The 4-vector under consideration will be the 4-velocity of a particle.

In the present section $u$ does not represent a rotation, but the algebraic value of the 3-velocity of a particle.

The hyperbolic rotation is a Lorentz boost that gives the change of coordinates between two inertial frames of reference, say ${\cal S}$ and ${\cal S}'$, in relative uniform translatory motion. The relative 3--velocity of ${\cal S}'$ with respect to ${\cal S}$ is $v\,{\qv e}_n$, and it is usual to denote $\beta =v/{\rm c}$ and
\begin{equation}
  \gamma_v ={1\over \sqrt{1-\displaystyle{v^2\over {\rm c}^2}}}\,.\end{equation}
The link with the rapidity $\delta$ is given by
\begin{equation}
  \cosh\delta =\gamma_v\,,\hskip .5cm \mbox{and}\hskip .5cm \sinh\delta =-\gamma_v\beta\,.\end{equation}
that is,   $\tanh \delta =-\beta=- v/{\rm c}$.

We consider a particle whose 3--velocity in ${\cal S}$ is $u\,{\qv e}_m$, so that the  minquat representation of its 4--velocity \cite{Gou,Mol}, with respect to ${\cal S}$ is
\begin{equation}
  U=\gamma_u ({\rm c}\,{\q e}_0+\I  u\,{\q e}_m)\,,\label{eq6.207}\end{equation}
where
\begin{equation}
  \gamma_u={1\over \sqrt{1-\displaystyle{u^2\over {\rm c}^2}}}\,.\end{equation}
The result of Section \ref{sect611} is applied by taking  the 4--velocity of the particle as $x$: we set $x=U$, $x_0=\gamma_u {\rm c}$ and $a= u/{\rm c}$.

The two cases of Section \ref{sect611} are the followings.
\begin{enumerate}
\item If ${\q e}_n={\q e}_m$. According to Eq.\ (\ref{eq6.203}), $x'$ takes the form $x'=x'_0({\q e}_0+\I a'\,{\q e}_m)$ with
  \begin{equation}
    x'_0=\gamma_u\gamma_v{\rm c}\left(1-{uv\over{\rm c}^2}\right)\,,\end{equation}
  and
  \begin{equation}
   x'_0 a'=\gamma _u\gamma_v(u-v)\,.
  \end{equation}
  (Since $|u|<{\rm c}$ and $|v|<{\rm c}$, we have $x'_0>0$.)
  We deduce
  \begin{equation}
    a'={u-v\over {\rm c}\left(1-\displaystyle{uv\over{\rm c}^2}\right)}
    \,,  \label{eq304}  \end{equation}
  so that the relative 3--velocity of the particle with respect to ${\cal S}'$ takes the form $u'{\qv e}_m$ with
  \begin{equation}
    u'=a'{\rm c}={u-v\over 1-\displaystyle{uv\over{\rm c}^2}}
    \,.  \label{eq6.212}  \end{equation}
  Since $|u|<1$, we remark that there is a real number $\alpha$ such that $a=u/{\rm c}=\tanh \alpha$. Equation (\ref{eq304}) becomes
  \begin{equation}
    a'={\tanh\alpha +\tanh\delta\over 1+\tanh\alpha\tanh\delta}=\tanh (\alpha +\delta)\,.\label{eq306}
  \end{equation}
  We write $a'=\tanh \alpha'$, and obtain $\alpha ' =\alpha +\delta$, which is the composition law of rapidities.
  (See Eq.\ \ref{eq118v} in Section \ref{sect258}.)
  
  Let $\gamma_{u'}$ be defined by
  \begin{equation}
    \gamma_{u'}={1\over \sqrt{1-\displaystyle{{u'}^2\over {\rm c}^2}}}\,.\end{equation}
  We have
  \begin{equation}
    {1\over \gamma_{u'}^2}=1-{(u-v)^2\over {\rm c}^2\left(1-\displaystyle{uv\over{\rm c}^2}\right)^2}=
    {{\rm c}^2+({u^2v^2/{\rm c}^2})-u^2-v^2\over  {\rm c}^2\left(1-\displaystyle{uv\over{\rm c}^2}\right)^2}
    ={1\over \gamma_u^2\gamma_v^2\left(1-\displaystyle{uv\over{\rm c}^2}\right)^2}\,,
  \end{equation}
  so that ($x'_0>0$)
  \begin{equation}
    x'_0=\gamma_{u'}{\rm c}\,.\end{equation}
 With respect to ${\cal S}'$, the four-velocity vector of the particle takes the form
  \begin{equation}
    U'=\gamma_{u'}{\rm c}\,{\q e}_0+\I\,\gamma_{u'} u'{\q e}_m\,,\end{equation}
  with is similar to Eq.\ (\ref{eq6.207}).

  Equation (\ref{eq6.212}) is the relativistic composition law of 3--velocities, when the particle velocity is along the relative velocity of ${\cal S}$ and ${\cal S}'$. It may be inverted in the form
    \begin{equation}
      u={u'+v\over 1+\displaystyle{u'v\over {\rm c}^2}}\,,\end{equation}
    which is exactly Eq.\ (\ref{eq119v}), in which $v_m$ is replaced with $v$, velocity $v_n$ is replaced with $u'$,  and $v$ is eventually replaced with $u$.

\item If ${\q e}_n\vec\cdot{\q e}_m=0$, Eq.\ (\ref{eq6.204}) gives
  \begin{equation}
    x'=\gamma_u\gamma_v{\rm c}\,{\q e}_0-\I\gamma_u\gamma_vv\,{\q e}_n+\I\gamma_u u\,{\q e}_m=x'_0{\q e}_0+\I x'_0a'\,{\q e}_u\,,
  \end{equation}
  that is
  \begin{equation}
    x'_0=\gamma_u\gamma_v{\rm c}=\gamma_vx_0\,,\label{eq6.312}\end{equation}
  \begin{equation}x'_0\,a'\,{\q e}_u={x'_0\,u'\over {\rm c}}\,{\q e}_u=\gamma_u(u\,{\q e}_m-\gamma_v v\,{\q e}_n)\,,
    \end{equation}
  and we deduce
  \begin{equation}
    u'\,{\q e}_u={u\over \gamma_v}{\q e}_m-v\,{\q e}_n\,.\label{eq314}\end{equation}
  Since ${\q e}_m$, ${\q e}_n$ and ${\q e}_u$ are unit pure quaternions and since ${\q e}_m$ and ${\q e}_n$ are orthogonal to each other, we obtain
  \begin{equation}
    u'=\sqrt{v^2+{u^2 \over {\gamma_v}^{\! 2}}}\,,\end{equation}
  that is
  \begin{equation}
    a'={u'\over {\rm c}}=\sqrt{\tanh^2\delta +{a^2\over \cosh^2\delta}}\,.\label{eq316}\end{equation}
\end{enumerate}

\subsubsection{Application to polarization optics}
The quaternion $\exp (\I\,{\q e}_n\delta /2)$ represents a dichroic device, whose dichroism is $\delta$ ($\delta >0$) and whose eigenvibrations are represented by the points $-{\qv e}_n$ and ${\qv e}_n$ on the Poincar\'e sphere. The minquat $x$ represents a partially polarized wave, and with the previous notations for polarization states we write $x=X=X_0(1+\I\rho\,{\qv e}_n)$, where $x_0=X_0$ is the lightwave power and $a=\rho$ is its degree of polarization.

We obtain the following results.
\begin{enumerate}
\item If ${\q e}_m={\q e}_n$, the polarized component of $X$ is polarized along an eigenvibration of the dichroic. According to Eq.\ (\ref{eq6.203}), the emerging state takes the form $X'=X_0'(1+\I \rho'\,{\q e}_m)$, with
  \begin{equation}
    X'_0=X_0(\cosh\delta +\rho\sinh\delta)\,,
  \end{equation}
  \begin{equation}\rho '={\sinh\delta +\rho\cosh\delta\over \cosh\delta +\rho\sinh\delta}
    \,.
  \end{equation}
  \begin{enumerate}
    \item
      If $\rho =1$, we obtain $\rho '=1$.  The minquat $X=X_0({\q e}_0+\I\,{\q e}_m)$ represents a completely polarized eigenstate and also does $X'$. Since $\delta >0$, we have $X'_0>X_0$ (the dichroic is a pure dichroic, so that the lightwave power is amplified; for a real case an isotropic absorption factor should be taken into account, see Section \ref{sect523}).
      \item
  For $0\le \rho <1$,  the degree of polarization $\rho$ can be written $\rho =\tanh \alpha$, with $\alpha \ge 0$. Then
  \begin{equation}
    \rho '={\rho +\tanh\delta \over 1+ \rho\tanh\delta}={\tanh\alpha +\tanh\delta \over 1+ \tanh\alpha\tanh\delta}=\tanh (\alpha+\delta )\,.\label{eq318}
  \end{equation}
  Equation (\ref{eq318}) is equivalent in polarization optics to Eq.\ (\ref{eq306}) in Relativity.
  
  Let us assume $\delta >0$. Then  $\tanh (\alpha +\delta)> \tanh\alpha$, which means $\rho '>\rho$. The degree of polarization is increased under a dichroic (with positive dichroism). That is why dichroics sometimes are called {\bf partial polarizers}.
  \end{enumerate}
%*******************************
\item  If ${\q e}_m$ and ${\q e}_n$ are orthogonal, according to Eq.\ (\ref{eq6.204}) we obtain
  \begin{equation}
    X'=X_0{\q e}_0\cosh\delta \left({\q e}_0+\I \,{\q e}_n\tanh\delta +{\I\rho\,{\q e}_m\over\cosh\delta}\right)\,,\end{equation}
  which can be written $X'=X'_0({\q e}_0+\I\rho '{\q e}_u)$. Since ${\q e}_m$ and ${\q e}_n$ are orthogonal, we obtain
  \begin{equation}
    \rho'=\sqrt{\tanh^2\delta +{\rho^2\over \cosh^2\delta}}\,,\label{eq6.321}
  \end{equation}
  which is equivalent to Eq.\ (\ref{eq316}). We also have
  \begin{equation}
    \rho '{\q e}_u={\q e}_n\tanh\delta +{\rho\,{\q e}_m\over\cosh\delta}\,,\end{equation}
  equivalent to Eq.\ (\ref{eq314});
  and $X'_0=X_0\cosh\delta$, equivalent to Eq.\ (\ref{eq6.312}).
  \begin{enumerate}
  \item If $\rho =1$, then $\rho '=1$: both $X$ and $X'$ correspond to completely polarized waves.
  \item If $0\le \rho <1$, Eq.\ (\ref{eq6.321}) gives  $\rho'^2\cosh^2\delta =\sinh^2\delta +\rho^2$, that is,
    $1-\rho^2=(1-\rho'^2)\cosh^2\delta$. Since $\rho <1$ and $\cosh^2\delta > 1$, we obtain
    $0<1-\rho'^2<1-\rho^2$ and eventually $\rho <\rho '<1$. The degree of polarization is increased by the  dichroic, which  operates as a partial polarizer.
    \end{enumerate}
  
\end{enumerate}

%****************
%WIGNERROT
%****************
\subsection{Wigner rotation and Thomas precession}\label{sect62}

\subsubsection{Geometric Wigner rotation}\label{sect621}

We will show that the product of two hyperbolic rotations, with distinct axes, does not reduce to a hyperbolic rotation, but involves an elliptic rotation, known as Wigner rotation.

\begin{proposition}[Wigner rotation]\label{prop61}
 {\its  Let $\exp (\I \,{\q e}_\ell \delta '/ 2)$ and $\exp (\I \,{\q e}_m\delta ''/ 2)$ represent two hyperbolic rotations, with $\delta '\delta ''\ne 0$ and let $\theta$ be the angle between ${\q e}_\ell$ and ${\q e}_m$, taken from ${\q e}_\ell$ to ${\q e}_m$ (we assume $0\le \theta\le \pi $).   Then $u=\exp (\I \,{\q e}_\ell \delta '/ 2)\,\exp (\I\, {\q e}_m\delta ''/ 2)$ takes the form $u=\exp ({\q e}_n\varepsilon/ 2)\, u'$, where $u'$ is a unit quaternion representing a hyperbolic rotation and, if $\theta\ne 0\mod \pi$, where  $\exp ({\q e}_n\varepsilon/ 2)$ represents an elliptic rotation (called Wigner rotation) whose axis is ${\q e}_n$, with ${\q e}_n\sin\theta ={\q e}_\ell\vec\times {\q e}_m$, and whose angle is $\varepsilon$  such that
  \begin{equation}
    \tan {\varepsilon\over 2}={\tanh\displaystyle{\delta '\over 2}\tanh\displaystyle{\delta ''\over 2}\sin\theta\over
      1+\tanh\displaystyle{\delta '\over 2}\displaystyle\tanh{\delta ''\over 2}\cos \theta}\,.\label{eq286}
  \end{equation}
  }
  \end{proposition}

\noindent{\its  Proof.}
Since $\delta'\delta ''\ne 0$, if $\theta\ne 0\mod \pi$, we have $\tan (\varepsilon /2)\ne 0$ and
$\varepsilon\ne 0\mod \pi$, so that $\exp ({\q e}_n\varepsilon/ 2)$ represents an actual elliptic rotation (not the identity).

To lighten notations we write
\begin{equation}
  r=\cos{\varepsilon\over 2}\,,\hskip .15cm
  s=\sin{\varepsilon\over 2}\,,\hskip .15cm
  r'=\cosh{\delta '\over 2}\,,\hskip .15cm
  s'=\sinh{\delta '\over 2}\,,\hskip .15cm
 r''=\cosh{\delta ''\over 2}\,,\hskip .15cm
 s''=\sinh{\delta ''\over 2}\,,\end{equation}
 so that Eq.\ (\ref{eq286}) becomes
  \begin{equation}
    {s\over r}={\displaystyle{s's''\over r'r''}\sin\theta\over 1+\displaystyle{s's''\over r'r''}\cos\theta}\,,
  \end{equation}
  and gives
  \begin{equation}
    sr'r'' + ss's''\cos\theta - rs's''\sin\theta =0\,.\label{eq289}
  \end{equation}

  For $\theta \ne 0$, the quaternion ${\q e}_n={\q e}_\ell \vec\times{\q e}_m/\sin\theta$ is unitary and orthogonal to the plane $({\q e}_\ell,{\q e}_m)$. We obtain a direct basis $\{ {\q e}_\ell, {\q e}_q,{\q e}_n\}$ of pure quaternions by introducing  ${\q e}_q={\q e}_n{\q e}_\ell ={\q e}_n\vec\times {\q e}_\ell$, so that ${\q e}_m={\q e}_\ell\cos\theta +{\q e}_q\sin\theta$. We define $u'=\exp(-{\q e}_n\varepsilon /2)\,u$ and  then derive
  \begin{eqnarray}
  u'\rap&=&\rap\exp \left(-{\q e}_n{\varepsilon\over 2}\right)\,u\nonumber \\
  &=&\rap\exp \left(-{\q e}_n{\varepsilon\over 2}\right)\;\exp \left(\I\, {\q e}_m {\delta ''\over  2}\right)\;\exp \left(\I\, {\q e}_\ell{\delta '\over 2}\right)\nonumber\\
   &=&\rap (r-s\,{\q e}_n)(r''+\I s''{\q e}_m)(r'+\I s'\,{\q e}_\ell)\nonumber \\
  &=&\rap (r-s\,{\q e}_n)\bigl[r''+\I s''({\q e}_\ell\cos\theta +{\q e}_q\sin\theta ) \bigr](r'+\I s'\,{\q e}_\ell)\nonumber \\
  &=&\rap  \I s''\cos\theta (r-s\,{\q e}_n)\,{\q e}_\ell(r'+\I s'\,{\q e}_\ell)
  + (r-s\,{\q e}_n)(r'' +\I s'' {\q e}_q\sin\theta )(r'+\I s'\,{\q e}_\ell)
  \nonumber \\
  &=&\rap rs's''\cos\theta +\I rr's''\,{\q e}_\ell\cos\theta- \I s r' s''\, {\q e}_q\cos\theta - s s' s'' \,{\q e}_n\cos\theta \nonumber \\
  & & \hskip 1cm
  + r r' r'' + s s' s''\sin\theta +\I (r s' r''+ s r' s''\sin\theta)\,{\q e}_\ell + \I (r r' s''\sin\theta- s s' r'')\,{\q e}_q\nonumber \\
  & &\hskip 1cm + (rs's''\sin\theta-sr'r'')\,{\q e}_n\,.\label{eq10q}\end{eqnarray}
  According to Eq.\ (\ref{eq289}), the coefficient of ${\q e}_n$ in Eq.\ (\ref{eq10q}) is zero, so that we obtain 
  \begin{eqnarray}
    u' \rap&=&\rap
  rs's''\cos\theta+ r r' r'' + s s' s''\sin\theta
  +\; \I (rr's''\cos\theta + r s' r''+ s r' s''\sin\theta)\,{\q e}_\ell \nonumber \\
  & &\hskip 1cm
  + \;\I (r r' s''\sin\theta- s s' r'' -  s r' s''\cos\theta )\,{\q e}_q\,.\label{eq326}
\end{eqnarray}
The quaternion $u'$ is a unit quaternion as product of three unit quaternions. Moreover, Eq.~(\ref{eq326}) shows that its vector part is imaginary: then $u'$ represents a hyperbolic rotation on the Minkowski space. The proof is complete. \qed

\begin{remark} {\rm In Proposition \ref{prop61}, we assume $0\le \theta\le \pi$. If $-\pi <\theta < 0$, we bring back to the previous assumption by changing ${\q e}_m$ into $-\,{\q e}_m$, that is, $\theta$ into $\pi +\theta$. We also change $\delta ''$ into $-\delta ''$, so that $u$ is not changed, because $\exp (\I\,{\q e}_m\delta ''/2)= \exp [\I (-\,{\q e}_m)(-\delta '')/2]$. 
  }
  \end{remark}

\subsubsection{Application to Relativity: Thomas precession \cite{Gou,Mis,Mol,Rou}}%***************************

The Thomas precession  is a relativistic effect that is no more than a physical achievement of the Wigner rotation. It is a consequence of the fact that, according to the previous section, the product of two boosts is not a boost, in general, but a boost plus an elliptic rotation (Wigner rotation).

We will only consider a particular case: quaternions ${\q e}_\ell$ and ${\q e}_m$ are orthogonal, that is $\theta =\pi /2$.  We have  ${\q e}_n={\q e}_\ell\vec\times {\q e}_m$ and ${\q e}_q={\q e}_m$. We deduce 
\begin{equation}
  \tan{\varepsilon\over 2}=\tanh{\delta '\over 2}\tanh{\delta ''\over 2}\,,\label{eq292}
\end{equation}
the axis of the Wigner rotation being ${\q e}_n$.

We consider three Galilean inertial systems ${\cal R}$, ${\cal R}'$ and ${\cal R}''$ with three respectives coordinates frames $\Omega$, $\Omega '$ and $\Omega ''$.  The system ${\cal R}'$ has a uniform translatory motion with respect to ${\cal R}$; and ${\cal R}''$ has a  uniform translatory motion with respect to ${\cal R}'$. Contrary to what would be expected according to the laws of classical mechanics, the connection between coordinates in $\Omega ''$ and coordinates in $\Omega$ does not correspond to a hyperbolic rotation: %motion of ${\cal R}''$ with respect to ${\cal R}$ is not a translatory motion: an additional rotation appears.
an additional rotation has to be taken into account.

We now assume that the relative velocity of ${\cal R}'$ with respect to ${\cal R}$ is $v'\,{\q e}_1$ and that the relative velocity of ${\cal R}''$ with respect to ${\cal R}'$ is $v''\,{\q e}_2$. The axis of the Wigner rotation is ${\q e}_3$.

In place of velocities, we use rapidities, that is $\delta '$ and $\delta ''$ such that
\begin{equation}
  \tanh\delta '=-{v'\over {\rm c}}\,,\hskip .5cm  \tanh\delta ''=-{v''\over {\rm c}}\,.\end{equation}

We assume $|v'|\ll{\rm c}$ and $|v''|\ll{\rm c}$, so that at first order
\begin{equation}
  \tanh{\delta '\over 2}\approx-{v'\over 2{\rm c}}\,,\hskip .5cm  \tanh{\delta ''\over 2}\approx -{v''\over 2{\rm c}}\,.\end{equation}
Equation (\ref{eq292}) gives
\begin{equation}
  \varepsilon\approx {v'v''\over 2{\rm c}^2}\,.
\end{equation}
Generally $v''$ is regarded like a variation of $v'$, namely $v''=\D v'$, and eventually
\begin{equation}
  {\D \varepsilon\over \D t}\approx {1\over 2{\rm c}^2}v'{\D v'\over \D t}\,,
\end{equation}
where $t$ denotes proper time in ${\cal R}$.
  
\subsubsection{Application to polarization optics}

In optics, the Wigner rotation causes the product of two dichroics not to be reduced to a dichroic device, in general, but to include a birefringent part. To provide an example we consider two rectilinear dichroics whose eigenvibrations are at $45^\circ$ to each other. On the Poincar\'e sphere the  axes of the two dichroics are along orthogonal unit pure quaternions ${\q e}_\ell$ and ${\q e}_m$ (their representative points are on the equator). Their respective dichroisms  are denoted $\delta '$ and $\delta ''$. 

We use notations of Section \ref{sect621} and assume $\theta =\pi /2$. The product of the two dichroics involves a rotatory power, because ${\q e}_n={\q e}_\ell\vec\times {\q e}_m={\q e}_\ell\,{\q e}_m={\q e}_3$, so that the Wigner rotation is around ${\q e}_3$. According to Proposition \ref{prop61} the angle of the rotation is $\varepsilon$ such that
\begin{equation}
  \tan{\varepsilon\over 2}=\tanh\delta '\tanh\delta '\,,\end{equation}
which correponds to a rotatory power $\varrho =\varepsilon/2$.

 The product of the two dichroic devices is represented by $u=\exp (\I\,{\q e}_3\varrho )\, u '$, with
\begin{eqnarray}
  u'\rap & = &\rap\cos\varrho \cosh{\delta '\over 2}\cosh{\delta ''\over 2}+
  \sin\varrho \sinh{\delta '\over 2}\sinh{\delta ''\over 2}\nonumber \\
  & & \hskip 1cm +\,\I\,{\q e}_\ell  \left(\cos\varrho \sinh{\delta '\over 2}\cosh{\delta ''\over 2}+
   \sin\varrho \cosh{\delta '\over 2}\sinh{\delta ''\over 2}\right)
   \nonumber \\
   & & \hskip 1cm + \,\I\,{\q e}_m  \left(\cos\varrho \cosh{\delta '\over 2}\sinh{\delta ''\over 2}-
   \sin\varrho \sinh{\delta '\over 2}\cosh{\delta ''\over 2}\right)\,,
\end{eqnarray}
because  ${\q e}_q$ in Eq.\ (\ref{eq326}) is equal to ${\q e}_m$ (we have ${\q e}_q={\q e}_3{\q e}_\ell= {\q e}_\ell\,{\q e}_m{\q e}_\ell={\q e}_m$). 
The unit quaternion $u'$ represents a rectilinear dichroic. It can be written
\begin{equation}
  u'=\cosh{\delta\over 2}+\I\,{\q e}_p\sinh{\delta \over 2}\,,
  \end{equation}
where
\begin{equation}
  \cosh{\delta\over 2}=\cos\varrho \cosh{\delta '\over 2}\cosh{\delta ''\over 2}+
  \sin\varrho \sinh{\delta '\over 2}\sinh{\delta ''\over 2}\,.
  \end{equation}

%****************
%FIN DE WIGNERROT
%****************

\subsection{Singular proper Lorentz rotations in Relativity\\ and in polarization optics}

\subsubsection{Geometrical analysis}
The composition of a hyperbolic rotation represented by $\exp (\I\,{\q e}_n\delta /2)$ and an elliptic rotation rerepresented by $\exp ({\q e}_m\varphi /2)$ is represented by
\begin{eqnarray}
  u\rap &=&\rap\exp ({\q e}_m\varphi /2)\exp (\I\,{\q e}_n\delta/2) \label{eq336}\\
  \rap &=&\rap \left(\cos{\varphi\over 2}+{\q e}_m\sin{\varphi\over 2}\right) \left(\cosh{\delta\over 2}+\I\,{\q e}_n\sinh{\delta\over 2}\right)\nonumber \\
  \rap &=&\rap  \cos{\varphi\over 2}\cosh{\delta\over 2}-\I\,{\q e}_m\vec\cdot{\q e}_n\sin{\varphi\over 2}\sinh{\delta\over 2} + {\q e}_m\sin{\varphi\over 2}\cosh{\delta\over 2}+ \I\, {\q e}_n\cos{\varphi\over 2}\sinh{\delta\over 2}\nonumber \\
     &&\ \hskip 8cm + \,\I\,{\q e}_m\vec \times {\q e}_n\sin{\varphi\over 2}\sinh{\delta\over 2}\,,\nonumber
\end{eqnarray}
which takes the form $u=\exp ({\q e_u}\psi /2)$,
where $\psi$ is a complex number and ${\q e}_u$ a complex pure quaternion. (We assume $\varphi\delta \ne 0$.)

In general, the quaternion ${\q e}_u$ is a complex unit pure quaternion ($N({\q e}_u)=1$), so that $u=\cos(\psi /2)+{\q e}_u\sin(\psi /2)$, and the corresponding proper Lorentz rotation is regular (in the meaning of Section \ref{sect256}): it admits two isotropic eigenminquats given by Eqs.\ (\ref{eq106}) and (\ref{eq107}).

But ${\q e}_u$ may also be a null quaternion ($N({\q e}_u)=0$), so that $u$ represents a singular proper Lorentz rotation. If $N({\q e}_u)=0$, we have $u=\exp ({\q e_u}\psi /2)= 1+({\q e}_u\psi /2)$, and for this to happen it is necessary that $\cos (\varphi/2)\cosh (\delta /2)=\pm 1$ and ${\q e}_m\vec\cdot {\q e}_n=0$ (because $\varphi\delta \ne 0$).

We conclude that if ${\q e}_m$ and ${\q e}_n$ are orthogonal to each other and if $\varphi$ and $\delta$ are such that
\begin{equation}
  \cos{\varphi\over 2}=\pm{1\over \cosh\displaystyle{\delta\over 2}}\,,\label{eq341}\end{equation}
then $u$ represents a singular proper Lorentz rotation. That rotation admits only one null-minquat eigendirection, which is along ${\q e}_0+\I\,{\q e}_m{\q e}_n$.

\subsubsection{Singular direction in Relativity}
Let ${\cal S}$ and ${\cal S}'$ be two Galilean inertial frames in uniform translatory motion. Let $\Omega$ be a spatial frame in ${\cal S}$ (coordinates $x_\mu$) and $\Omega '$ in ${\cal S}'$ (coordinates $x'_\mu$). In general the previous frames are such that the $x'_j$--axis is parallel to the $x_j$--axis; moreover $x'_0=0$, if and only if, $x_0=0$; the connection between those coordinates in given by an hyperbolic rotation, in the form $x'=\exp (\I\,{\q e}_n\delta/2)\,x\,\exp (\I\,{\q e}_n\delta/2)$, where $\delta$ is the rapidity (the relative velocity of ${\cal S}'$ with respect to ${\cal S}$ is $v\,{\q e}_n$ and $\tan\delta = -v/{\rm c}$).

Let us assume now that the axes in ${\cal S}'$ are rotated with respect to the $x'_j$--axes. We denote by $x''_j$ the rotated axes. The passage from $\Omega '$ to $\Omega ''$ is a pure rotation and takes the form $x''=\exp ({\q e}_m\varphi /2)\,x'\,\exp (-{\q e}_m/2)$.
The passage from ${\Omega}$ to $\Omega ''$ is then
\begin{equation}
  x''=\exp ({\q e}_m\varphi /2)\exp (\I\,{\q e}_n\delta/2)\,x\,\exp (\I\,{\q e}_n\delta/2)\exp (-{\q e}_m\varphi /2)\,,\end{equation}
  and corresponds to a proper Lorentz rotation, represented by $u$ of the form
  $u=\exp ({\q e}_u\psi /2)$.

  In general the proper Lorentz rotation $u$ admits two null-minquat eigendirections. That means that an observer in ${\cal S}$ and an observer in ${\cal S}$ will find two directions in space that are identical for them \cite{Syn}. In the singular case,  the two directions reduce to a single one \cite{Syn}.

  \subsubsection{Singular axis in polarization optics}\label{sect633}
  In polarization optics, the quaternion $u$ in Eq.\ (\ref{eq336}) represents the succession (or product, or composition) of a dichroic  (dichroism $\delta$, axis ${\q e}_n$) and a  birefringent (birefringence $\varphi$, axis ${\q e}_m$), in this order. In general the resulting device admits two  unit polarization eigenstates (they are not orthogonal, see Section \ref{sect526}). If the birefringent and the dichroic are rectilinear, however, the birefringent eigenvibrations being at 45$^\circ$ of the dichroic eigenvibrations, so that ${\q e}_m$ and ${\q e}_n$ are orthogonal, and if moreover $\varphi$ and $\delta$ satisfy Eq.\ (\ref{eq341}), the device represented by $u$ corresponds to a singular case:  there is only one unit polarization eigenstate, that is, only one null-minquat eigendirection.
 (Not only rectilinear devices may generate singular cases, elliptical ones also may.)

  A singular-axis device  is also obtained through the superposition of birefringence and dichroism. We consider a crystal of optical length $\ell$ that exhibits birefringence and dichroism at the same time (each infinitesimal sample of the crystal exhibits both properties). We use notations of Section \ref{sect255} and denote $\eta$ the birefringence and $\zeta$ the dichroism by unit length, the birefringence axis being ${\q e}_m$ and the dichroism axis being ${\q e}_n$. We define  $\varphi =\eta\ell$ and $\delta =\zeta \ell$, so that the crystal is thus the superposition of a birefringent represented by $\exp ({\q e}_m\varphi /2)$ and a dichroic, represented by $\exp (\I\,{\q e}_n\delta /2)$. The crystal is a pure dephasor represented by
  \begin{equation}
    u=\exp {{\q e}_m\varphi +\I \,{\q e}_n\delta\over 2}\,.\end{equation}
  The singular case occurs if $N({\q e}_m\varphi +\I \,{\q e}_n\delta )=0$, that is if $\delta =\pm\varphi$ and ${\q e}_m\vec\cdot{\q e}_n=0$. Only one polarization is unchanged under propagation in the crystal. Such an effect has been observed on iolite crystals and the single eigenpolarization has been experimentally revealed  by Pancharatnam \cite{Pan1,Pan2,Pan3} (the result is also reported by Ramachandran and Ramaseshan in {\its Crystal optics} \cite{Ram} p.\ 143--144). Iolite is an absorbing biaxial crystal that exhibits  rectilinear birefringence and rectilinear dichroism at the same time. Near an optical axis there is a direction for which birefringence and dichroism are equal with  rectilinear eigenvibrations at $45^\circ$: conditions for having a singular axis are fulfilled. The rectilinear eigenvibrations are represented by ${\q e}_m$ (birefringence) and ${\q e}_n$ (dichroism) on the equator (Poincar\'e sphere), so that the unit eigenminquat is $X=1-\I \,{\q e}_m{\q e}_n=1+\I\,{\q e}_3$ (or $1-\I\,{\q e}_3$, depending on the orientations of  ${\q e}_m$ and  ${\q e}_n$), which means that the eigenpolarization is left-circular (or right-circular).
 
%***************
%FIN DE RELAPOL
%***************

%*****************
%INEQUALITY
%*****************
\section{Triangle inequality and applications}\label{sect7}

\subsection{Triangle inequality}

A geometric property of Minkowski space is that the triangle inequality takes a form inverse from its (well known) Euclidean form \cite{Man,Rou}.

\begin{proposition}[Triangle inequality]\label{prop71}
{\its Let $\vec x=(x_0,x_1,x_2,x_3)$ and $\vec y=(y_0,y_1,y_2,y_3)$ be two vectors of the Minkowski space that may independently be isotropic or timelike. We assume $x_0\,y_0>0$. Then (triangle inequality)
  \begin{equation}
    \sqrt{Q(\vec x+\vec y)}\ge \sqrt{Q(\vec x)}+\sqrt{Q(\vec y)}\,,\label{eq164}\end{equation}
  where $Q$ is the Lorentz quadratic form. The equality holds if, and only if, $\vec x$ and $\vec y$ are collinear.}
\end{proposition}

\medskip
A nice and concise proof is given by  Manoukian \cite{Man}. Here we will prove the triangle inequality for minquats. We replace the Lorentz form $Q$ with $N$, the norm of quaternions, and since $Q(\vec x)=N(x)$, if $\vec x=(x_0,x_1,x_2,x_3)$ and $x=x_0+\I (x_1e_1+x_2e_2+x_3e_3)$, Proposition \ref{prop71} is equivalent to the following proposition: 

\begin{proposition}[Triangle inequality for minquats] {\its Let $x$ and $y$ be two minquats that may independently be isotropic (null) or scalarlike (timelike). We denote $x_0$ and $y_0$ the respective scalar components of $x$ and $y$, and assume $x_0\,y_0>0$. Then (triangle inequality)
  \begin{equation}
    \sqrt{N(x+y)}\ge \sqrt{N(x)}+\sqrt{N(y)}\,,\label{eq164b}\end{equation}
  where $N$ denotes the norm of quaternions. The equality holds if, and only if, $x$ and $y$ are collinear.}
  \end{proposition}

\medskip
\noindent{\its Proof.} {\its (i) Inequality.} We write $x=x_0+\I x_m\,{\q e}_m$ and $y=y_0+\I y_n\,{\q e}_n$, where ${\ q }e_m$ and ${\q e}_n$ are real unit pure quaternions. Since $x$ and $y$ are isotropic or scalarlike, we have $N(x)\ge 0$ and $N(y)\ge 0$; then $|x_0|\ge |x_m|$, and $|y_0|\ge |y_n|$, and since $x_0\,y_0>0$, we have $x_0\,y_0\ge |x_m|\,|y_n|$.

Since $x$ and $y$ are   minquats, %we have $Q(x)=N(x)$ and $Q(y)=N(y)$. Moreover
$x+y$ is also a minquat, and
\begin{eqnarray}
  N(x+y)\rap &=&\rap (x+y)(x^*\!+y^*)=xx^*+yy^*\!+xy^*\!+yx^* \nonumber \\
 \rap &=& \rap N(x)+N(y) +(x_0+\I x_m\,{\q e}_m)(y_0-\I y_n\,{\q e}_n)+(y_0+\I y_n\,{\q e}_n)(x_0-\I x_m\,{\q e}_m)\nonumber \\
  \rap &=&\rap N(x)+N(y)+2x_0y_0+x_my_n\,{\q e}_m\,{\q e}_n+x_my_n\,{\q e}_n\,{\q e}_m\nonumber \\
  \rap &=&\rap N(x)+N(y)+2x_0y_0-2x_my_n\,{\q e}_m\vec \cdot \,{\q e}_n\,.\label{eq165}
\end{eqnarray}
Since $x_my_n\,{\q e}_m\vec\cdot \,{\q e}_n\le |x_m|\,|y_n|$ (which also holds if  $x_my_n\,{\q e}_m\vec\cdot \,{\q e}_n\le 0$), we obtain
\begin{equation}
  N(x+y)\ge N(x)+N(y)+2x_0y_0-2|x_m|\,|y_n|.\label{eq166}\end{equation}
From $x_0y_0\ge |x_m|\,|y_n|$ and Eq.\ (\ref{eq166}) we deduce $N(x+y)\ge 0$, so that $x+y$ is a scalarlike or isotropic minquat, and $\sqrt{N(x+y)}$ makes sense.

From $(x_0\,|y_n|-y_0\,|x_m|)^2\ge 0$\,, we deduce
\begin{equation}
  (x_0y_0)^2+(x_my_n)^2-2x_0y_0|x_m|\,|y_n|\ge (x_0y_0)^2+(x_my_n)^2-(x_0y_n)^2-(y_0x_m)^2\,,\label{eq174}\end{equation}
that is
\begin{equation}
(x_0y_0-|x_m|\,|y_n|)^2\ge \left[(x_0)^2-(x_m)^2\right]\left[(y_0)^2-(y_n)^2\right]=N(x)\,N(y)\,.\end{equation}
Since $x_0y_0\ge |x_m|\,|y_n|\ge 0$, we obtain
\begin{equation}
  x_0y_0-|x_m|\,|y_n|\ge \sqrt{N(x)}\,\sqrt{N(y)}\,,\label{eq176}\end{equation}
and Eq. (\ref{eq166}) leads to
\begin{equation}
  N(x+y)\ge N(x)+N(y)+2 \sqrt{N(x)}\,\sqrt{N(y)}=  \left(\sqrt{N(x)}+\sqrt{N(y)}\right)^2,\end{equation}
which gives Eq.\ (\ref{eq164b}).

\medskip
\noindent{\its (ii) Equality.} 
We  first assume $x$ and $y$ collinear. There exists a real number $\alpha$ such that $y=\alpha x$.   Then $N(y)=\alpha^2N(x)$. Since $x_0y_0>0$ and $y_0=\alpha x_0$, necessarily $\alpha >0$, and we obtain
\begin{equation}
  \sqrt{N(x+y)}= \sqrt{N[(1+\alpha)x]}
  =(1+\alpha )\sqrt{N(x)}
  =\sqrt{N(x)}+\sqrt{N(y)}\,.\label{eq172}\end{equation}

Conversely, if Eq.\ (\ref{eq172}) holds, 
we have
\begin{eqnarray}
  x\vec\cdot y\rap &=&\rap{1\over 2}\bigl[N(x+y)-N(x)-N(y)\bigr]=\sqrt{N(x)N(y)} \nonumber \\
  \rap &=&\rap
  \sqrt{\bigr[ (x_0)^2-(x_m)^2\bigr]\bigl[ (y_0)^2-(y_n)^2\bigr]}\,.\end{eqnarray}
We also have
\begin{equation}
  x\vec\cdot y=(x_0+\I x_m\,{\q e}_m)\vec\cdot (y_0+\I y_n\,{\q e}_n)=
  x_0y_0-x_my_n\,{\q e}_m\vec\cdot \,{\q e}_n\,,\end{equation}
and Eq. (\ref{eq176}) gives
\begin{equation}
  x_0y_0-|x_m|\,|y_n|\ge x_0y_0-x_my_n\,{\q e}_m\vec\cdot \,{\q e}_n\,,\end{equation}
so that, since $x_0y_0\ge |x_m|\,|y_n|$, we necessarily have
\begin{equation}
  |x_m|\,|y_n|\le x_my_n\,{\q e}_m\vec\cdot \,{\q e}_n\,.\end{equation}
But in general $ |x_m|\,|y_n|\ge x_my_n\,{\q e}_m\vec\cdot \,{\q e}_n$, and we conclude
\begin{equation}
  |x_m|\,|y_n| =x_my_n\,{\q e}_m\vec\cdot \,{\q e}_n\,,\label{eq183}\end{equation}
which can be obtained only if ${\q e}_m$ and ${\q e}_n$ are collinear (${\q e}_m\vec\cdot \,{\q e}_n=1$). 

Then Eq.\ (\ref{eq174}) becomes an equality so that
\begin{equation}
  (x_0|y_n|-y_0|x_m|)^2=0\,,\end{equation}
Since $x_0y_0>0$, the number $x_0$ is not zero and we write
\begin{equation}
  |y_n|={y_0\over x_0}|x_m|\,,\end{equation}
and since $|x_m|\,|y_n|\ge 0$, Eq. (\ref{eq183}) leads to
\begin{equation}
y_n\,{\q e}_n={y_0\over x_0}x_m\,{\q e}_m\,, \end{equation}
and finally
\begin{equation}
  y=y_0+\I y_n\,{\q e}_n={y_0\over x_0}(x_0+\I x_m\,{\q e}_m)={y_0\over x_0}\,x\,,\end{equation}
which means that $x$ and $y$ are collinear.
\hfill $\qed$

\subsection{Application to Relativity: twin paradox  \cite{Gou,Bri,Cho,Sch}}%*******************************************

The so called ``twin paradox'' can be explained by applying the triangle inequality to two observers, one of whom is making a round trip.
We assume that the notions of proper time and worldline are known.

\begin{figure}[h]
  \begin{center}
  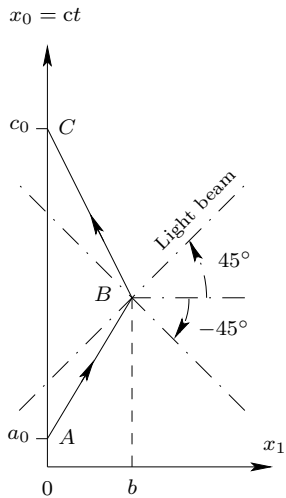
  \caption{\small A schematic diagram for analyzing the twin paradox. Both $\vec {AB}$ and $\vec {BC}$ are timelike vectors: angles of $\vec {AB}$ and $\vec{BC}$ with the $x_0$--axis are less than 45$^\circ$.\label{fig8}}
\end{center}
  \end{figure}

The effect is analyzed on Fig.\ \ref{fig8}, where only one spatial direction is considered. An observer ${\cal O}$ stays at abscissa $x_1=0$, while another observer ${\cal O}'$ travels from abscissa $x_1=0$ to  abscissa $x_1=b$ on the $x_1$--axis and then travels back to abscissa $x_1=0$.  A Galilean inertial frame ${\cal S}$ is attached to ${\cal O}$. Both motions of ${\cal O}'$ are accomplished in uniform translation with respect to ${\cal O}$.

The worldline of ${\cal O}$ is $\vec{AC}$. 
The spatio-temporal  coordinates of $A$ in ${\cal S}$ are $(a_0,0,0,0)$ and  
those of $C$ are $(c_0,0,0,0)$. Then
\begin{equation}
  \vec {AC}=(c_{0}-a_{0},0,0,0)\,.\end{equation}
The proper time-length of ${\cal O}$ between events $A$ and $C$ is ($Q$ denotes the Lorentz quadratic form)
\begin{equation}
  \Delta t= {1\over {\rm c}}\sqrt{Q(\vec {AC})}\,,\end{equation}
which is  the time-length of the ${\cal O}'$--travel as seen by  ${\cal O}$ in ${\cal S}$.

The wordline of ${\cal O}'$ is $\vec{AB}+\vec{BC}$. For the traveling observer the duration of the travel is
\begin{equation}\Delta t'={1\over {\rm c}}\left(\sqrt{Q(\vec{AB})}+\sqrt{Q(\vec{BC})}\right)\,.\end{equation}
The triangle inequality leads to
\begin{equation}
  \sqrt{Q(\vec {AC})}>\sqrt{Q(\vec {AB})}+\sqrt{Q(\vec {BC})}\,,\end{equation}
that is $\Delta t >\Delta t'$.
The travel duration  measured by ${\cal O}$ (in ${\cal S}$) is longer than the travel duration measured by ${\cal O}'$ (on his own clock). If the two observers are twins, ${\cal O}$ is older than ${\cal O}'$ when they meet again.

%$$$$$$$$$$$$$$$$$$$$$$$$$$$$$$$$$$$$$$$$$$
%$$$$$$$$$$$$$$$$$$$$$$$$$$$$$$$$$$$$$$$$$$

\subsection{Application to polarization optics}

\subsubsection{Upper bounds of the degrees of polarization of incoherent superpositions of waves}\label{sect731}

The incoherent superposition of two partially polarized states, represented by minquats $X$ and $Y$, is represented by the sum $X+Y$. Since polarization states are represented by null or scalarlike minquats, we conclude that
$\sqrt{N(X+Y)}\ge \sqrt{N(X)}+\sqrt{N(Y)}$.  
The result may be expressed with the corresponding degrees of polarization. For that, we
consider a partially polarized state $X$, whose degree of polarization is $\rho_{_X}$. The state can be written in the form $X=X_0(1+\I\rho_{_X} \,{\q e}_m)$ ($X_0> 0$), so that $N(X)=(X_0)^2(1-{\rho_{_X}}^{\!\! 2})$.
For physical states, we have $0\le \rho_{_X} \le 1$, then $N(X)\ge 0$, and
\begin{equation}
  \sqrt{N(X)}=X_0\sqrt{1-{\rho_{_X}}^{\!\! 2}}\,.\end{equation}

If we consider the incoherent superposition of $X=X_0(1+\I\rho_{_X} \,e_m)$ and $Y=Y_0(1+\I\rho_{_Y} \,e_n)$, we obtain
\begin{equation}
  X+Y=X_0+Y_0+\I (X_0\rho_{_X}\,{\q e}_m+Y_0\rho_{_Y}\,{\q e}_n)=(X_0+Y_0)(1+\I \rho \,{\q e}_\ell)\,,\end{equation}
where $\rho$ is the degree of polarization of $X+Y$, and ${\q e}_\ell$ is a real unit pure quaternion.
The triangle inequality leads to
\begin{equation}
  \sqrt{1-\rho^2}\ge {X_0\sqrt{1-{\rho_{_X}}^2}+Y_0\sqrt{1-{\rho_{_Y}}^2}\over X_0+Y_0}\,.\label{eq361}\end{equation}

\bigskip
\noindent {\bf Examples.}
We consider two waves ${\cal X}$ and ${\cal Y}$ whose polarization states are represented by minquats $X$ and $Y$. We symbolically write ${\cal X}+{\cal Y}$ the incoherent superposition of waves ${\cal X}$ and ${\cal Y}$, whose polarization state is represented by $X+Y$.

\begin{enumerate}

\item  $X$ and $Y$ are collinear:  $X=X_0(1+\I\rho_{_X} {\q e}_m)$ and $Y=(Y_0/X_0)X=Y_0(1+\I\rho_{_X} {\q e}_m)$, that is $\rho_{_X}=\rho_{_Y}$. The triangle inequality becomes an equality. Thus:
  \begin{enumerate}
  \item Since $\rho_{_X} =\rho_{_Y}$, by Eq.\ (\ref{eq361}) we obtain $\rho =\rho_{_X}$.
  \item In particular, if $\rho_{_X}=\rho_{_Y}=1$ (${\cal X}$ and ${\cal Y}$ are completely polarized), then $\sqrt{1-\rho^2}=0$, so that $\rho =1$. The wave ${\cal X}+{\cal Y}$ is completely polarized.
    \item If ${\cal X}$ and ${\cal Y}$ are unpolarized, then $\rho_{_X}=\rho_{_Y}=0$ and   $\sqrt{1-\rho^2}=1$, so that  ${\cal X}+{\cal Y}$ is  unpolarized.
     \end{enumerate}

\item  $X$ and $Y$ are not collinear.
  \begin{enumerate}
  \item If  ${\cal X}$ and ${\cal Y}$ have equal degrees of polarization (not zero): $\rho_{_X}=\rho_{_Y}\ne 0$, we obtain $\sqrt{1-\rho^2}>\sqrt{1-{\rho_{_X}}^2}$ : the degree of polarization of $X+Y$ is smaller than the degree of polarization of both $X$ and $Y$. In particular, if $\rho_{_X}=\rho_{_Y}=1$, we obtain $\sqrt{1-\rho^2}>0$, which provides no additional information on $\rho$, since we a priori knew that $\rho <1$.  A more precise result is obtained if we take the angle between ${\q e}_m$ and ${\q e}_n$ into account, as done in Section \ref{sect732}.
  \item For $\rho_{_X}=0.80$, $\rho_{_Y}=0.60$ and $X_0=Y_0$, we obtain   $\sqrt{1-\rho^2}>0.70$ and $\rho < 0.714$.

  \end{enumerate}

  \end{enumerate}

\subsubsection{Explicit expression of the degree of polarization of the incoherent superposition of two waves}\label{sect732}

We write $X=X_0(1+\I\rho_{_X}{\q e}_m)$ and  $Y=Y_0(1+\I\rho_{_Y}{\q e}_n)$, and then
\begin{eqnarray}
  N(X+Y)\rap&=&\rap(X+Y)(X^*+Y^*)= XX^*+YY^*+XY^*+YX^* \nonumber \\
  &=&\rap N(X)+N(Y)+ X_0Y_0\bigl[(1+\I\rho_{_X}{\q e}_m)(1-\I\rho_Y{\q e}_n)
    +(1+\I\rho_{_Y}{\q e}_n)(1-\I\rho_{_X}{\q e}_m)\bigr] \nonumber \\
  &=&\rap  N(X)+N(Y)+2X_0Y_0(1- \rho_{_X}\rho_{_Y}{\q e}_m\vec\cdot{\q e}_n)\,,
\end{eqnarray}
that is
\begin{equation}
  {1-\rho^2}={{X_0}^2(1-{\rho_{_X}}^2)+{Y_0}^2(1-{\rho_{_Y}}^2)+2X_0Y_0(1- \rho_{_X}\rho_{_Y}{\q e}_m\vec\cdot{\q e}_n)\over (X_0+Y_0)^2}\,.\end{equation}

\bigskip
\noindent {\bf Examples.}
\begin{enumerate}
\item Both ${\cal X}$ and ${\cal Y}$ are completely polarized ($\rho_{_X}=\rho_{_Y}=1$) and $X_0=Y_0$. Then
\begin{equation}
  1-\rho^2= {1\over 2}(1-{\q e}_m\vec\cdot{\q e}_n)\,.\end{equation}
We obtain the following results:
\begin{enumerate}
  \item  If moreover the waves ${\cal X}$ and ${\cal Y}$ are  orthogonally  polarized, that is if  ${\q e}_n=-\,{\q e}_m$,
    then $\rho =0$, which means that ${\cal X}+{\cal Y}$ is unpolarized. The incoherent superposition of two orthogonal completely polarized waves of equal powers is unpolarized. 
    \item If $X$ and $Y$ are collinear, that is  ${\q e}_n={\q e}_m$, then $\rho =1$, and ${\cal X}+{\cal Y}$ is completely polarized (along ${\q e}_m$ on the Poincar\'e sphere). This is example 1-b of Section \ref{sect731}.
\item If ${\cal X}$ and ${\cal Y}$ are rectilinearly polarized at 45$^\circ$ to each other (in the wave plane), then 
${\q e}_m$ and ${\q e}_m$ are orthogonal on the Poincar\'e sphere and ${\q e}_m\vec\cdot {\q e}_n=0$. We obtain  $\rho =\sqrt{2}/2\approx 0.707$. The same result is obtained if ${\cal X}$ is rectilinearly polarized and ${\cal Y}$ circularly polarized. The incoherent superposition of the two waves is partially polarized.
\end{enumerate}
\item We assume $\rho_{_X}=\rho_{_Y}=0$. Then $1-\rho^2=1$ and $\rho =0$. This is example 1-c of Section \ref{sect731}.
\item Let us assume $\rho_{_X}=0.80$,  $\rho_{_Y}=0.60$ and $X_0=Y_0$.
  \begin{enumerate}
     \item If ${\q e}_n=-{\q e}_m$, we obtain  $1-\rho^2=0.99$ and then $\rho =0.10$.
     \item If ${\q e}_m\vec\cdot {\q e}_n=0$, we obtain  $1-\rho^2=0.75$ and then $\rho = 0.50$.
     \item If ${\q e}_n={\q e}_m$, we obtain  $1-\rho^2=0.51$ and then $\rho =0.70$.
        
  \end{enumerate}
  The first two examples above show that the estimate of the example 2-b of Section \ref{sect731} ($\rho  < 0.714$) may be a rather rough estimate. Moreover, the value $\rho=0.714$ cannot be reached, whatever ${\q e}_m$ and ${\q e}_n$. To obtain $\rho =0.714$, we should have $1-\rho^2=0.49$, and then, with $X_0=Y_0$,
  \begin{equation}
    0.49={1\over 4}(0.36+0.64+2-0.96\,{\q e}_m\vec\cdot{\q e}_n)\,,\end{equation}
  that is
  \begin{equation}
    0.24 \,{\q e}_m\vec\cdot\,{\q e}_n=0.26\,,
  \end{equation}
  which is not possible, because ${\q e}_m$ and ${\q e}_n$ are unit pure quaternions (for which $|{\q e}_m\vec\cdot\,{\q e}_n|\le 1$).
\end{enumerate}

%*****************
%FIN DE INEQUALITY
%*****************

%*******************************
\section{Conclusion}\label{conc}
%*******************************

Without doubt the special theory of relativity constitutes the main application of the Minkow\-ski space. Moreover the fundamental role of the special theory of relativity in physics in defining   the framework of space and time leads physicists to speak of Minkowski  spacetime.  But Minkowski  space also plays a role in another branch of physics, explicitly, in polarization optics, involving physical magnitudes not directly related to time nor to space. There is then an interest in studying the Minkowski  space structure from a geometrical point of view, without referring to time and space.  The present article is an attempt in such a way.

Quaternion algebra appears like an appropriate algebraic structure in representing both  Minkowski  space and proper Lorentz rotations, regardless of the intended application. That is the point of view developed in the present article. Thus both polarization optics and special theory of relativity should appear like concrete and physical implementations of an abstract geometrical structure, that of the Minkowski  space.

\appendix

%APPENDIX A**************************************
\section{Quaternionic trigonometry}\label{appenA}
%************************************************

\noindent{\bf Usual trigonometry.}
Before we give some  trigonometric formulas that include quaternions, we recall  usual relations between elliptical sine and cosine functions and their hyperbolic counterparts.

If $x$ is a real number, we start with
\begin{equation}
  \exp x =\cosh x+\sinh x\,,\label{eq261a}\end{equation}
where \begin{equation}
  \cosh x={\exp x+\exp (-x)\over 2}\,,\hskip.5cm \mbox{and}\hskip .5cm
  \sinh x={\exp x-\exp(-x)\over 2}\,.\label{eq262a}\end{equation}
We also have
\begin{equation}
  \exp\I x=\cos x+\I\sin x\,,\label{eq263a}\end{equation}
where
\begin{equation}
  \cos x={\exp \I x+\exp (-\I x)\over 2}\,,\hskip.5cm \mbox{and}\hskip .5cm
  \sin x={\exp \I x-\exp(-\I x)\over 2\I}\,.\label{eq264a}\end{equation}
From Eqs.\ (\ref{eq262a}) and (\ref{eq264a}) we deduce
\begin{equation}
  \cosh \I x =\cos x\,,\hskip 1cm \sinh\I x=\I \sin x\,,\label{eq265a}\end{equation}
and
\begin{equation}
  \cos\I x= \cosh x\,,\hskip 1cm \sin \I x=\I\sinh x\,.\label{eq266a}\end{equation}
We also have
\begin{equation}
  1=\cos^2x+\sin^2x=\cosh^2\I x-\sinh^2\I x\,,\end{equation}
and
\begin{equation}
  1=\cosh^2x-\sinh^2x=\cos^2\I x+\sin^2\I x\,.\end{equation}

\bigskip \noindent {\bf Hyperbolic functions with quaternion arguments.}
 If $q$ is a quaternion, the general formula for the exponential is
\begin{equation}
  \exp q=1+q+{q^2\over 2!}+{q^3\over 3!}+ {\dots} +{q^j\over j!}+{\dots}
  =\sum_{j=0}^{+\infty}{q^j\over j!}\,. \label{eq267a}
  \end{equation}
We define hyperbolic functions by
\begin{equation}
  \cosh q={\exp q+\exp(-q)\over 2}=\sum_{j=0}^{+\infty}{q^{2j}\over (2j)!}\,,\label{eq269a}\end{equation}
\begin{equation}\sinh q ={\exp q-\exp(-q)\over 2}=\sum_{j=0}^{+\infty}{q^{2j+1}\over (2j+1)!}\,,\label{eq270a}\end{equation}
so that
\begin{equation}
 \exp q=\cosh q+\sinh q\,.\end{equation}

 Quaternions $\exp q$, $\cosh q$ and $\sinh q$ commute, because they involve only powers of $q$ in their series expansions. From $\exp q\exp (-q)=1$, we deduce
\begin{equation}
  \cosh^2q-\sinh^2q =(\cosh q+\sinh q)(\cosh q-\sinh q)=\exp q\exp (-q)=1\,.\end{equation}

\bigskip \noindent {\bf Example.}
Let $q=1+\I \,{\q e}_1$. (We note that $N(q)=0$.) Then
  \begin{eqnarray}
    q^2\rap &=&\rap(1+\I\,{\q e}_1)(1+\I\,{\q e}_1)=2(1+\I\,{\q e_1})=2q\,, \nonumber \\
    q^3\rap &=&\rap qq^2=4q \,,\nonumber \\
    q^4\rap&=&\rap  qq^3=8q \,,\nonumber \\
    \hskip -.1cm\dots\rap&&\rap\dots  \nonumber  \\
    q^j \rap &=&\rap qq^{j-1}=2^{j-1}q\,.
  \end{eqnarray}
  We obtain (with $\E =\exp 1$)
  \begin{equation}
    \exp q=1+ q\sum_{j=1}^{+\infty}{2^{j-1}\over j!}=1+{q\over 2}\sum_{j=1}^{+\infty}{2^j\over j!}
    =1+{q\over 2}(\E^2-1)\,.\label{eqA309}
  \end{equation}
  We also have
  \begin{eqnarray}
    (-q)^2\rap&=&\rap (1+\I\,{\q e}_1)(1+\I\,{\q e}_1)=2(1+\I\,{\q e_1})=2q\,, \nonumber \\
    (-q)^3\rap &=&\rap  -qq^2=-4q \,,\nonumber \\
    (-q)^4\rap &=&\rap  (-q)(-q)^3=8q \,,\nonumber \\
    \dots&&\dots  \nonumber  \\
    (-q)^j \rap&=&\rap (-q)(-q)^{j-1}=(-1)^j2^{j-1}q\,,
  \end{eqnarray}
  so that
   \begin{equation}
    \exp (- q)=1+ q\sum_{j=1}^{+\infty}(-1)^j{2^{j-1}\over j!}=1+{q\over 2}\sum_{j=1}^{+\infty}(-1)^j{2^j\over j!}
    =1+{q\over 2}(\E^{-2}-1)\,.\label{eqA311}
   \end{equation}
  Since $q^2=2q$, we check
   \begin{equation}
     \exp q\exp (-q)= \left(1+{q\over 2}\E^2-{q\over 2}\right)\left(1+{q\over 2}\E^{-2}-{q\over 2}\right)=1\,.\end{equation} 

   From Eqs.\ (\ref{eqA309}) and (\ref{eqA311}),  we deduce
   \begin{equation} \cosh q=1+{q\over 2}(\cosh 2 -1)\,, \hskip .5cm \mbox{and}\;\;\;
     \sinh q={q\over 2}\sinh 2\,,\end{equation}
   or more explicitly
    \begin{equation} \cosh (1+\I\,{\q e}_1)=1+{1+\I\,{\q e}_1\over 2}(\cosh 2 -1)\,, \hskip .5cm \mbox{and}\;\;\;
      \sinh  (1+\I\,{\q e}_1)={ 1+\I\,{\q e}_1\over 2}\sinh 2\,.\end{equation}
    We remark that $\cosh (1+\I\,{\q e}_1)$ and $\sinh (1+\I\,{\q e}_1)$ may also be written as follows
    \begin{eqnarray}
      &&\hskip -.7cm\cosh (1+\I\,{\q e}_1)=1+(1+\I\,{\q e}_1)\sinh^21=\cosh^21+\I\,{\q e}_1\sinh^21\,,\label{eqA315}\\
      &&\hskip -.7cm\sinh (1+\I\,{\q e}_1)=(1+\I\,{\q e}_1)\cosh 1\sinh 1\,.\label{eqA316}
      \end{eqnarray}
We use $q^2=2q$ and  eventually check
   \begin{equation} \cosh^2q -\sinh^2q =1+ {q^2\over 4}(1+\cosh^2 2-2\cosh 2)+ q(\cosh2 -1)-{q^2\over 4}\sinh^2 2=1\,.\end{equation}

\bigskip  \noindent {\bf Pure-quaternion trigonometry \cite{Gir}.}
We assume that $q={\q e}_q\varphi$, where ${\q e}_q$ is a real pure unit quaternion and $\varphi$ a real number. Since ${{\q e}_q}^{\! 2}=-1$, we have
\begin{equation}
  \cosh {\q e}_q\varphi =\sum_{j=0}^{+\infty}{({\q e}_q\varphi )^{\! 2j}\over (2j)!}=\sum_{j=0}^{+\infty}(-1)^j{\varphi ^{2j}\over (2j)!}=\cos\varphi\,,\label{eq271a}\end{equation}
and
\begin{equation}
  \sinh {\q e}_q\varphi =
  \sum_{j=0}^{+\infty}{({\q e}_q\varphi )^{\! 2j+1}\over (2j+1)!}={\q e}_q\sum_{j=0}^{+\infty}(-1)^j{\varphi ^{2j+1}\over (2j+1)!}={\q e}_q\sin\varphi\,,\label{eq272a}\end{equation}
so that
  \begin{equation}
 \exp{\q e}_q\varphi = 
     \cosh {\q e}_q\varphi +\sinh{\q e}_q\varphi =\cos\varphi +{\q e}_q\sin\varphi\,.\label{eq273a}
  \end{equation}
  If we replace $\varphi$ with ${\q e}_q\varphi$ in Eqs.\ (\ref{eq271a}) and (\ref{eq272a}) we obtain
  \begin{eqnarray}
    &&\hskip -.6cm \cos {\q e}_q\varphi =\cosh (-\varphi)=\cosh\varphi\,,\label{eq274a}\\
  &&\hskip -.6cm\sin {\q e}_q\varphi =-\,{\q e}_q\sinh (-\varphi )={\q e}_q\sinh\varphi\,.\label{eq275a}
\end{eqnarray}
Equations (\ref{eq271a}), (\ref{eq272a}), (\ref{eq274a}) and (\ref{eq275a}) are  similar to Eqs. (\ref{eq265a}) and (\ref{eq266a}) where $\I$ has been changed  into ${\q e}_q$.

If we replace $\varphi$ with $\I\varphi$ in Eq.\ (\ref{eq273a}), we obtain
\begin{equation}
  \exp \I\,{\q e}_q\varphi = \cosh \I\,{\q e}_q\varphi +\sinh\I\,{\q e}_q\varphi =\cos {\q e}_q\varphi +\I\sin{\q e}_q\varphi =\cosh\varphi+ \I\,{\q e}_q\sinh\varphi\,,
\end{equation}
that is
\begin{equation}
  \cosh \I\,{\q e}_q\varphi =\cosh \varphi\,,\hskip 1.6cm \sinh\I\,{\q e}_q\varphi =\I\,{\q e}_q\sinh\varphi\,,
  \label{eqA324}
\end{equation}
and then
 \begin{equation}
   \cosh^2\I\,{\q e}_q\varphi-\sinh^2\I\,{\q e}_q\varphi =\cosh^2\varphi -\sinh^2\varphi =1\,.
    \end{equation}

If we replace $\varphi$ with $\I\varphi$ in Eqs.\ (\ref{eq274a}) and (\ref{eq275a}), we obtain
  \begin{equation}
  \cos \I\,{\q e}_q\varphi =\cosh \I\varphi =\cos\varphi\,,\hskip .5cm \sin\I\,{\q e}_q\varphi ={\q e}_q\sinh\I\varphi =\I{\q e}_q\sin\varphi\,,\label{eqA326}
  \end{equation}
  and
  \begin{equation}
   \cos^2\I\,{\q e}_q\varphi+\sin^2\I\,{\q e}_q\varphi =\cos^2\varphi +\sin^2\varphi =1\,.
  \end{equation}

  Eventually we point out that similar results are obtained for $\exp {\q e}_u\psi$, where $\psi$ is a complex number and ${\q e}_u$ a complex unit pure quaternion ($N({\q e}_u)=1$ and ${{\q e}_u}^{\!\! 2}=-1$).  The reason is that Eqs.\ (\ref{eq271a}) and (\ref{eq272a}) hold if we replace ${\q e}_q$ (a real unit pure quaternion) with ${\q e}_u$, and $\varphi$ (a real number) with $\psi$ (a complex number). Then
  \begin{equation}
  \exp {\q e}_u\psi =\cos\psi +{\q e}_u\sin\psi\,,\label{eq387}
  \end{equation}
  and
  \begin{equation}
  N(\exp {\q e}_u\psi )=(\cos\psi +{\q e}_u\sin\psi)(\cos\psi -{\q e}_u\sin\psi)=1\,.\label{eq388}
  \end{equation}
  (Equation (\ref{eq93n}) provides an  example of a complex unit pure quaternion).

  \bigskip \noindent {\bf Example.}
 We apply Eq.\ (\ref{eqA324}) and obtain
  \begin{eqnarray}
    &&\hskip -.6cm\cosh (1+\I\,{\q e}_1)=\cosh 1\cosh \I\,{\q e}_1 +\sinh 1\sinh \I\,{\q e}_1
    =\cosh^21 +\I\,{\q e}_1\sinh^2 1\,,\\
 &&\hskip -.6cm\sinh (1+\I\,{\q e}_1)=\sinh 1\cosh \I\,{\q e}_1 +\cosh 1\sinh \I\,{\q e}_1
    =(1+\I\,{\q e}_1)\cosh 1\sinh1 \,,
    \end{eqnarray}
which are Eqs.\ (\ref{eqA315}) and (\ref{eqA316}) once more.

  \bigskip
  \noindent {\bf  Zero-square quaternions.}
Equation (\ref{eq267a}) holds for every quaternion $q$. Let us assume now that $q$ is such that $q^2=0$. Then
  $\exp q=1+q$, by Eq.\ (\ref{eq267a}). We apply Eqs.\ (\ref{eq269a}) and (\ref{eq270a}) and obtain
  \begin{equation}
    \cosh q=1\,,\hskip 1cm \sinh q=q\,,\end{equation}
  so that
  \begin{equation}
    \exp q=1+q=\cosh q+\sinh q\,.\end{equation}
Since $q^2=0$, we obtain $\cosh^2q-\sinh^2q=1$.
  
  For example, let $q={\q e}_1+\I\,{\q e}_2$. Then $q^2=({\q e}_1+\I\,{\q e}_2)({\q e}_1+\I\,{\q e}_2)=0$, and
  \begin{equation}
    \cosh ({\q e}_1+\I\,{\q e}_2)=1\,,\hskip 1cm
    \sinh ({\q e}_1+\I\,{\q e}_2)={\q e}_1+\I\,{\q e}_2\,,\end{equation}
  so that
  \begin{equation}
    \exp ({\q e}_1+\I\,{\q e}_2)=1+{\q e}_1+\I\,{\q e}_2\,.
    \end{equation}

  More generally, let ${\q e}_u={\q e}_m+\I\,{\q e}_n$, where ${\q e}_m$ and ${\q e}_n$ are real unit pure quaternion, with ${\q e}_m\vec\cdot\, {\q e}_n=0$, so that $N({\q e}_u)=-{{\q e}_u}^{\! 2}=0$. Let $q={\q e}_u\psi$, where $\psi$ is a complex number. Then $q^2={{\q e}_u}^{\! 2}\psi^2=0$, as above,
  and we obtain
  \begin{equation}
    \exp {\q e}_u\psi =\exp q=1+q=\cosh q+\sinh q=\cosh {\q e}_u\psi +\sinh{\q e}_u\psi=
    1 + {\q e}_u\psi\,.
  \end{equation}
  We remark that   $\exp {\q e}_u\psi$ cannot be written as in Eq.\ (\ref{eq387}), if $N({\q e}_u)=0$.

%APPENDIX  B**********************************************************************
\section{Proof of $V={\q e}_u\,\sin (\psi /2)$ (Sect. \ref{sect258})}\label{appenB}
%**********************************************************************************

Notations are those of Section \ref{sect258}, related to the  composition of two elliptic rotations.
We assume ${\q e}_n\ne \pm {\q e}_m$.  We recall that $\theta$, $\varphi$ belong to the open interval $]-\pi, \pi [$.

    Let us assume $\psi =0\mod 2\pi$. Then $u_0=\pm 1$ and since ${u_0}^2+\|V\|^2=N(u)=1$, we conclude that $\| V\|=0$ and then $V=0$. Since ${\q e}_n\ne \pm \, {\q e}_m$, vectors ${\qv e}_m$, ${\qv e}_n$ and ${\qv e}_n\vec\times {\qv e}_m$ are linearly independant, and if $V=0$, then  necessarily $\varphi =0=\theta$, according to Eq.\ (\ref{eq64a}).  Since $\varphi\ne 0\ne \theta$, we conclude that $\cos (\psi/2)\ne \pm 1$ and then $\sin (\psi /2)\ne 0$.

The angle $\alpha$ between ${\q e}_m$ and ${\q e}_n$ is such that $\cos\alpha ={\q e}_m\vec\cdot {\q e}_n$ ($\alpha\ne 0\mod \pi$, because ${\q e}_m\ne \pm \,{\q e}_n$).
Let ${\q e}_p$ be the unit pure quaternion defined by ${\q e}_p\sin\alpha={\q e}_m\times {\q e}_n$, and let ${\q e}_q$ be the unit pure quaternion defined by ${\q e}_q={\q e}_m\times {\q e}_p$. The unit quaternion ${\q e}_n$ is written
${\q e}_n={\q e}_m\cos\alpha -{\q e}_q\sin\alpha$.
According to Eq.\ (\ref{eq64a}), we have
\begin{eqnarray}
V \rap &=&\rap
{\q e}_n\cos{\theta\over 2}\sin{\varphi\over 2}+{\q e}_m\cos{\varphi\over 2}\sin{\theta\over 2}
+{\q e}_n\times {\q e}_m\sin{\theta\over 2}\sin{\varphi\over 2} \\
\rap & =&\rap
{\q e}_m\left(\cos{\varphi\over 2}\sin{\theta\over 2}
+\cos\alpha  \cos{\theta\over 2}\sin{\varphi\over 2}\right)
+{\q e}_p\sin\alpha \sin{\theta\over 2}\sin{\varphi\over 2}
-{\q e}_q\sin\alpha \cos{\theta\over 2}\sin{\varphi\over 2}\,.\nonumber
\end{eqnarray}

Since ${\qv e}_m$, ${\qv e}_p$ and ${\qv e}_q$ form an orthogonal direct basis of ${\mathbb E}_3$, we have
\begin{eqnarray}
\|V\|^2\rap &=&\rap \cos^2{\varphi\over 2}\sin^2{\theta\over 2}+ \cos^2\alpha  \cos^2{\theta\over 2}\sin^2{\varphi\over 2}
+2\cos\alpha  \cos{\varphi\over 2}\sin{\theta\over 2}\cos{\theta\over 2}\sin{\varphi\over 2}\nonumber \\
& & \hskip 1cm + \sin^2\alpha \sin^2{\theta\over 2}\sin^2{\varphi\over 2}+ \sin^2\alpha \cos^2{\theta\over 2}\sin^2{\varphi\over 2}\nonumber \\
\rap &=&\rap  \cos^2{\varphi\over 2}\sin^2{\theta\over 2}+\cos^2{\theta\over 2}\sin^2{\varphi\over 2}
+  \sin^2\alpha \sin^2{\theta\over 2}\sin^2{\varphi\over 2}
+2\cos\alpha  \cos{\varphi\over 2}\sin{\theta\over 2}\cos{\theta\over 2}\sin{\varphi\over 2}
\nonumber \\
\rap &=&\rap \cos^2{\varphi\over 2}-\cos^2{\varphi\over 2}\cos^2{\theta\over 2}+\sin^2{\varphi\over 2}-\sin^2{\varphi\over 2}\sin^2{\theta\over 2} \nonumber \\
& &\hskip 1cm  +  \sin^2\alpha \sin^2{\theta\over 2}\sin^2{\varphi\over 2}
+2\cos\alpha  \cos{\varphi\over 2}\sin{\theta\over 2}\cos{\theta\over 2}\sin{\varphi\over 2}
\nonumber \\
\rap&=&\rap  1-\cos^2{\varphi\over 2}\cos^2{\theta\over 2} -\cos^2\alpha \sin^2{\varphi\over 2}\sin^2{\theta\over 2}
+2\cos\alpha  \cos{\varphi\over 2}\sin{\theta\over 2}\cos{\theta\over 2}\sin{\varphi\over 2}
\nonumber \\
\rap &=&\rap  1- \left(\cos{\varphi\over 2}\cos{\theta\over 2}-{\q e}_m\vec\cdot{\q e}_n\sin{\varphi\over 2}\sin{\theta\over 2}\right)^2\nonumber \\
\rap &=&\rap  1-\cos^2{\psi\over 2}=\sin^2{\psi \over 2}\,.
\end{eqnarray}
Since $\sin (\psi /2)\ne 0$, we conclude that $V/\sin(\psi /2)$ is a unit pure quaternion: it can be written ${\q e}_u$. Eventually we obtain
\begin{equation}V={\q e}_u\sin {\psi \over 2}\,.\label{eq312}\end{equation}

We remark that Eq.\ (\ref{eq312}) leads to $V=0$, if $\psi =0$, so that if both $\theta$ and $\varphi$ are $0$, Eq.\ (\ref{eq312}) still holds true.

%APPENDIX C************************************************************************************
\section{Mappings $q\longmapsto u\,q\,u^*$ are rotations on ${\mathbb H}_{\rm c}$}\label{appenC}
%**********************************************************************************************

Let $R$ be the mapping $R : {\mathbb H}_{\rm c}\longrightarrow {\mathbb H}_{\rm c}$ defined by $q\longmapsto R(q)=u\,q\,u^*$, where $u$ is a unit quaternion. Since $N(u\,q\,u^*)=N(q)$, we conclude that $R$ is an isometry on ${\mathbb H}_{\rm c}$. To prove that $R$ is a rotation, we have to prove that $\det R=1$.

\medskip
\noindent {\its i-- Real elliptic rotations.} Let $u=\exp {\q e}_n\varphi$, where $\varphi$ is  a real number and ${\q e}_n$ a real unit  pure quaternion.
If ${\q e}_n={\q e}_1$, we obtain the matrix of $R$ by replacing $\varphi$ with $2\varphi$ in Eq.\ (\ref{eq30}). More generally the matrix of $R$ in the canonical basis $\{ {\q e}_0,{\q e}_1,{\q e}_2,{\q e}_3\}$ of  ${\mathbb H}_{\rm c}$ is given by Eq.\ (\ref{eq65}). 
Then 
${\rm det}\, R=1$, and $R$ is a rotation. The rotation angle is $2\varphi$.

\medskip
\noindent {\its ii-- Imaginary elliptic  rotations of axis ${\q e}_1$.} We assume first that $u$ takes the form $u=\exp {\I\,\q e}_1\theta$, where $\theta$ is a real number. 
We have $R({\q e}_0)={\q e}_0$, and
\begin{eqnarray}
  R({\q e}_1)\rap &=&\rap (\cosh \theta +\I \,{\q e}_1\sinh\theta )\,{\q e}_1(\cosh \theta -\I\, {\q e}_1\sinh\theta )
  \nonumber \\
  \rap &=&\rap (-\I \sinh\theta +{\q e}_1\cosh\theta )(\cosh \theta -\I\, {\q e}_1\sinh\theta )\nonumber \\
  \rap &=&\rap {\q e}_1(\cosh^2\theta-\sinh^2\theta )={\q e}_1\,.
\end{eqnarray}
We also have
\begin{eqnarray}
  R({\q e}_2)\rap &=&\rap (\cosh \theta +\I\, {\q e}_1\sinh\theta )\,{\q e}_2(\cosh \theta -\I \,{\q e}_1\sinh\theta )
  \nonumber \\
  \rap &=&\rap ({\q e}_2\cosh\theta +\I \,{\q e}_3\sinh\theta)(\cosh \theta -\I\, {\q e}_1\sinh\theta )\nonumber \\
  \rap &=&\rap {\q e}_2(\cosh^2\theta+\sinh^2\theta )+2\I\,{\q e}_3\cosh\theta\sinh\theta \nonumber \\
  \rap &=&\rap {\q e}_2\cosh2\theta +\I \,{\q e}_3\sinh 2\theta\,,
\end{eqnarray}
and
\begin{eqnarray}
  R({\q e}_3)\rap &=&\rap (\cosh \theta +\I \,{\q e}_1\sinh\theta )\,{\q e}_3(\cosh \theta -\I \,{\q e}_1\sinh\theta )
  \nonumber \\
  \rap &=&\rap ({\q e}_3\cosh\theta -\I\,{\q e}_2\sinh\theta ) (\cosh \theta -\I \,{\q e}_1\sinh\theta )\nonumber \\
  \rap &=&\rap -2\I\,{\q e}_2\cosh\theta\sinh\theta +{\q e}_3(\cosh^2\theta+\sinh^2\theta ) \nonumber \\
  \rap &=&\rap -\I \,{\q e}_2\sinh2\theta + {\q e}_3\cosh  2\theta\,.
\end{eqnarray}
The matrix of $R$ is then
\begin{equation}
  {\cal R}= \begin{pmatrix} 1 & 0 & 0 &0 \\
    0 & 1 & 0 & 0 \\
    0 & 0 & \cosh 2\theta & -\I\sinh 2\theta \\
    0 & 0 & \I\sinh 2\theta & \cosh2\theta \end{pmatrix}
  \,,\end{equation}
and $\det R =\det {\cal R} =1$: the mapping $R$ is a rotation on ${\mathbb H}_{\rm c}$. The rotation angle is $2\I\theta$.

\medskip
\noindent {\its iii-- Imaginary elliptic  rotations of arbitrary axis ${\q e}_n$.} We assume $u=\exp \I\,{\q e}_n\theta$ with $\theta$ a real number and ${\q e}_n$ a real unit pure quaternion (${\q e}_n\ne {\q e}_1$). Let ${\q e}_p$ be the unit pure quaternion such that
\begin{equation}
  {\q e}_1\vec\times {\q e}_n={\q e}_p\sin\alpha\,,\end{equation}
where $\alpha$ is the angle between ${\q e}_1$ and ${\q e}_n$ ($-\pi <\alpha <\pi$, $\alpha\ne 0$). Then ${\q e}_n$ can be deduced from ${\q e}_1$ under the rotation $R_\alpha$ of angle $\alpha$ around ${\q e}_p$, that is
\begin{equation}
  {\q e}_n=\E^{{\q e}_p\alpha /2}\,{\q e}_1\,\E^{-{\q e}_p\alpha /2}\,.\end{equation}
Let $w=\exp ({\q e}_p\alpha /2)$. Then (this is Lemma 1 of \ref{appenD})
\begin{eqnarray}
  w\,\exp (\I\,{\q e}_1\theta )\,w^*\rap&=&\rap w ({\q e}_0\cosh\theta +\I\,{\q e}_1\sinh\theta )\,w^*
  ={\q e}_0\cosh\theta +\I\,w\,{\q e}_1\,w^*\sinh\theta )\nonumber \\
  &=&\exp(\I \,w\,{\q e}_1\,w^*\theta)= \exp (\I\,{\q e}_n\theta )\,,
  \end{eqnarray}
that is
\begin{equation}
  \exp (\I\,{\q e}_n\theta )=\exp \left({\q e}_p{\alpha\over 2}\right) \exp (\I\,{\q e}_1\theta ) \exp \left(-{\q e}_p{\alpha\over 2}\right)\,,\end{equation}
  and
  \begin{equation}
  \exp (-\I\,{\q e}_n\theta )=\exp \left({\q e}_p{\alpha\over 2}\right) \exp (-\I\,{\q e}_1\theta ) \exp \left(-{\q e}_p{\alpha\over 2}\right)\,.
\end{equation}
Let $R_1$ be the rotation represented by $\exp {\q e}_1\theta$, and $R_{-\alpha}$ be the rotation of angle $-\alpha$ around ${\q e}_p$.
We derive
\begin{eqnarray}
  R(q)\rap &=&\rap u\,q\,u^*= \exp (\I\, {\q e}_n\theta)\,q \exp (-\I\,{\q e}_n\theta )\\
  &=&\rap \exp \left({{\q e}_p\alpha\over 2}\right)\,\exp (\I \,{\q e}_1\theta )\,\exp \left(-{{\q e}_p\alpha\over 2}\right)\, q\, \exp \left({{\q e}_p\alpha\over 2}\right)\,\exp (-\I \,{\q e}_1\theta)\exp \left(-{{\q e}_p\alpha\over 2}\right)\,,\nonumber \end{eqnarray}
so that
\begin{equation}
   R(q)=R_\alpha \circ R_1\circ R_{-\alpha} (q)\,,\end{equation}
and $R$ is a rotation as product of rotations on ${\mathbb H}_{\rm c}$.

\medskip
\noindent {\its iv-- Complex elliptic  rotations.} The unit quaternion $u$ takes the form
$u=\exp {\q e}_n\psi$,  with ${\q e}_n$ a real unit pure quaternion and $\psi =\varphi +\I\theta$. We write $\exp {\q e}_n\psi =\exp{\q e}_n\varphi \, \exp\I\,{\q e}_n\theta$ and deduce that $R$ is the (commutative)  product of a real rotation $R_{\rm r}$ and an imaginary rotation $R_{\I}$ so that $\det R=\det R_{\rm r} \det R_{\I}=1$.

\medskip
\noindent {\its v-- Complex-axis elliptic rotations.} The unit quaternion $u$ takes the form $u=\exp {\q e}_u\psi$, where ${\q e}_u$ is a complex unit pure quaternion and $\psi$ a complex number.

We consider the group SO$_+(1,3)$ of proper Lorentz rotations. Those rotations are re\-pre\-sented by quaternions as $u$ above. Since SO$_+(1,3)$ is generated by elliptic and hyperbolic rotations, the quaternion $u$ is a product of the form $u=vw$ (or $u=wv$) with $v=\exp {\q e}_m\varphi$ and $w=\exp {\q e}_n\theta$. But $v$ and $w$ also represent elliptic rotations on ${\mathbb H}_{\rm c}$: their products ($vw$ or $wv$)  then represent  rotations on ${\mathbb H}_{\rm c}$. These rotations operates, on ${\mathbb H}_{\rm c}$, according to $q'=w\,q\,w^*$ and $q''=v\,q'\,v^*$, so that
\begin{equation}
  q''=v\,q'\,v^*=v(w\,q\,w^*)v^*=(vw)\,q\,(vw)^*=u\,q\,u^*\,.
\end{equation}

%APPENDIX D************************************************
\section{Two lemmas on unit pure quaternions}\label{appenD}
%**********************************************************

\begin{figure}[b]%$$$$$$$$$$$$$$$$$$$$$$$$$$$$$$$$$$$$$$$$$$$$$$$$$$$$$$$$$$$$$$
  \begin{center}
    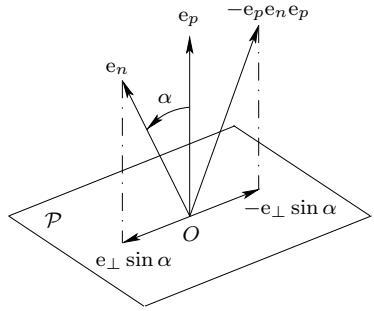
    \caption{\small Elements for the proof of Lemma 2. The plane ${\cal P}$ is orthogonal to ${\q e}_p$. The orthogonal projection of ${\q e}_n$ on ${\cal P}$ is ${\q e}_\perp \sin\alpha$, and $-{\q e}_\perp \sin\alpha$ is the orthogonal projection of $-{\q e}_p{\q e}_n{\q e_p}$.\label{figC}}
    \end{center}
  \end{figure}%$$$$$$$$$$$$$$$$$$$$$$$$$$$$$$$$$$$$$$$$$$$$$$$$$$$$$$$$$$$$$$$$$

\noindent{\bf Lemma 1.}  {\its Let ${\q e}_u$ be a complex unit  pure quaternion and $\psi$ a complex number. Then for every  real unit quaternion $w$ 
\begin{equation}
  w\,\exp({\q e}_u\psi )\,w^*=\exp (w\,{\q e}_u\,\psi \,w^*)\,.
\end{equation}
}
\medskip
 \noindent {\its Proof.} Since $w$ is a unit quaternion, we have $w\,w^*=1$, so that
 \begin{equation}
   w\,\exp({\q e}_u\psi )\,w^*= w\,({\q e}_0\cos\psi +{\q e}_u\sin\psi)\,\,w^*
   ={\q e}_0\cos\psi +w\,{\q e}_u\,w^*\sin\psi\,.
   \end{equation}
 To conclude, we remark that $(w\,{\q e}_u\,w^*)^*=(w^*)^*\,{{\q e}_u}^{\!\! *}\,w^*=-w\,{\q e}_u\,w^*$, which means that $w\,{\q e}_u\,w^*$ is a unit pure quaternion, so that
 \begin{equation}
   w\,\exp({\q e}_u\psi )\,w^*= \exp(w\,{\q e}_u\,w^*\psi )=\exp(w\,{\q e}_u\,\psi\,w^* )\,.
   \end{equation}
 The proof is complete. \qed
 \goodbreak

 \medskip
 Most often in this article Lemma 1 is applied for ${\q e}_u$ a real unit pure quaternion (generally denoted ${\q e}_n$). 

\bigskip
\noindent{\bf Lemma 2.} {\its Let ${\q e}_n$ and ${\q e}_p$ be two real unit  pure quaternions and let $\alpha$  be the angle between them. Then}
  \begin{equation}
    {\q e}_n-{\q e}_p{\q e}_n{\q e}_p=2\,({\q e}_n\vec\cdot {\q e}_p) \,{\q e}_p=2\,{\q e}_p\cos\alpha\,.
    \end{equation}

  \medskip
  \noindent {\its Proof.} We consider real pure quaternions as vectors of ${\mathbb E}_3$, with origin at $O$: we write ${\q e}_n$ in place of ${\qv e}_n$, and  ${\q e}_j$  in place of ${\qv e}_j$ ($j=1,2,3$).

Let ${\cal P}$ be the plane orthogonal to ${\q e}_p$ and passing through $O$ (see Fig.\ \ref{figC}). Then ${\q e}_n$ can be written ${\q e}_n={\q e}_p\cos\alpha+ {\q e}_\perp\sin\alpha$, where ${\q e}_\perp\sin\alpha$ is the (orthogonal) projection of ${\q e}_n$ on ${\cal P}$ and is perpendicular to ${\q e}_p$.  The  pure quaternion (or vector) ${\q e}_\perp$ becomes $-\,{\q e}_\perp$ under the rotation of angle $\pi$ around ${\q e}_p$ (Fig.\ \ref{figC}), so that
\begin{equation}
  -\,{\q e}_\perp=\exp\left({\q e}_p{\pi\over 2}\right)\,{\q e}_\perp\exp\left(-{\q e}_p{\pi\over 2}\right)
  =-\,{\q e}_p\,{\q e}_\perp{\q e}_p\,,\end{equation}
that is 
\begin{equation}
  {\q e}_p\,{\q e}_\perp{\q e}_p={\q e}_\perp\,.\label{eq406}\end{equation}
We deduce then
\begin{eqnarray}
  {\q e}_n-{\q e}_p{\q e}_n{\q e}_p \!\!\!\! &=&\!\!\!\! {\q e}_p\cos\alpha + {\q e}_\perp\sin\alpha - {\q e}_p ({\q e}_p\cos\alpha + {\q e}_\perp\sin\alpha ){\q e}_p\nonumber \\
  \!\!\!\! &=&\!\!\!\!  2 {\q e}_p\cos\alpha + ({\q e}_\perp - {\q e}_p\,{\q e}_\perp {\q e}_p)\sin\alpha\nonumber \\
  \!\!\!\! &=&\!\!\!\!   2 {\q e}_p\cos\alpha \,.\label{eq407}
\end{eqnarray}
The proof is complete. \qed

\bigskip
\goodbreak
The following corollary is a straightforward consequence of Lemma 2.

\bigskip
\noindent{\bf Corollary.} {\its If ${\q e}_n$ and ${\q e}_p$ are orthognal real unit pure  quaternions, then
  ${\q e}_p\,{\q e}_n\,{\q e}_p={\q e}_n$. Moreover, if $V=\alpha {\q e}_n$ ($\alpha$ a real number), then  ${\q e}_p\,V\,{\q e}_p=V$.}

\bigskip
\noindent{\bf Remark D.1 [Another proof of Lemma 2]}
We begin with a direct proof of the previous corollary. If ${\q e}_n$ and ${\q e}_p$ are ortho\-go\-nal  we have ${\q e}_n\,{\q e}_p=-{\q e}_p\,{\q e}_n$ and then
${\q e}_p\,{\q e}_n\,{\q e}_p=-({\q e}_p)^2\,{\q e}_n={\q e}_n$, that is ${\q e}_n-{\q e}_p\,{\q e}_n\,{\q e}_p=0$. The proof of the corollary is complete.

To prove Lemma 2, we first consider than ${\q e}_n$ and ${\q e}_p$ are orthogonal. Then ${\q e}_n-{\q e}_p\,{\q e}_n\,{\q e}_p=0$, as we have just proved, and Lemma 2 is proved for that case.

If ${\q e}_n$ is not  orthogonal to ${\q e}_p$, let ${\q e}_\perp$ be such that
${\q e}_\perp\sin\alpha={\q e}_n-{\q e}_p\cos\alpha$. We have ${\q e}_p\vec\cdot \,{\q e}_\perp=0$, and ${\q e}_\perp$ is orthognal to ${\q e}_p$; we may write
${\q e}_\perp-{\q e}_p\,{\q e}_\perp\,{\q e}_p=0$, which is Eq.\ (\ref{eq406}). The end of the proof is given by Eq.\ (\ref{eq407}).

%APPENDIX E%***************************************************************
\section{Regular rotations: eigenquaternions and\\ eigenvalues}\label{appenE}
%**************************************************************************

\noindent{\bf Eigenquaternions and eigenvalues.}
We will only prove Eqs.\ (\ref{eq106})  and (\ref{eq110}). 
Let us denote $C=\cos (\psi /2)$ and $S=\sin (\psi /2)$, so that $u=C+S\,{\q e}_u$ and $\overline{u}^{\, *}=\overline{C}-\overline{S}\;\overline{{\q e}_u}$. Let $x'=(1+\I\,{\q e}_u)(1+\I\,\overline{{\q e}_u})$. We have
\begin{equation}
  \Bigl(\overline{1+\I\,{\q e}_u}\Bigr)^{\! *}=(1-\I\,\overline{{\q e}_u})^*=1+\I\,\overline{{\q e}_u}\,,
    \end{equation}
and
\begin{equation}
  \Bigl(\overline{1+\I\,\overline{{\q e}_u}}\Bigr)^{\! *}=(1-\I\,{\q e}_u)^*=1+\I\,{\q e}_u\,,
    \end{equation}
so that
\begin{equation}
  \overline{x'}^{\, *}=\left[\overline{(1+\I\,{\q e}_u)(1+\I\,\overline{{\q e}_u})}\right]^*
  =\Bigl(\overline{1+\I\,\overline{{\q e}_u}}\Bigr)^{\! *}
  \Bigl(\overline{1+\I\,{\q e}_u}\Bigr)^{\! *}=
  (1+\I\,{\q e}_u)(1+\I\,\overline{{\q e}_u})=x'\,,
  \end{equation}
and $x'$ is a minquat.

Then
\begin{eqnarray}
  u\,x'\,\overline{u}^{\, *}\rap &=&\rap (C+S\,{\q e}_u)(1+\I\,{\q e}_u)(1+\I\,\overline{{\q e}_u})(\overline{C}-\overline{S}\;\overline{{\q e}_u})\nonumber \\
  \rap &=&\rap \bigl[C-\I\, S+ (S+\I \,C)\,{\q e}_u)\bigr]\bigl[\,\overline{C}+\I\, \overline{S}- (\overline{S}-\I \,\overline{C}\,)\,\overline{{\q e}_u}\bigr]\nonumber \\
  \rap &=&\rap (C-\I\,S)(1+\I\,{\q e}_u)(\overline{C}+\I\,\overline{S})(1+\I\,\overline{{\q e}_u})\nonumber \\
  \rap &=&\rap (C-\I\,S)(\overline{C}+\I\,\overline{S})(1+\I\,{\q e}_u)(1+\I\,\overline{{\q e}_u})\nonumber \\
    \rap &=&\rap (C-\I\,S)(\overline{C}+\I\,\overline{S})\, x'\,,
\end{eqnarray}
and $x'$ is an eigenminquat of the rotation represented by $u$.  The correponding eigenvalue is
\begin{equation}
  \Lambda= (C-\I\,S)(\overline{C}+\I\,\overline{S})= C\,\overline{C}+ S\,\overline{S} +\I\,C\,\overline{S}-\I\,\overline{C}\,S\,.
\end{equation}

%APPENDIX F%*****************************************************
\section{Singular Lorentz rotations: eigenminquats}\label{appenF}
%****************************************************************

We  shall prove that ${\q e}_0+\I\,{\q e}_m{\q e}_n$ is an eigenminquat of the mapping  $x\longmapsto u_{_+}\,x\,\overline{u_{_+}\!\!}^{\, *}$.
 For the sake of simplicity we set $\phi =\varphi /2$ in Eq.\ (\ref{eq109s}). Since ${\q e}_m$ and ${\q e}_n$ are orthogonal, we have ${\q e}_m\,{\q e}_n\,{\q e}_m={\q e}_n$, and  ${\q e}_n\,{\q e}_m\,{\q e}_n={\q e}_m$, according to the corollary of \ref{appenD}. The following derivation shows that ${\q e}_0+\I\,{\q e}_m{\q e}_n$ is an eigenminquat of the proper Lorentz rotation  $x\longmapsto u_{_+}x\,\overline{u_{_+}\!\!}^{\, *}$\,:
\begin{eqnarray}
  u_{_+}\,({\q e}_0+\I\,{\q e}_m{\q e}_n)\, \overline{u_{_+}\!\!}^{\, *}
  \rap &=&\rap
  ({\q e}_0+{\q e}_m\phi+\I\,{\q e}_n\phi )
  ({\q e}_0+\I\,{\q e}_m{\q e}_n)
   ({\q e}_0-{\q e}_m\phi+\I\,{\q e}_n\phi )\nonumber
  \\
  \rap &=&\rap
  ({\q e}_0+ \I\,{\q e}_m{\q e}_n+{\q e}_m\phi -\I\,{\q e}_n\phi + \I\,{\q e}_n\phi -{\q e}_n\,{\q e}_m\,{\q e}_n \phi) ({\q e}_0-{\q e}_m\phi+\I\,{\q e}_n\phi )\nonumber
  \\
  \rap &=&\rap
  ({\q e}_0+ \I\,{\q e}_m{\q e}_n)({\q e}_0-{\q e}_m\phi+\I\,{\q e}_n\phi )\nonumber
  \\
  \rap &=&\rap
  {\q e}_0 -{\q e}_m\phi +\I\,{\q e}_n\phi +\I\,{\q e}_m{\q e}_n-\I\,{\q e}_m\,{\q e}_n\,{\q e}_m \phi +{\q e}_m\phi\nonumber \\
   \rap &=&\rap
  {\q e}_0+ \I\,{\q e}_m{\q e}_n\,.
\end{eqnarray}
The corresponding eigenvalue is $1$.

A similar derivation holds for $u_{_-}\!\! ={\q e}_0+({\q e}_m-\I\,{\q e}_n)\phi$ and  shows that ${\q e}_0-\I\,{\q e}_m{\q e}_n$ is an eigenmiquat for the mapping  $x\longmapsto  u_{_-} x\, \overline{u_{_-}\!\!}^{\, *}$ (just replace $\I$ with $-\I$ in the previous derivation).

That there is only one eigendirection for null minquats  may be understood as follows.
We consider the case ${\q e}_m={\q e}_1$ and ${\q e}_n={\q e}_2$, so that ${\q e}_m\,{\q e}_n={\q e}_3$.
Let $x={\q e}_0+\I\,{\q e}_p$ be a null minquat, with $x_0=1$ and
\begin{equation}
  {\q e}_p=x_1\,{\q e}_1+x_2\,{\q e}_2+x_3\,{\q e}_3\,,\hskip .5cm  (x_1)^2+(x_2)^2+(x_3)^2=1\,.\label{eq409}\end{equation}
The image of $x$
by the considered proper Lorentz rotation is
\begin{equation}
  x'=x'_0{\q e}_0+\I x'_1{\q e}_1+\I x'_2{\q e}_2+\I x'_3{\q e}_3= u_{_+} x\, \overline{u_{_+}\!\!}^{\, *}\,,
\end{equation}
that is
\begin{eqnarray}
 \!\!\left[\begin{matrix}x'_0\cr x'_1 \cr x'_2\cr x'_3
  \end{matrix}\right]
  \rap &= &\rap  \left[\begin{matrix}1\cr \phi \cr \I\phi \cr 0
  \end{matrix}\right] \left[\begin{matrix}1\cr x_1 \cr x_2\cr x_3
  \end{matrix}\right] \left[\begin{matrix}1\cr -\phi \cr \I\phi \cr 0
    \end{matrix}\right]=
  \left[\begin{matrix}1-\I\phi x_1+\phi x_2 \\ \I x_1+\phi -\phi x_3\\ \I x_2+\I\phi -\I\phi x_3 \\ \I x_3+\I\phi x_2+\phi x_1
    \end{matrix}\right]
  \left[\begin{matrix}1\cr -\phi \cr \I\phi \cr 0
    \end{matrix}\right]=
  \left[\begin{matrix}1+ 2\phi x_2 +2\phi^2(1-x_3)\\ x_1\\ (1+2\phi^2)x_2 \\ x_3+2\phi x_2+2\phi^2 (1-x_3)
    \end{matrix}\right]\,.\nonumber \\
  & &\label{eq446}
  \end{eqnarray}
For $1+\I\,{\q e}_p$ to be an eigenvector, we should have
\begin{equation}
  x'_0={x'_1\over x_1}={x'_2\over x_2}={x'_3\over x_3}\,,\label{eq447}\end{equation}
with $x'_j=0$ if $x_j=0$.

The reasoning as follows.
\begin{itemize}
\item Let us assume $x_1x_2\ne 0$. Then $x'_1/x_1=1$ does not depend on $\phi$, so that $x'_2/x_2$ should not depend on $\phi$, which would be possible only if $x_2=0$.

  We conclude that necessarily $x_1x_2=0$.

  \begin{itemize}
\item
Let us assume $x_1\ne 0$ and $x_2=0$. Then $x'_1/x_1=1$ and $x'_3/x_3$ should not depend on $\phi$\,: necessarily $x_3=1$, which leads to $(x_1)^2+(x_3)^2=(x_1)^2+1>1$, a contradiction with Eq.\  (\ref{eq409}).

\item
  Let us assume $x_1=0$ and $x_2\ne 0$.
  \begin{itemize}
    \item If $x_3\ne 1$, Eq.\ (\ref{eq447}) gives
\begin{equation}
  x'_0={x'_3\over x_3}={x'_0-x'_3\over 1-x_3}={1-x_3\over 1 - x_3}=1\,,\label{eq448}\end{equation}
and necessarily $x'_2/x_2$ should not depend on $\phi$, that is, $x_2$ should be zero, a contradiction with the assumption.
\item If $x_3=1$, then $(x_2)^2+(x_3)^2=(x_2)^2+1>1$, a contradiction with Eq.\ (\ref{eq409}). 
  
\item We conclude that $x_1=0$ and $x_2\ne 0$ is not admissible.
  \end{itemize}
\end{itemize}

\item Eventually, necessarily $x_1=x_2=0$, and $x_3=1$. There is only one null-minquat eigendirection.
\end{itemize}

The reasoning holds  in the more general orthonormal basis $\{ {\q e}_m,{\q e}_n, {\q e}_m\,{\q e}_n\}$ in place of $\{ {\q e}_1,{\q e}_2, {\q e}_3\}$.

\bigskip
\noindent{\bf Remark F.1}  To establish the second equality in Eq.\ (\ref{eq448}) we use the following identity (with $abcd\ne 0$)
\begin{equation}
  {a\over b}={c\over d}={a-c\over b-d}\,.\end{equation}
To establish the third equality we report the values of $x'_0$ and $x'_3$ given by Eq.\ (\ref{eq446}).

%APPENDIX G**********************************************************
% POUVOIR ROTATOIRE SUIVI DU BIREFRINGENT****************************
\section{Equivalent birefringence and optical activity}\label{appenG}
%********************************************************************

The superposition of the rectilinear birefringent represented by $\exp ({\q e}_1\varphi /2)$ and the circular birefringent (optical activity) represented by $\exp ({\q e}_3\theta /2)$ is an elliptical birefringent represented by $u=\exp [({\q e}_1\varphi + {\q e}_3\theta /2)]$ (we assume $\varphi\theta\ne 0$). The birefringence is $\psi =(\varphi^2+\theta^2)^{1/2}\ne 0$.

The elliptical birefringent may be seen as the succession (in the order) of an equivalent optical activity, represented by $\exp ({\q e}_3\theta '/2)$, and an equivalent rectilinear birefringent, represented by  $\exp ({\q e}_m\varphi ' /2)$. We write then
\begin{equation}
  \exp {{\q e}_m\varphi '\over 2}\exp {{\q e}_3\theta '\over 2}=\exp{ {\q e}_n \psi \over 2}
  =\exp{ {\q e}_1 \varphi +{\q e}_3\theta \over 2}\,.\label{eq438}
\end{equation}
The unit quaternion ${\q e}_m$ takes the form
\begin{equation}{\q e}_m={\q e}_1\cos\alpha ' +{\q e}_2\sin\alpha '\,,\end{equation}
because the equivalent birefringent is rectilinear.

The equivalent parameters are $\varphi '$, $\theta '$ and $\alpha '$ (which provides ${\q e}_m$). To obtain the link between them and the intrinsic parameters ($\varphi$ and $\theta$), we proceed as follows.

We first derive
\begin{eqnarray}
  \exp {{\q e}_m\varphi '\over 2}\exp {{\q e}_3\theta '\over 2}\rap &=&\rap
   \left(\cos{\varphi '\over 2}+{\q e}_1\cos\alpha '\sin{\varphi '\over 2}+{\q e}_2\sin\alpha '\sin{\varphi '\over 2}\right) \left(\cos{\theta '\over 2}+{\q e}_3\sin{\theta '\over 2}\right) \nonumber \\
  \rap &=&\rap
  \left[\begin{matrix}\cos (\varphi '/2) \cr \cos\alpha '\sin (\varphi '/ 2) \cr  \sin\alpha '\sin (\varphi '/2) \cr 0\end{matrix} \right]
  \left[\begin{matrix}\cos (\theta '/2) \cr 0 \cr 0\cr \sin (\theta '/2)\end{matrix} \right]\nonumber \\
  \rap &=&\rap
  \left[\begin{matrix} \cos (\varphi '/2)\cos (\theta '/2)\cr
      \cos\alpha '\sin (\varphi '/ 2) \cos (\theta '/2)+
     \sin\alpha '\sin (\varphi '/ 2) \sin (\theta '/2)\cr
     \sin\alpha '\sin (\varphi '/ 2) \cos (\theta '/2)
      - \cos\alpha '\sin (\varphi '/ 2)\sin (\theta '/2) \cr
     \cos (\varphi '/2) \sin (\theta '/2) \cr
      \end{matrix} \right]\,, \nonumber
  \end{eqnarray}
so that Eq.\ (\ref{eq438}) is equivalent to the system
    \begin{subequations}
  \begin{numcases}{}
   \cos {\varphi '\over 2} \cos \displaystyle{\theta '\over 2}=\cos {\psi \over 2}\,,\label{eq439a}\\
      \sin \displaystyle{\varphi '\over 2}\cos \left(\alpha '-\displaystyle{\theta '\over 2}\right)={\cos\chi}\sin {\psi\over 2}\,, \label{eq439b}\\
      \sin \displaystyle{\varphi '\over  2}\sin \left(\alpha '-\displaystyle{\theta '\over 2}\right)=0\,,\label{eq439c}\\
       \cos {\varphi '\over 2}\sin \displaystyle{\theta '\over 2}= {\sin\chi}\sin {\psi\over 2}\,.\label{eq439d}
    \end{numcases}
    \end{subequations}
    According to Eq.\ (\ref{eq439d}), we remark that $\theta '\ne 0$ (because $\chi\psi \ne 0$). Equations (\ref{eq439a}) and (\ref{eq439d}) give $\cos^2(\varphi '/2) =\cos^2 (\psi /2)+\sin^2\chi\sin^2(\psi /2)<1$, so that $\varphi '\ne 0$. 
Equation (\ref{eq439c}) gives $\alpha '-(\theta '/2)=0\mod \pi$, so that $\cos[\alpha '-(\theta '/2)]=\pm 1$. If we choose  $\alpha '-(\theta '/2)=0$,  we obtain
    \begin{subequations}
  \begin{numcases}{}
    \cos {\varphi '\over 2}\cos \displaystyle{\theta '\over 2}=\cos {\psi \over 2}\,,\label{eq440a}\\
    \sin \displaystyle{\varphi '\over 2}={\cos\chi}\sin {\psi\over 2}\,,\label{eq440b} \\
    \alpha '-\displaystyle{\theta'\over 2}=0\,,\label{eq440c}\\ 
    \cos {\varphi '\over 2}\sin \displaystyle{\theta '\over 2}= {\sin\chi}\sin {\psi\over 2}\,,\label{eq440d}
    \end{numcases}\label{s440}
    \end{subequations}
and if we choose  $\alpha '-(\theta '/2)=\pi$, the system becomes
       \begin{subequations}
  \begin{numcases}{}
    \cos {\varphi '\over 2}\cos \displaystyle{\theta '\over 2}=\cos {\psi \over 2}\,,\label{eq441a}\\
    \sin \displaystyle{\varphi '\over 2}=-{\cos\chi}\sin {\psi\over 2}\,,\label{eq441b} \\
    \alpha '-\displaystyle{\theta'\over 2}=\pi\,,\label{eq441c}\\ 
    \cos {\varphi '\over 2}\sin \displaystyle{\theta '\over 2}= {\sin\chi}\sin {\psi\over 2}\,.\label{eq441d}
    \end{numcases}\label{s441}
       \end{subequations}
In both cases we have $\tan (\theta '/2)=\sin\chi \tan (\psi /2)$, so that $\theta '$ is defined modulo $2\pi$.

We remark that the sign of $\varphi '$ is given by Eqs. (\ref{eq440b}) and (\ref{eq441b}), so that if $\varphi '$ is solution of the system (\ref{s440}), then $-\varphi '$ is solution of (\ref{s441}). And if $\alpha '$ is solution of (\ref{s440}) then $\alpha '-\pi$ is solution of (\ref{s441}). Then if ${\q e}_m$ is the axis for the solution of (\ref{s440}), the unit quaternion $-\,{\q e}_m$ is the axis for (\ref{s441}). The equivalent birefringent is represented by $\exp ({\q e}_m\varphi '/2)$ and is the same for both systems. There is a unique solution for both (\ref{s440}) and (\ref{s441}).

Eventually, the equivalent parameters are as follows
\begin{subequations}
  \begin{numcases}{}
    \sin{\varphi '\over 2}=\cos\chi\sin{\psi\over 2}={\varphi \over\sqrt{\varphi^2+\theta^2}}\sin{\sqrt{\varphi^2+\theta^2}\over 2}\,, \\
    \tan{\theta '\over 2}=\sin\chi\tan{\psi\over 2}={\theta \over\sqrt{\varphi^2+\theta^2}}\tan{\sqrt{\varphi^2+\theta^2}\over 2}\,,
    \\
    \alpha ' ={\theta '\over2}\,.\label{eq442c}
     \end{numcases}\label{s442}
\end{subequations}

As explained in Section \ref{sect544} expressing the intrinsic parameters as functions of the equi\-va\-lent ones is interesting for their measurements. The elliptic birefringence $\psi$ is given by Eq.~(\ref{eq439a})
\begin{equation}
  \cos{\psi\over 2}=\cos{\theta '\over 2}\cos{\varphi '\over 2}\,,\end{equation}
and from Eqs. (\ref{eq439b}) and (\ref{eq439d}) we deduce
\begin{equation}
  \tan\chi =\sin{\theta '\over 2} \cot {\varphi '\over 2}\,.\end{equation}

\bigskip
\noindent{\bf Remark G.1} If the equivalent rectilinear birefringent is chosen to be crossed by light before the equi\-valent circular birefrigent (optical activity), that is, if $u$ is
\begin{equation}
  u=\exp {{\q e}_3\theta '\over 2}\exp {{\q e}_m\varphi '\over 2}\,,\end{equation}
the only change in the result is that Eq.\ (\ref{eq442c}) is replaced with $\alpha' =-\theta '/2$.  Only the direction of ${\q e}_m$ is changed.

%**************************************
%BIBLIOGRAPHY**************************


\begin{thebibliography}{99}\label{ref2}
%**************************************
%**************************************


\parskip -.05cm
{\small

\bibitem{Sil1} {\sc Silberstein (L.)}, Quaternionic form of relativity, {\its Phil. Mag.} {\bf 23} (1912) 790--809.

  \bibitem{Con} {\sc Conway (A. W.)}, The quaternionic form of relativity, {\its Phil. Mag.} {\bf 24} (1912) 208--208.

  \bibitem{Sil2} {\sc Silberstein (L.)}, Second memoir on quaternionic relativity, {\its Phil. Mag.} {\bf 25} (1913) 135--144.

  \bibitem{Dir} {\sc Dirac (P. A. M.)}, Application of quaternions to Lorentz transformations, {\its Proc. Irish Acad.} {\bf A 50} (1945) 261--270.

    \bibitem{Kwa1} {\sc Kwal (B.)}, La th\'eorie des \'equations de Maxwell et le calcul des op\'erateurs matriciels, {\its Journal de physique} {\bf 8}, (1934) 445--448.

    \bibitem{Kwa2} {\sc Kwal (B.)}, Sur la description spatio-temporelle des ph\'enom\`enes quantiques, {\its Le Journal de physique et le radium}, {\bf VII} Tome VIII (1937) 81--87.

    \bibitem{Cer} {\sc Cernosek (J.)}, Simple geometrical method for analysis of elliptical polarization, {\its J. Opt. Soc. Am.} {\bf 61} (1970) 324--327.
      
  \bibitem{The} {\sc Theocaris (P. S.), Gdoutos (E. E.)}, {\its Matrix theory of
        photoelasticity}, Springer Verlag, Berlin, 1979.

   \bibitem{Syn} {\sc Synge (J. L.)}, Quaternions, Lorentz transformations, and the Conway-Dirac-Eddington matrices, Dublin Institute for Advanced Studies, Serie {\bf A 21} (1972) 1--67.

  \bibitem{PPF0} {\sc Pellat-Finet (P.)}, {\its De la bir\'efringence elliptique}, Th\`ese de Docteur-ing\'enieur, Universit\'e d'Aix-Marseille {\sc iii} and Universit\'e de Toulon et du Var, 1983.

 \bibitem{PPF2} {\sc Pellat-Finet (P.)}, Repr\'esentation des \'etats et des op\'erateurs de
          polarisation de la lumi\`ere par des quaternions, {\its Optica Acta}, {\bf 31} (1984) 415--434.

 \bibitem{PPF5} {\sc Pellat-Finet (P.)}, An introduction to a  vectorial calculus for polarization optics, {\its Optik}, {\bf 84} (1990) 169--175.

   \bibitem{Bruy} {\sc  Bruy\`ere (F.)}, Effets de polarisation dans les syst\`emes \`a amplification optique
     de longue distance, Thesis, Universit\'e Paris-Sud, Orsay, 1994.

     \bibitem{PPF11} {\sc Pellat-Finet (P.)},  Iterative experimental method for generating eigenstates and principal states of polarization, {\its Appl. Opt.} {\bf 51} (2012) 4403--4408.

   \bibitem{Vig} {\sc Vign\'eras (M.-F)},  {\its Arithm\'etique des alg\`ebres de quaternions}, Springer, Berlin, 1980.

   \bibitem{Voi} {\sc Voight (J.)}, {\its Quaternion Algebras}, Springer, 2021.

  \bibitem{Alt} {\sc Altmann (S. L.)}, {\its Rotations, quaternions and double groups}, Clarendon Press, Oxford, 1986.


  \bibitem{Kam} {\sc Karney (C. F. F.)},  Quaternions in molecular modelling, {\its J. Molecular Graphics and Modelling}, {\bf 25} (2007) 595--604.

  \bibitem{Adl} {\sc Adler (S. L.)}, {\its Quaternionic quantum mechanics and quantum fields}, Oxford University Press, New York, 1995.

  \bibitem{Mah} {\sc Mahecha (J.)}, {\its Mec\'anica cl\'asica avanzada}, Editorial Universidad de Antioquia, Medell\'in, 2006.

  \bibitem{Gir} {\sc Girard (P. R.)}, {\its Quaternions, Clifford Algebras and Relativistic Physics}, Birkha\"user, Basel, 2007.

  \bibitem{Ozc} {\sc \"Ozcan (B.),  Kinli (F.),   Kira\c c (F.)},    Quaternion Capsule Networks, arxiv 2007.04389v1 (2020) 1--8.

    \bibitem{PPF3} {\sc Pellat-Finet (P.)}, {\its Repr\'esentation g\'eom\'etrique de la lumi\`ere polaris\'ee}, Th\`ese de doctorat d'\'Etat \`es Sciences physiques, Universit\'e de Toulon et du Var, 1987.

\bibitem{PPF4} {\sc Pellat-Finet (P.)}, Sur la g\'eom\'etrie de la lumi\`ere polaris\'ee,
  {\its C. R. Ac. Sc. Paris}, {\bf 308} {\sc II} (1989) 827--830.

\bibitem{PPF6} {\sc Pellat-Finet (P.)}, Geometrical approach to  polarization
  optics. {\sc I}-- Geometrical structure of polarized light,
  {\its Optik}, {\bf 87} (1991) 27--33.

\bibitem{PPF7} {\sc Pellat-Finet (P.)}, Geometrical approach to  polarization
  optics. {\sc II}-- Quaternionic representation  of polarized light,
  {\em Optik}, {\bf 87} (1991) 68--76.

  \bibitem{PPF10} {\sc Pellat-Finet (P.), Bausset (M.)}, What is common to both
    polarization optics and relativistic kinematics?, {\its Optik}, {\bf 90} (1992) 101--106.

  \bibitem{Deh} {\sc Deheuvels (R.)}, {\its Formes quadratiques et groupes classiques}, P.U.F., Paris, 1981.

  \bibitem{Gou} {\sc Gourgoulhon (\'E.)}, {\its Special Relativity in General Frames. From Particles to Astrophysics}, Springer, Berlin, 2013.  

\bibitem{Pen} {\sc Penrose (R.), Rindler (W.)}, {\its Spinors and space-time}, Cambridge University Press, Cambridge, 1984.

\bibitem{Land} {\sc Landau (L.), Lifchitz (E.)}, {\its Th\'eorie des champs} in {\its Physique th\'eorique}, Vol. 2, 4th edn, Mir, Moscow, 1989. (Translated from Russian. Published in English: {\its The classical theory of fields}, in {\its Course of Theoretical physics}, Vol. 2, Pergamon Press, Oxford.)

   \bibitem{Mol} {\sc M\o ller (C.)}, {\its The theory of relativity}, Oxford University Press, London, 1962.

   \bibitem{Rou} {\sc Roug\'e (A.)}, {\its Introduction \`a la relativit\'e}, \'Editions de l'\'Ecole polytechnique, Palaiseau, 2008.

   \bibitem{Mis} {\sc Misner (C. W.), Thorne (K. S.), Wheeler (J. A.)}, {\its Gravitation}, W. H. Freeman and Company, New York, 1973.

    \bibitem{Bor} {\sc Born (M.), Wolf (E.)}, {\its Principles of Optics}, 7th  edn.,
        Cambridge University Press, Cambridge, 1999.

\bibitem{Cla} {\sc Clarke (D.), Grainger (J.\ F.)}, {\its Polarized light and optical
  measurement}, Pergamon Press, Oxford, 1971.

 \bibitem{Gol} {\sc Goldstein (D.)}, {\its Polarized Light}, 2nd edn., CRC Press, Boca Raton, 2003.

\bibitem{Ram} {\sc Ramachandran (G.\ N.), Ramaseshan (S.)}, {\its Crystal Optics}, in
  Handbuch der Physik {\sc XXV}/1, S. Fl\"ugge editor, Springer Verlag,
  Berlin, 1961.
  
\bibitem{Shu} {\sc Shurcliff (W.)}, {\its Polarized light}, Harvard University Press,
  Cambridge, 1962.

\bibitem{Sim} {\sc Simmons (J. W.), Gutmann (M. J.)}, {\its States,  Waves and Photons: a
  modern introduction to light}, Addison Wesley, Reading, 1970.

 \bibitem{Bru} {\sc Bruhat (G.), Kastler (A.)}, {\its Optique}, 6th edn., Masson,
    Paris, 1992.

      \bibitem{Her} {\sc Herriau (J.-P.), Huignard (J.-P.)}, Some polarization properties of volume holograms in Bi$_{12}$SiO$_{20}$ crystals and applications, {\its Appl. Opt.} {\bf 17} (1978) 1851--1852.

      \bibitem{Hui} {\sc Huignard (J.-P.), G\"unter (P.) (Eds.)}, {\its Photorefractive materials and their applications}. {\its I: Fundamental phenomena}, and  {\its II: Surveys and applications}, Springer, Berlin, 1988, 1989.

      
\bibitem{PPF13} {\sc Pellat-Finet (P.)},  Measurement of the electro-optic coefficient of BSO crystals,
  {\its  Opt. Comm.} {\bf 50} (1984) 275-280.

  \bibitem{Pet} {\sc Petrov (M. P.), Stepanov (S. L.), Khomenko (A. V.)}, {\its Photorefractive crystals in coherent optical systems}, Springer, Berlin, 1991.
        
        
  \bibitem{PPF12} {\sc Pellat-Finet (P.)}, Quaternionic representation and design of depolarizers, arxiv 2405.04196 (2024) 1--9.

    \bibitem{Pan1} {\sc Pancharatnam (S.)}, The propagation of light in absorbing biaxal crystals. I--Theoretical, {\its Proc. Ind. Acad. Sci. A}, {\bf 42} (1955) 86--109. 


    \bibitem{Pan2} {\sc Pancharatnam (S.)}, The propagation of light in absorbing biaxal crystals. II--Experimental, {\its Proc. Ind. Acad. Sci. A}, {\bf 42} (1955) 235--248. 

 \bibitem{Pan3} {\sc Pancharatnam (S.)}, General theory of interference and its applications. Part IV--Interference figures in absorbing biaxal crystals, {\its Proc. Ind. Acad. Sci. A}, {\bf 46} (1957) 1--18. 

\bibitem{Man} {\sc Manoukian (E.\ B.)}, On the reversal of the triangle inequality
      in Minkowski spacetime in relativity, {\its Eur. J. Phys.} {\bf 14} (1993) 43--44.

\bibitem{Bri} {\sc Brillouin (L.)}, {\its Relativity Reexamined}, Academic Press, New York, 1970.

  \bibitem{Cho} {\sc Choquet-Bruhat (Y.)}, {\its Introduction to General Relativity, Black Holes and Cosmology}, Oxford University Press, New York, 2015.

  \bibitem{Sch} {\sc Schutz (B.)}, {\its A First Course in  General Relativity}, Cambridge University Press, Cambridge, 1985.

}
\end{thebibliography}
\end{document}